\documentclass[twocolumn]{aastex62}

\usepackage{newtxtext,newtxmath}

\usepackage[T1]{fontenc}
\usepackage{ae,aecompl}

\usepackage{graphicx}	
\usepackage{epstopdf}
\usepackage{amsmath}	
\usepackage{amssymb}	
\usepackage{hyperref}
\usepackage{longtable}

\newcommand{\kd}{KMOS$^{\rm 3D}$\,}
\newcommand{\sig}{$\sigma_0$\,}

\begin{document}

\title[]{The Evolution and Origin of Ionized Gas Velocity Dispersion from $z\sim2.6$ to $z\sim0.6$ with \kd \footnote{Based on observations collected at the Very Large Telescope (VLT) of the European Southern Observatory (ESO), Paranal, Chile, under ESO program IDs 092.A-0091, 093.A-0079, 094.A-0217, 095.A-0047, 096.A-0025, 097.A-0028, 098.A-0045, 099.A-0013, 0100.A-0039, and 0101.A-0022.} }
\shorttitle{Evolution of Gas Velocity Dispersion}

\author{H.~\"Ubler}\affiliation{Max-Planck-Institut f\"ur extraterrestrische Physik, Giessenbachstr.\ 1, D-85748 Garching, Germany}
\correspondingauthor{H.~\"Ubler}
\email{hannah@mpe.mpg.de}

\author{R.~Genzel}\affiliation{Max-Planck-Institut f\"ur extraterrestrische Physik, Giessenbachstr.\ 1, D-85748 Garching, Germany}\affiliation{Departments of Physics and Astronomy, University of California, Berkeley, CA 94720, USA}

\author{E.~Wisnioski}\affiliation{Research School of Astronomy and Astrophysics, Australian National University, Canberra, ACT 2611, Australia}\affiliation{ARC Centre of Excellence for All Sky Astrophysics in 3 Dimensions (ASTRO 3D)}

\author{N.~M.~Förster~Schreiber}\affiliation{Max-Planck-Institut f\"ur extraterrestrische Physik, Giessenbachstr.\ 1, D-85748 Garching, Germany}

\author{T.~T.~Shimizu}\affiliation{Max-Planck-Institut f\"ur extraterrestrische Physik, Giessenbachstr.\ 1, D-85748 Garching, Germany}

\author{S.~H.~Price}\affiliation{Max-Planck-Institut f\"ur extraterrestrische Physik, Giessenbachstr.\ 1, D-85748 Garching, Germany}

\author{L.~J.~Tacconi}\affiliation{Max-Planck-Institut f\"ur extraterrestrische Physik, Giessenbachstr.\ 1, D-85748 Garching, Germany}

\author{S.~Belli}\affiliation{Max-Planck-Institut f\"ur extraterrestrische Physik, Giessenbachstr.\ 1, D-85748 Garching, Germany}

\author{D.~J.~Wilman}\affiliation{Universit\"ats-Sternwarte Ludwig-Maximilians-Universit\"at M\"unchen, Scheinerstr.\ 1, D-81679 M\"unchen, Germany}\affiliation{Max-Planck-Institut f\"ur extraterrestrische Physik, Giessenbachstr.\ 1, D-85748 Garching, Germany}

\author{M.~Fossati}\affiliation{Institute for Computational Cosmology and Centre for Extragalactic Astronomy, Department of Physics, Durham University, South Road, Durham DH1 3LE, UK}

\author{J.~T.~Mendel}\affiliation{Research School of Astronomy and Astrophysics, Australian National University, Canberra, ACT 2611, Australia}\affiliation{ARC Centre of Excellence for All Sky Astrophysics in 3 Dimensions (ASTRO 3D)}

\author{R.~L.~Davies}\affiliation{Max-Planck-Institut f\"ur extraterrestrische Physik, Giessenbachstr.\ 1, D-85748 Garching, Germany}

\author{A.~Beifiori}\affiliation{Universit\"ats-Sternwarte Ludwig-Maximilians-Universit\"at M\"unchen, Scheinerstr.\ 1, D-81679 M\"unchen, Germany}\affiliation{Max-Planck-Institut f\"ur extraterrestrische Physik, Giessenbachstr.\ 1, D-85748 Garching, Germany}

\author{R.~Bender}\affiliation{Universit\"ats-Sternwarte Ludwig-Maximilians-Universit\"at M\"unchen, Scheinerstr.\ 1, D-81679 M\"unchen, Germany}\affiliation{Max-Planck-Institut f\"ur extraterrestrische Physik, Giessenbachstr.\ 1, D-85748 Garching, Germany}

\author{G.~B.~Brammer}\affiliation{Cosmic Dawn Center, Niels Bohr Institute, University of Copenhagen, Juliane Maries Vej 30, DK-2100 Copenhagen, Denmark}

\author{A.~Burkert}\affiliation{Universit\"ats-Sternwarte Ludwig-Maximilians-Universit\"at M\"unchen, Scheinerstr.\ 1, D-81679 M\"unchen, Germany}\affiliation{Max-Planck-Institut f\"ur extraterrestrische Physik, Giessenbachstr.\ 1, D-85748 Garching, Germany}

\author{J.~Chan}\affiliation{Department of Physics and Astronomy, University of California, Riverside, CA 92521, USA}

\author{R.~I.~Davies}\affiliation{Max-Planck-Institut f\"ur extraterrestrische Physik, Giessenbachstr.\ 1, D-85748 Garching, Germany}

\author{M.~Fabricius}\affiliation{Max-Planck-Institut f\"ur extraterrestrische Physik, Giessenbachstr.\ 1, D-85748 Garching, Germany}

\author{A.~Galametz}\affiliation{Observatoire de Genève, Université de Genève, 51 Ch.\ des Maillettes, CH-1290 Versoix, Switzerland}

\author{R.~Herrera-Camus}\affiliation{Departamento de Astronomia, Universidad de Concepción, Avenida Esteban Iturra s/n, Casilla 160-C, Concepción, Chile}\affiliation{Max-Planck-Institut f\"ur extraterrestrische Physik, Giessenbachstr.\ 1, D-85748 Garching, Germany}

\author{P.~Lang}\affiliation{Max-Planck-Institut f\"ur Astronomie, K\"onigstuhl 17, D-69117 Heidelberg, Germany}

\author{D.~Lutz}\affiliation{Max-Planck-Institut f\"ur extraterrestrische Physik, Giessenbachstr.\ 1, D-85748 Garching, Germany}

\author{I.~G.~Momcheva}\affiliation{Space Telescope Science Institute, 3700 San Martin Drive, Baltimore, MD 21218, USA}

\author{T.~Naab}\affiliation{Max-Planck-Institut f\"ur Astrophysik, Karl-Schwarzschild-Str.\ 1, D-85748 Garching, Germany}

\author{E.~J.~Nelson}\affiliation{Institute for Theory and Computation, Harvard-Smithsonian Center for Astrophysics, 60 Garden St., MS 51 Cambridge, MA 02138, USA}

\author{R.~P.~Saglia}\affiliation{Max-Planck-Institut f\"ur extraterrestrische Physik, Giessenbachstr.\ 1, D-85748 Garching, Germany}\affiliation{Universit\"ats-Sternwarte Ludwig-Maximilians-Universit\"at M\"unchen, Scheinerstr.\ 1, D-81679 M\"unchen, Germany}

\author{K.~Tadaki}\affiliation{National Astronomical Observatory of Japan, 2-21-1 Osawa, Mitaka, Tokyo 181-8588, Japan}

\author{P.~G.~van~Dokkum}\affiliation{Department of Astronomy, Yale University, New Haven, CT 06511, USA} 

\author{S.~Wuyts}\affiliation{Department of Physics, University of Bath, Claverton Down, Bath, BA2 7AY, UK}


\begin{abstract}
	We present the $0.6<z<2.6$ evolution of the ionized gas velocity dispersion in 175 star-forming disk galaxies based on data from the full \kd integral field spectroscopic survey.
	In a forward-modelling Bayesian framework including instrumental effects and beam-smearing, we fit simultaneously the observed galaxy velocity and velocity dispersion along the kinematic major axis to derive the intrinsic velocity dispersion $\sigma_0$.
	We find a reduction of the average intrinsic velocity dispersion of disk galaxies as a function of cosmic time, from $\sigma_0\sim45$~km~s$^{-1}$ at $z\sim2.3$ to $\sigma_0\sim30$~km~s$^{-1}$ at $z\sim0.9$. 
	There is substantial intrinsic scatter ($\sigma_{\sigma_0, {\rm int}}\approx10$~km~s$^{-1}$) around the best-fit $\sigma_0-z$-relation beyond what can be accounted for from the typical measurement uncertainties ($\delta\sigma_0\approx12$~km~s$^{-1}$), independent of other identifiable galaxy parameters. This potentially suggests a dynamic mechanism such as minor mergers or variation in accretion being responsible for the scatter. 
	Putting our data into the broader literature context, we find that ionized and atomic+molecular velocity dispersions evolve similarly with redshift, with the ionized gas dispersion being $\sim10-15$~km s$^{-1}$ higher on average.
	We investigate the physical driver of the on average elevated velocity dispersions at higher redshift, and find that our galaxies are at most marginally Toomre-stable, suggesting that their turbulent velocities are powered by gravitational instabilities, while stellar feedback as a driver alone is insufficient. This picture is supported through comparison with a state-of-the-art analytical model of galaxy evolution.
\end{abstract}




\section{Introduction}\label{intro}

Extragalactic surveys over the last decades have produced thousands of spectrally and spatially resolved observations of galaxies from the present day out to $z\sim4$. For massive galaxies on the star-forming main sequence, these efforts resulted in two main findings regarding their kinematic evolution: (i) already by $z\sim2$, the majority of star-forming galaxies (SFGs) show ordered rotation, and (ii) their velocity dispersions are higher by factors of 2-5 compared to local SFGs (\citealp{Labbe03, FS06b, FS09, FS18, Genzel06, Genzel08, Genzel14b, Cresci09, Epinat09, Epinat12, Law09, Jones10, Gnerucci11, Wisnioski11, Wisnioski15}, in prep.; \citealp{Miller12, Swinbank12, Stott16, Simons17}). The redshift evolution of the ionized gas velocity dispersion has captured a lot of attention through its potential to constrain feedback and star formation models \citep{FS06b, Genzel06, Genzel08, Genzel11, Weiner06a, Kassin07, Kassin12, Epinat09, Epinat12, Law09, Lehnert09, Lehnert13, Gnerucci11, Wisnioski12, Wisnioski15, Swinbank12b, Newman13, Simons16, Simons17, Turner17, Mason17, Zhou17, Johnson18, Girard18}. 

Starting from small scales in the Milky Way, the velocity dispersion in molecular clouds is proportional to cloud size and mass, in a way that suggests molecular clouds are turbulent, with kinetic and gravitational energy being in near equipartition \citep[][and references therein]{Larson81, McKee07, Heyer15}. However, the lack of dependence of the turbulence level on factors such as environment or local star formation activity points towards larger scale drivers (\citealp{Heyer04, Brunt09}; but see \citealp{Heyer15} for extreme environments). 

In nearby galaxies, velocity dispersions of atomic gas are $\sigma_{\rm HI}\approx10-12$~km s$^{-1}$ on scales of $\sim100$~pc \citep[][]{Dib06, Tamburro09, Ianjamasimanana12, Fukui09, CalduPrimo13, Mogotsi16, Koch19}. Molecular gas velocity dispersions are typically lower, with reported ratios in the range $\sigma_{\rm CO}/\sigma_{\rm HI}\approx0.3-1$ \citep[][]{Tamburro09, Ianjamasimanana12, Fukui09, Wong09, CalduPrimo13, Druard14, Mogotsi16, Levy18, Koch19}. Ionized gas velocity dispersions are substantially higher, with $\sigma_{\rm H\alpha}\approx24$~km s$^{-1}$ \citep{Epinat10}.

At high redshift, most measurements of gas velocity dispersion are based on ionized gas, which is accessible from the ground in the near-infrared through strong rest-frame optical lines. Typical values are $\sigma=25-100$~km s$^{-1}$ for disk galaxies. It is smore challenging to measure accurate velocity dispersions at high redshift because of the combined effects of beam-smearing and limited instrumental spectral resolution \citep[see][]{Davies11}. The former can be corrected for instance by using the velocity field and the spatial resolution of the observations to create a beam-smearing map \citep[e.g.][]{Green10, Gnerucci11, Epinat12}, through model-based look-up tables \citep[e.g.][]{Burkert16, Johnson18}, or through forward-modelling \citep[e.g.][]{Cresci09, Genzel11, DiTeodoro16, WuytsS16, Varidel19}. 
Typical spectral resolutions of near-infrared spectroscopic observations at $z\sim1-3$ correspond to velocity dispersions of $\sigma_{\rm instrumental}\approx30-40$~km s$^{-1}$. However, depending on the signal-to-noise ratio (S/N), it is possible to recover velocity dispersions through forward-modelling down to 1/3 of the instrumental resolution. 

It is well established that the galactic gas velocity dispersion is correlated with redshift \citep[e.g.\ review by][]{Glazebrook13}, but the physical processes responsible for driving and maintaining the dispersions are still debated.
It has been shown theoretically that constant energy input is necessary to maintain turbulence in the interstellar medium (ISM) because it will otherwise decay within a few Myr \citep[e.g.][]{MacLow98, Stone98}. A number of potential drivers has been identified, with two main classes: (i) the conversion of kinetic energy through stellar feedback in the form of winds, expanding H{\sc{ii}} regions, and supernovae, and (ii) the release of gravitational energy through clump formation, radial flows within the disk, and accretion from the cosmic web. Other possible sources include effects of galactic rotation, fluid instabilities, and galaxy interactions \citep[see][for a review]{ElmegreenB04}.
Generally, the different scales on which the proposed mechanisms operate present a challenge to simulations \citep[see][for a review]{Naab17}.

In this paper, we investigate the intrinsic velocity dispersion of the ionized gas phase in rotation-dimonated, star-forming galaxies from our \kd survey at $0.6<z<2.6$. 
In Section~\ref{data} we briefly present the \kd data set. 
Our modelling and sample selection is discussed in Section~\ref{modelling}.
In Section~\ref{redshift} we investigate the evolution of the intrinsic velocity dispersion with redshift and put it into the broader context of multi-phase literature values from $z=4$ to $z=0$. 
In Section~\ref{drivers} we discuss possible drivers of turbulence, particularly gravitational instabilities and stellar feedback, and compare our data to a state-of-the-art analytical model by \cite{Krumholz18}.
We conclude our study in Section~\ref{conclusion}.

Throughout, we adopt a \cite{Chabrier03} initial mass function and a flat $\Lambda$CDM cosmology with $H_{0}=70 {\rm \,km\, s^{-1}\, Mpc^{-1}}$, $\Omega_{\Lambda}=0.7$, and $\Omega_{m}=0.3$.\\

\section{The \kd survey}\label{data}

Our study is based on data from the \kd survey, targeting the H$\alpha$ line emission of primarily main sequence galaxies in three redshift bins centered at $z\sim0.9$, $z\sim1.5$, and $z\sim2.3$. The survey is presented by \cite{Wisnioski15} and Wisnioski et al., in prep., to which we refer the reader for details. Below, we summarize its main characteristics.

The \kd galaxies were selected from the 3D-HST survey \citep{Brammer12, Skelton14, Momcheva16}, providing optical-to-8$\mu$m photometry and, importantly, secure spectroscopic or grism redshifts, so that bright OH skylines at the location of the H$\alpha$ line emission could be avoided. In addition, high-resolution imaging for all galaxies is available through CANDELS \citep{Grogin11, Koekemoer11, vdWel12}, and further multi-wavelength coverage through photometry from {\it Spitzer}/MIPS and {\it Herschel}/PACS \citep[][and references therein]{Lutz11, Magnelli13, Whitaker14}.

For the \kd survey, we selected galaxies with stellar masses log$(M_*/M_{\odot})>9$ and $K_{\rm AB}\lesssim23$. 
The selection aimed to provide a homogeneous coverage of the star formation main sequence across stellar mass in the three redshift slices, thus ensuring near equal statistical coverage up to the highest masses. In addition, \kd also extends below the main sequence regime where galaxies are `quiescent', and it contains starburst outliers above the main sequence.

Stellar masses were derived following \cite{WuytsS11a}, by fitting the broad- and medium-band optical-to-mid-infrared spectral energy distribution with \cite{Bruzual03} stellar population synthesis models, adopting a \cite{Calzetti00} extinction law, solar metallicity, and a range of star formation histories. 
Gas mass measurements are not available for most of our galaxies. We exploit the scaling relation by \cite{Tacconi18} which depends on redshift, offset from the main sequence, and stellar mass, with the main sequence prescription by \cite{Whitaker14}, to estimate the molecular gas masses ($M_{\rm gas}$) of our galaxies. We don't account for atomic gas in this study. 
The derivation of star-formation rates (SFRs) followed the ladder of SFR indicators as described by \cite{WuytsS11a}. 

Structural properties such as the axis ratio $q=b/a$, the disk effective radius $R_e$, and the bulge-to-total stellar mass fraction $B/T$ are based on two-dimensional Sérsic models to the stellar light distribution high-resolution $H-$band images from {\it HST} observations \citep{vdWel12, Lang14}. For the effective radius we apply a color correction following \cite{vdWel14a}.

The survey was conducted during the years 2013 to 2018 with the multiplexing near-infrared integral field spectrograph KMOS \citep{Sharples04, Sharples13} at the {\it{Very Large Telescope}}.
The full \kd sample consists of 740 targeted galaxies (Wisnioski et al., in prep.).\\

\section{Dynamical Modelling and Sample Selection}\label{modelling}

We constrain the intrinsic velocity dispersions by forward-modelling the observed one-dimensional velocity and velocity dispersion profiles extracted from the data cubes.
For this work, we use the two-dimensional kinematic information to determine the kinematic major axis, and to distinguish rotation-dominated, dispersion-dominated, and disturbed systems. 
The full kinematic information on the motions of stars or gas in the plane of a rotating disk can be extracted along its kinematic major axis. Modelling the one-dimensional kinematics instead of the two- or three-dimensional data increases the S/N of our measurements, and thus allows us to study a larger sample of galaxies with reliable modelling.
We have verified that this has only a minor impact on the derived dynamical parameters, with an average, non-systematic difference of one-dimensional {\it vs.\ }two-dimensional intrinsic velocity dispersion of $\sim5-10$\%.\\

\subsection{One-Dimensional Kinematic Profiles}\label{extractions}

We derive two-dimensional projected H$\alpha$ velocity and velocity dispersion fields for all \kd galaxies using {\sc{linefit}} \citep{Davies09, Davies11, FS09}, a code that takes into account the instrument line spread function and fits a Gaussian model for each spaxel of the reduced data cube after continuum subtraction. 
From these maps we exclude spaxels with $S/N\leq2$, uncertainties on the velocity or velocity dispersion of $\geq100$~km~s$^{-1}$, as well as off-source fits to noise features. 
We determine the maximum and minimum of the velocity map through a weighted average of either the 5~\% of spaxels of both the highest and lowest velocity values for galaxies with $\geq50$ suitable spaxels, or otherwise of the five spaxels with highest and lowest velocities.
The kinematic major axis is defined as the line going through the maximum and minimum of the velocity map. 
The kinematic center is defined as the midpoint on the kinematics major axis connecting the maximum and minimum of the velocity map. 
This method follows the procedures outlined by \cite{Wisnioski15}, and in the \kd data release and final survey paper by Wisnioski et al., in prep. 

Along the kinematic major axis, we then extract spectra in circular apertures of diameter $2\times$FWHM of the model-independent point spread function (PSF) associated with each individual galaxy. 
Here, the flux from all spaxels within an aperture is integrated to create a single spectrum. For the dynamical modelling of our galaxies (see Section~\ref{dysmal}), we repeat this same procedure for each iteration of the model fitting to properly account for any effects related to this integration process.  
We consider a galaxy to be spatially resolved if we can measure its kinematics over a total of at least $3\times$PSF$_{\rm FWHM}$ along the kinematic major axis. 
We fit the H$\alpha$ velocity and velocity dispersion from the resulting spectra, providing us with the one-dimensional rotation curve $v_{\rm rot}(r)\cdot\sin(i)$ and dispersion profile $\sigma(r)$, uncorrected for beam-smearing. Uncertainties for each data point are derived using Monte Carlo analysis and have typical values of 6~km s$^{-1}$ and 10~km s$^{-1}$ for the velocity and dispersion values, respectively.

With this methodology we have successfully extracted kinematic profiles for all 535 \kd H$\alpha-$detected galaxies with secure redshifts.\\

\subsection{Dynamical Modelling with {\sc dysmal}}\label{dysmal}

As a next step, we consider all galaxies with kinematic profile extractions that are resolved, a total of 456 SFGs. We further exclude targets for which we detect multiple systems within the IFU, and we eliminate merging or potentially interacting systems with larger separations based on projected distances, redshift separations, and mass ratios, as informed through the 3D-HST catalog (Mendel et al., in prep.). Galaxies that are strongly contaminated by sky features, have prominent broad line regions, or have very strong outflows affecting the recovery of the galaxies' velocity and dispersion, are also excluded. This results in 356 galaxies.

We exploit the dynamical fitting code {\sc{dysmal}} \citep{Cresci09, Davies11, WuytsS16, Uebler18} to model our galaxies. 
{\sc{dysmal}} is a forward-modeling code that allows for a flexible number of components (disk, bulge, halo, etc.) and free parameters. It accounts consistently for finite scale heights and flattened spheroidal potentials \citep{Noordermeer08}, and it includes the effects of pressure support on the rotation velocity. It also accounts for the instrument line spread function, and for beam-smearing effects by convolving with the two-dimensional PSF of each galaxy.

For our modelling, we assume a velocity dispersion that is isotropic and constant throughout the disk, motivated by deep adaptive optics imaging spectroscopy on kpc scales of 35 $z=1-2.6$ SFGs in the SINS/zC-SINF sample \citep[][see also Section~\ref{feedback}]{Genzel06, Genzel08, Genzel11, Genzel17, Cresci09, FS18}. 
We note that for nearby galaxies radially declining velocity dispersions have been observed for atomic and molecular gas \citep{vdKruit84, Dickey90, Boulanger92, Kamphuis93, Meurer96, Petric07, Tamburro09, Wilson11, CalduPrimo13, Mogotsi16, Sun18, Koch19}, where the velocity dispersion usually reaches a constant level only in the disk outskirts. The observed radial changes in velocity dispersion are however rarely larger than $10-20$~km s$^{-1}$, and such variations on small scales are likely washed out through the coarser spatial resolution of typical high$-z$ observations (but see Section~\ref{feedback} for a high-resolution example).  

We create a three-dimensional mass model of each galaxy consisting of an exponential disk with the effective radius $R_e$ adopted from the $H-$band measurements, with ratio of scale height to scale length $q_0=0.25$, and with a central bulge ($R_{e,{\rm bulge}}=1$~kpc, S\'{e}rsic index $n_{\rm S,bulge}=4$, e.g.\ \citealp{Lang14, Tacchella15a}). 
The value of $q_0=0.25$ is motivated by the fall-off in the $q=b/a$ distribution of SFGs at the mass and redshift of our sample \citep{vdWel14b}.
For galaxies without an $H-$band based measurement of the bulge mass (see Section~\ref{data}; ca.\ 30\%) we use average values of $B/T=[0.25; 0.35; 0.45; 0.5]$ for total stellar masses of log$(M_{\star}/M_{\odot})=[<10.8;10.8-11;11-11.4;>11.4]$, following \cite{Lang14}.  
We fix the physical size of the bulge because individual measurements of $R_{e,{\rm bulge}}$ are very uncertain, in contrast to measurements of $B/T$ \citep[see][]{Tacchella15a}. In a population-averaged sense, however, $R_{e,{\rm bulge}}=1$~kpc is a robust choice \citep[see][]{Lang14}. 
We calculate the galaxy inclination $i$ as $\cos(i)=[(q^2-q_0^2)/(1-q_0^2)]^{1/2}$. The mass model is then rotated to match the galaxy's observed orientation in space, convolved with the line spread function and the PSF of the observation to take into account beam-smearing, and subsequently pixelated to resemble the spatial sampling of the observation. We approximate the PSF as a two-dimensional Moffat function that has been fitted to the standard star observations associated with each KMOS detector and pointing. 
For our modelling, we assume that light traces mass.

Using {\sc{dysmal}}, we simultaneously fit the one-dimensional velocity and velocity dispersion profiles of our galaxies in observed space. The best-fitting intrinsic rotation velocity, $v_{\rm rot}$, is constrained both through the mass model and the intrinsic velocity dispersion via pressure support. 
We apply Markov chain Monte Carlo (MCMC) sampling to determine the model likelihood based on comparison to the observed one-dimensional kinematic profiles, and assuming Gaussian measurement noise. 
To ensure convergence of the MCMC chains, we model each galaxy with 400 walkers, a burn-in phase of 50-100 steps, followed by a running phase of another 50-100 steps (>10 times the maximum autocorrelation time of the individual parameters). For each free parameter, we adopt the median of all model realizations as our best fit value, with asymmetric uncertainties corresponding to the 1$\sigma$ confidence ranges of the one-dimensional marginalized posterior distributions.

In order to recover the intrinsic velocity dispersion as best as possible, we consider a total of three setups with varying free parameters and treatment of the kinematic profiles: 

\begin{enumerate}

	\item In our first setup, we feed the kinematic profiles obtained as described in Section~\ref{extractions}, with free parameters being the total dynamical mass in the range log($M_{\rm tot}/M_{\odot})=[9; 13]$, and the intrinsic velocity dispersion in the range $\sigma_0=[5; 300]$~km s$^{-1}$. $M_{\rm tot}$ is the total mass distributed in the three-dimensional disk plus bulge structure necessary to reproduce the observed kinematics. Other parameters are fixed, specifically $i$, $R_e$, and $B/T$.

	\item Due to extinction, skyline contamination, and noise limitations, some galaxies display asymmetric kinematic profiles. 
Therefore, we employ a symmetrization technique in a second setup, where the one-dimensional profiles are folded (for the dispersion profile) or rotated (for the rotation curve) around the kinematic centre, interpolated onto a common grid, and averaged by calculating the mean at each radial grid point to obtain symmetric profiles, with uncertainties added in quadrature. 
Again, free parameters are $M_{\rm tot}$ and $\sigma_0$, allowed to vary within the same ranges as for setup~1. 

	\item As noted in Section~\ref{data}, $R_e$ and $B/T$ of our galaxies are derived from $H-$band imaging. It is known that the mass distribution derived from the $H-$band light might differ from the corresponding  H$\alpha$ flux profiles (\citealp[e.g.][]{WuytsS12, Tacchella15, NelsonE16b}; Wilman et al., in prep.). In particular the dispersion profiles can be sensitive to the central mass concentration. 
In the third setup we therefore proceed as in setup~2, but additionally leave the disk effective radius $R_e$ and the bulge-to-total fraction $B/T$ as free parameters. 
For $R_e$ we use a truncated Gaussian prior centered on the fiducial value with a standard deviation of 1~kpc, and truncated at $\pm2.5$~kpc from the peak value, with hard bounds of $R_e=[0.1; 20]$~kpc. 
For $B/T$ we use a Gaussian prior centered on the fiducial value with a standard deviation of 0.2 and hard bounds of $B/T=[0; 1]$.

\end{enumerate}

\begin{deluxetable}{lccc}
 \tablecaption{Comparison of modelling results from the three setups (S1, S2, S3) described in Section~\ref{dysmal}.
 \label{tab:setups}}
\tablehead{ \colhead{comparison} & \colhead{quantity} & \colhead{mean} & \colhead{std.\ dev.} }   
\startdata
	S1 -- S2 & $\Delta\sigma_0$ [km s$^{-1}$] & 0.9 & 6.0 \\
	 & $\Delta$log($M_{\rm tot})$ [dex of $M_{\odot}$] & -0.01 & 0.03 \\
	S1 -- S3 & $\Delta\sigma_0$ [km s$^{-1}$] & 0.5 & 7.4 \\
	 & $\Delta$log($M_{\rm tot})$ [dex of $M_{\odot}$] & -0.06 & 0.11 \\
	S2 -- S3 & $\Delta\sigma_0$ [km s$^{-1}$] & -1.4 & 5.3 \\
	 & $\Delta$log($M_{\rm tot})$ [dex of $M_{\odot}$] & -0.04 & 0.10 \\ \hline
	S3: $H-$band -- H$\alpha$\tablenotemark{a} & $\Delta R_e$ [kpc] & -0.6 & 1.0 \\
	 & $\Delta B/T$ & 0.03 & 0.14 \\
\enddata
\tablenotetext{a}{Comparison of the fiducial $R_e$ and $B/T$ as derived from the stellar light $H-$band images (see Section~\ref{data}) to the modelling results from setup 3, where we fit for $R_e$ and $B/T$ as detailed in Section~\ref{dysmal}.}
\end{deluxetable}

Comparing results from the three setups, we generally find good agreement for both the derived intrinsic dispersions and the dynamical masses, as listed in Table~\ref{tab:setups}. For setup~3, the model-derived (mass/H$\alpha$) effective radii are systematically higher compared to the $H-$band measurements by $\sim0.6$~kpc. 
For the range of $R_e\approx2-10$~kpc and log$(M_*/M_{\odot})\approx9.2-11.5$ in our kinematic sample, this is agreement with the results by \cite{NelsonE16b} and Wilman et al., in prep., who find $R_{e,{\rm H\alpha}}/R_{e,H}\approx1.1-1.2$ from high-resolution {\it HST} observations and from our full \kd sample, respectively. The average agreement between the $H-$band-derived $B/T$ and the model-derived $B/T$ is better, however the model-derived value is likely more realistic for cases with only a grid-based $B/T$.

We tested a fourth setup for a subset of our sample, including not only the bulge and disk components but in addition an NFW halo \citep{NFW96}, with a prior on the expected dark matter halo mass \citep{Moster18} and the concentration parameter fixed to the theoretically expected value \citep{Dutton14}. The resulting best-fit velocity dispersions are robust in that they agree within the uncertainties with the results from the other three setups with a standard deviation of 5.9~km s$^{-1}$, and there are no systematic effects. 
However, the limited field-of-view of KMOS (compared to e.g.\ SINFONI) together with our typical integration times of $5-9$~h per target constrain our ability to map the faint outskirts of galaxies where the kinematics are most sensitive to additional dynamical components with a different mass distribution. Therefore, we do not include fits from this fourth setup in our final sample.\\

\subsection{The Kinematic sample}\label{sample}

We inspect the fits from all three model setups to create our best-fit sample. By default, we choose the fit to setup 1, but if it is bad or poorly constrained, we consider setups 2 and 3 in this order. Galaxies with poor fits in all setups are excluded. With this strategy we stay as closely as possible to the original data, but at the same time do not need to disregard galaxies with one-sided extinction or skyline contamination that otherwise show good data quality, and we can choose fits from setup 3 with a more appropriate mass distribution, if necessary. 

Finally, we impose a $v_{\rm rot}/\sigma_0\geq1$ cut to focus on rotation-dominated systems. Here, $v_{\rm rot}$ is the model intrinsic rotation velocity at $1.38~R_e$, which is the location of the peak of the rotation curve for a Noordermeer disk with $n_{\rm S}=1$. 
Our final sample consists of 175 galaxies, with 80, 47, and 48 galaxies in the redshift slices $z=0.6-1.1$, $z=1.2-1.7$, and $z=1.9-2.6$. Of those galaxies, 56~\% are from setup 1, 31~\% from setup 2, and 13~\% from setup 3. We show examples of galaxies and their fits from different setup in Appendix~\ref{append}. 
The averaged uncertainties on our derived $\sigma_0$ values cover the range $\delta\sigma_0=2-29$~km s$^{-1}$, with $68^{\rm th}$ percentiles of $\delta\sigma_0=5-15$~km s$^{-1}$, and mean values in the three redshift slices $z\sim0.9; 1.5; 2.3$ of $\delta\sigma_0=8; 10; 13$~km s$^{-1}$. Asymmetric uncertainties can be as high as $\delta\sigma_0=37$~km s$^{-1}$.

\begin{figure}
	\centering
	\includegraphics[width=\columnwidth]{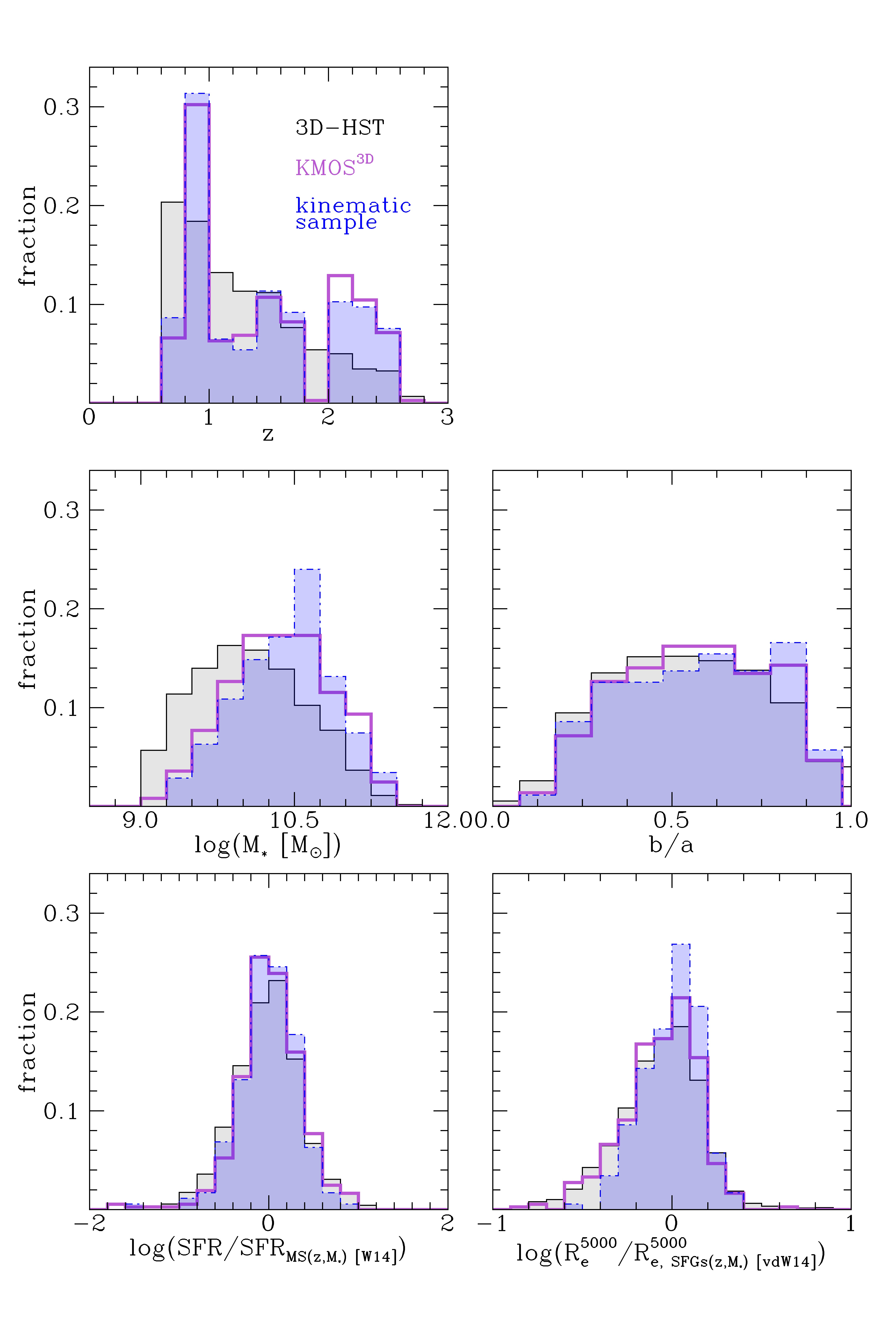}
    \caption{Distribution of physical properties of our kinematic sample (blue shading) compared to the full \kd survey (pink lines) and the underlying star-forming galaxy population at $0.6<z<2.7$ taken from the 3D-HST source catalog (grey shading) with log($M_*/M_{\odot})\geq9$, $K_{\rm AB}<23$~mag, and ${\rm SFR}/M_*>0.7/t_{\rm Hubble}$. We show redshift $z$ (top left), stellar mass (middle left), axis ratio $b/a$ (middle right), offset from the main sequence (bottom left), and offset from the mass-size relation (bottom right).
The SFR is normalized to the main sequence as derived by \cite{Whitaker14} at the redshift and stellar mass of each galaxy, using the redshift-interpolated parametrization by \cite{Wisnioski15}. The effective radii as measured from the $H-$band are corrected to the rest-frame 5000~{\AA} and normalized to the mass-size relation of SFGs as derived by \cite{vdWel14a} at the redshift and stellar mass of each galaxy.
    For our kinematic sample, there is no selection bias in redshift $z$, axis ratio $b/a$, or offset from the main sequence. Due to the $K_{\rm AB}<23$~mag cut for our \kd survey, \kd galaxies have higher stellar masses compared to the 3D-HST sample. Galaxies in our kinematic sample have on average larger sizes compared to all \kd galaxies as well as the 3D-HST sample. This is due to our conservative definition of resolved kinematics (see Section~\ref{extractions}).}
    \label{fig:parent}
\end{figure}

In Figure~\ref{fig:parent}, we compare physical properties of our final sample (blue shading) to the underlying representative population of star-forming galaxies from the 3D-HST survey (grey shading) and to the full \kd sample (pink lines).
Compared to our full \kd sample, we have not selected preferentially in redshift. In terms of stellar mass, both our full \kd sample and our kinematic sample include fewer lower mass systems compared to the 3D-HST galaxies, such that our sample is not mass-complete. This is mainly a consequence of the $K_{\rm AB}\lesssim23$ cut. 
With respect to the main sequence of star-forming galaxies, however, our kinematic sample follows the distribution of both the full \kd and the 3D-HST sample.
The fraction of systems with effective radii below the population average is smaller for our kinematic sample compared to the 3D-HST and \kd samples. This is due to our conservative definition of resolved kinematics, where we request measurements over at least $3\times{\rm PSF}_{\rm FWHM}$, with the primary effect of reducing the number of galaxies with $R_e<2$~kpc.
Generally, for very small systems it is more challenging to recover the intrinsic velocity dispersion, because the kinematics are often unresolved \citep[but see][for a detailed study of the kinematics of compact galaxies in the \kd survey]{Wisnioski18}.
Axis ratios of our galaxies are homogeneously distributed, following the \kd and 3D-HST parent samples (see also Section~\ref{inclination}).

\begin{figure}
	\centering
	\includegraphics[width=\columnwidth]{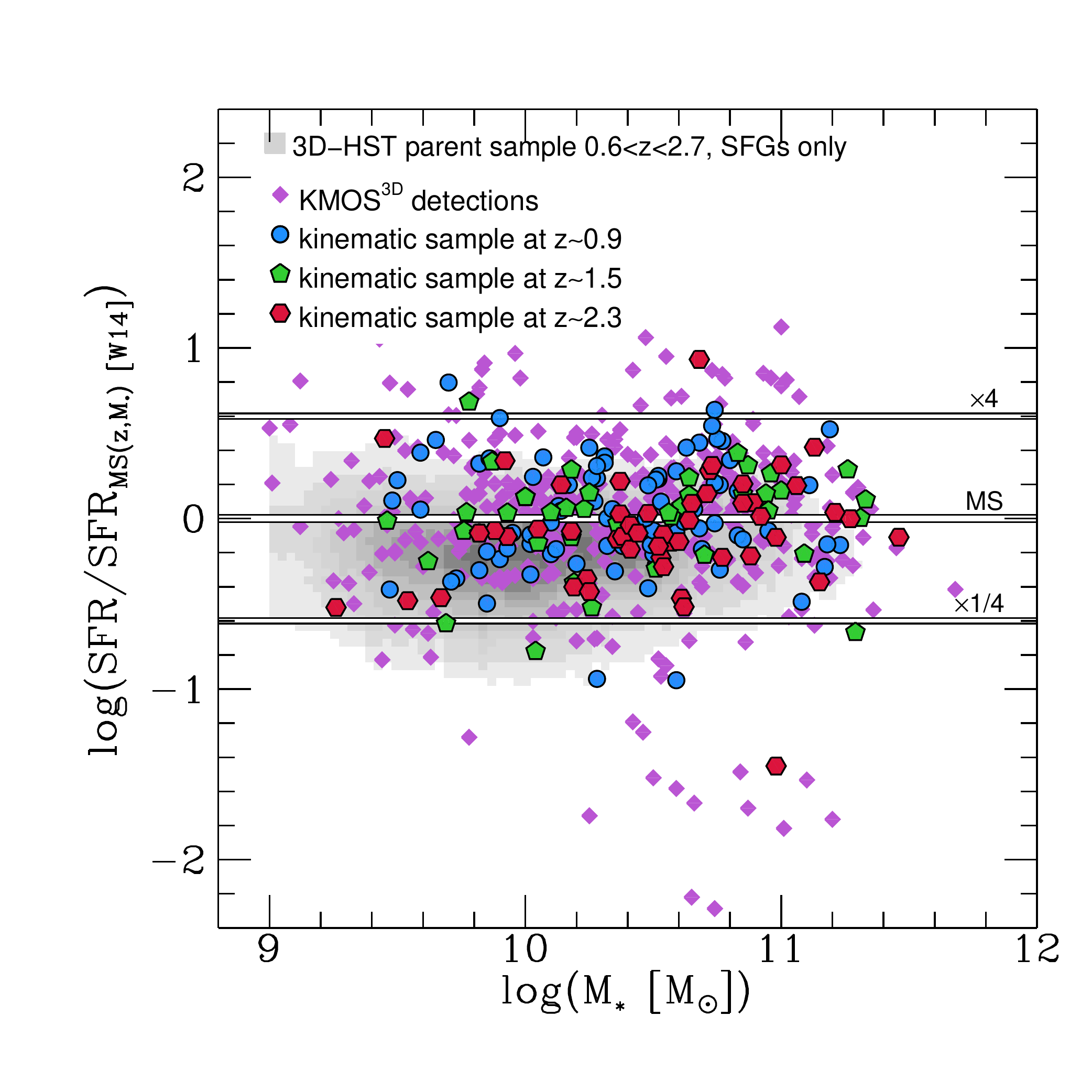}
	\includegraphics[width=\columnwidth]{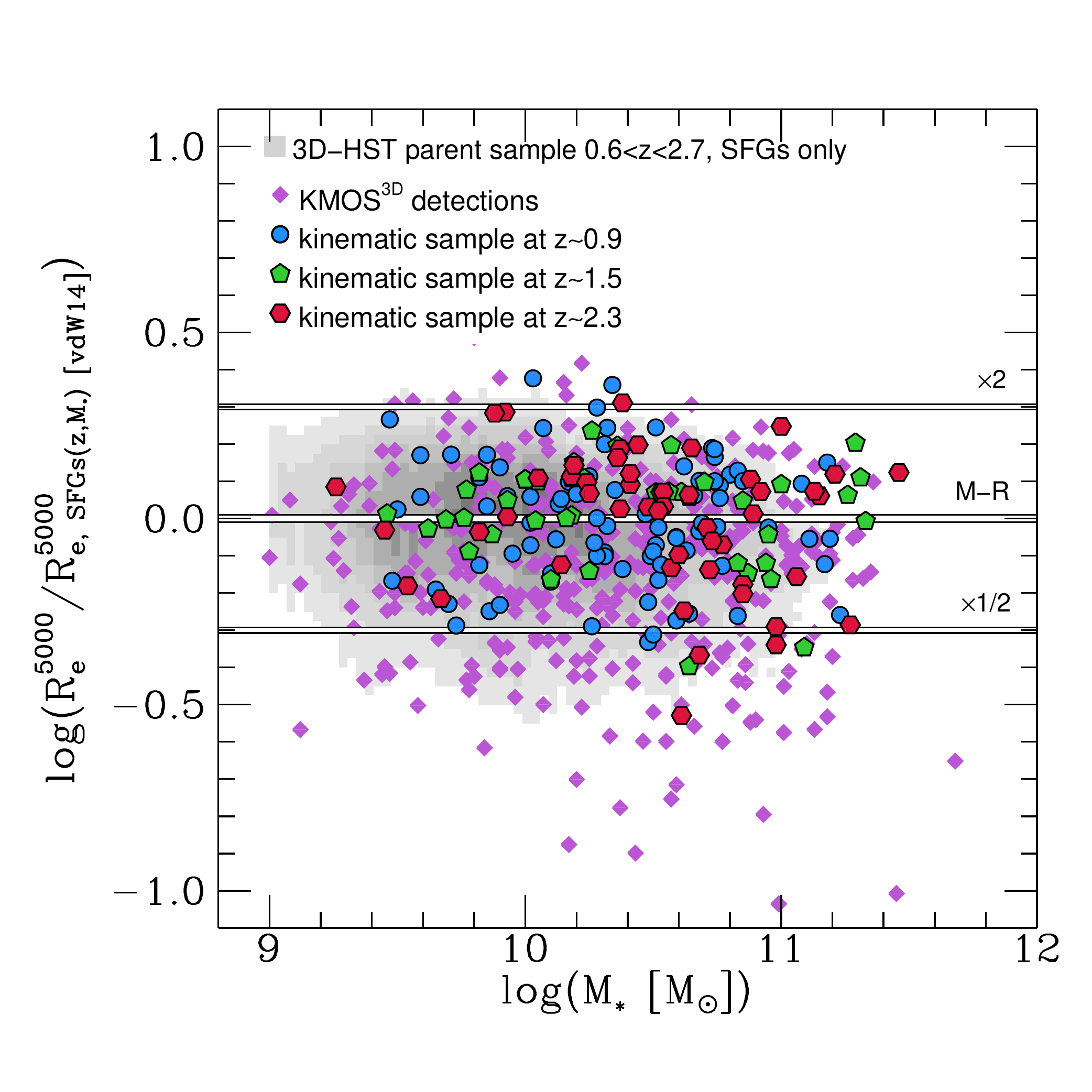}
    \caption{Location of our kinematic sample in the $M_*-$SFR (top) and $M_*-R_e$ (bottom) planes as compared to all detected \kd galaxies (pink diamonds) and the underlying star-forming galaxy population at $0.6<z<2.7$ taken from the 3D-HST source catalog (greyscale) with log($M_*/M_{\odot})\geq9$, $K_{\rm AB}<23$~mag, and ${\rm SFR}/M_*>0.7/t_{\rm Hubble}$. In the top panel, the SFR is normalized to the main sequence as derived by \cite{Whitaker14} at the redshift and stellar mass of each galaxy, using the redshift-interpolated parametrization by \cite{Wisnioski15}. In the bottom panel, the effective radii as measured from the $H-$band are corrected to the rest-frame 5000~{\AA} and normalized to the mass-size relation of SFGs as derived by \cite{vdWel14a} at the redshift and stellar mass of each galaxy.
    The galaxies in our kinematic sample are distributed along the main sequence, and have typical sizes for their redshifts. However, the size distribution of our targets is biased towards higher-than-average sizes, also compared to our \kd parent sample. This is introduced through our selecting only galaxies with resolved kinematics (see Section~\ref{extractions}).}
    \label{fig:msmr}
\end{figure}

In Figure~\ref{fig:msmr}, we show SFR (top) and size (bottom) both as a function of stellar mass for the 3D-HST parent sample  (grey density histogram), the full \kd sample (purple diamonds), and our final kinematic sample at redshifts $z\sim0.9$ (blue circles), $z\sim1.5$ (green pentagons), and $z\sim2.3$ (red hexagons). The figure illustrates the homogeneous coverage of the \kd survey of typical main sequence galaxies over more than two orders of magnitude in stellar mass. 
Similarly, the galaxies from our final sample are distributed along the main sequence and have typical sizes for their redshifts, with a tendency towards higher-than-average sizes particularly at $z\sim2.3$. This bias at the highest redshifts is introduced through our conservative definition of resolved galaxies, and ensures robust $\sigma_0$ measurements even at these high redshifts.\\

\subsection{Upper Limit Cases}\label{limits}

Our final sample contains 28 galaxies for which the best-fit \sig value within the 1$\sigma$ uncertainties is lower than 10~km s$^{-1}$. In using the H$\alpha$ line we are supposedly tracing emission originating from ionized H{\sc{ii}} regions. Due to thermal broadening ($\sigma_{\rm th}\approx10$~km s$^{-1}$) as well as the expansion of H{\sc{ii}} regions ($v_{\rm ex}\gtrsim10$~km s$^{-1}$), we expect some minimum velocity dispersion for the average galaxy of $\sigma_0\approx10-15$~km s$^{-1}$ \citep{Shields90}. 

This minimum value is lower than the typical spectral resolution of KMOS: the effective FWHM spectral resolution at the H$\alpha$ line measured from the reduced data of galaxies in our \kd survey is $\Delta R=\lambda/\Delta\lambda\sim3515; 3975; 3860$ in the $YJ$, $H$, and $K$ bands, respectively (Wisnioski et al., in prep.). 
For our kinematic sample, the corresponding mean spectral resolutions are $\sigma_{\rm instrumental}\sim37; 32; 34$~km s$^{-1}$. 
However, as discussed in more detail in Wisnioski et al., in prep., within the bands there are variations of the spectral resolution of up to $\Delta R=1000$ for individual IFUs. It is therefore crucial to measure the associated spectral resolution at H$\alpha$ for each individual galaxy from sky or arc lines in order to reliably recover the velocity dispersion, as it is done for KMOS$^{\rm 3D}$.

Our line fitting procedure can recover intrinsic velocity dispersions that are a fraction of the instrumental resolution. However, these measurements get increasingly uncertain for decreasing intrinsic velocity dispersions. For galaxies for which the best-fit \sig value within the 1$\sigma$ uncertainties is lower than 10~km s$^{-1}$, 
we adopt as a conservative upper limit the upper 2$\sigma$ boundary of the marginalized posterior distribution derived from the MCMC chain. The resulting upper limits lie between 18 and 53~km s$^{-1}$.\\

\subsection{Validation of Point Spread Function and Line Spread Function Corrections}\label{obsprop}

\begin{deluxetable}{lcc}
\tablewidth{0.9\columnwidth}
 \tablecaption{Spearman rank correlation coefficients, $\rho_{\rm S}$,  and their significance $\sigma_\rho$, between \sig and respectively $R_e$, $R_{\rm max}/R_e$, $\sigma_{\rm instrumental}$, and $b/a$.
 \label{tab:obspropcorr}}
\tablehead{ \colhead{quantity} & \colhead{\hspace{1cm}$\rho_{\rm S}$}\hspace{1cm} & \colhead{$\sigma_\rho$} }   
\startdata
	$R_e$ & 0.01 & 1.2 \\
	$R_{\rm max}/R_e$ & -0.05 & 0.7 \\
	$\sigma_{\rm instrumental}$ & -0.07 & 0.9 \\
	$b/a$ & -0.04 & 0.5 \\
\enddata
\end{deluxetable}

Before we investigate in detail the redshift evolution of \sig and its potential drivers, we want to exclude any residual effects of beam-smearing. Therefore we consider \sig as a function of the effective radius, $R_e$, and of the ratio of the outermost measured data point to the effective radius, $R_{\rm max}/R_e$. 

We do not find significant correlations with $R_e$ or $R_{\rm{max}}/R_e$, as listed in Table~\ref{tab:obspropcorr}  (for $R_e$ see also Figure~\ref{fig:sig0_trends}). 
We would expect correlations with these parameters if unresolved rotation enters our measure of velocity dispersion. As mentioned in Sections~\ref{extractions} and \ref{sample}, we only consider galaxies for our final sample for which we can extract kinematics over a distance of at least $3\times{\rm PSF}_{\rm FWHM}$, with a mean value of $4\times {\rm PSF}_{\rm FWHM}$. However, the extracted kinematics can still be affected by beam-smearing even in the outer parts of the galaxies. The fact that we do not find correlations with size implies that our forward-modelling procedure properly accounts for beam-smearing even for the smaller systems we include. 

Similarly, we test for correlations of \sig with instrumental resolution and again we do not find a significant correlation, indicating that both our kinematic fitting code and forward-modelling procedure properly account for the instrumental line-spread function (see Table~\ref{tab:obspropcorr}).\\

\subsection{Vertical {\it vs.\ }Radial Velocity Dispersion}\label{inclination}

\begin{figure}
	\centering
	\includegraphics[width=\columnwidth]{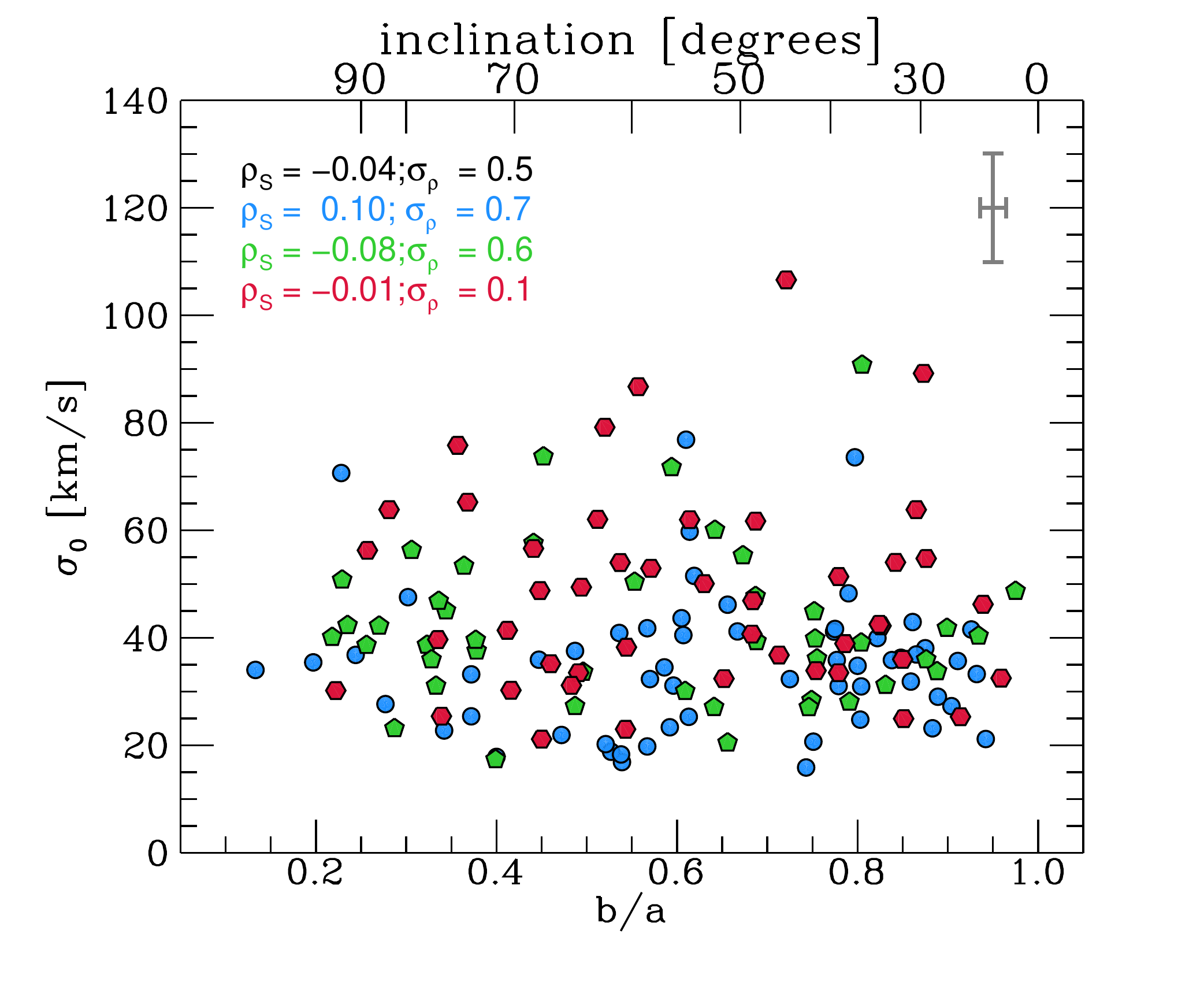}
    \caption{Intrinsic velocity dispersion $\sigma_0$ as a function of axis ratio $b/a$ as measured from the $H-$band.  Spearman rank correlation coefficients $\rho_{\rm S}$ and their significance $\sigma_\rho$ are given in the panel for the full sample (black) and the redshift slices at $z\sim 0.9$ (blue), $z\sim1.5$ (green), and $z\sim2.3$ (red). A typical error bar is shown in the top right corner. We do not find significant correlations between \sig and $b/a$ for the full sample nor the individual redshift bins.}
    \label{fig:inc}
\end{figure}

For local galaxies there exists a correlation between galaxy inclination and line-of-sight velocity dispersion. This is due to the transition from measuring predominantly vertical velocity dispersion in face-on systems to measuring predominantly radial velocity dispersion in edge-on systems, with a typical ratio of $\sigma_z/\sigma_r\sim0.6$ \citep{vdKruit11, Glazebrook13}. For instance, \cite{Leroy08} find for the THINGS sample that the H{\sc i} line-of-sight velocity dispersion increases for galaxies with $i>60^{\circ}$ ($b/a<0.5$), as does the variation of velocity dispersion in individual galaxies. 
Intriguingly, evidence for higher velocity dispersions in more edge-on systems has been found in the higher resolution $z\sim1-2$ data from the SINS survey \citep{Genzel11}.
For our \kd kinematic sample, and in agreement with the earlier results by \cite{Wisnioski15}, we do not find a correlation between $\sigma_0$ and $b/a$, as shown in Figure~\ref{fig:inc} and Table~\ref{tab:obspropcorr}, possibly due to the coarser spatial resolution of our data.\\

\section{Velocity Dispersion Increases with Redshift}\label{redshift}

Previous studies have shown that the velocity dispersion of star-forming galaxies increases with  redshift \citep{FS06b, Genzel06, Genzel11, Weiner06a, Kassin07, Kassin12, Wisnioski12, Wisnioski15, Newman13, Simons16, Simons17, Mason17, Turner17, Johnson18}, albeit with large uncertainties and scatter. 
In the following, we confirm and increase the robustness of this conclusion with the highest quality IFU data now available with \kd on sub-galactic scales, over a wider redshift and mass range than previously, and using a sample purely selected on the basis of disk galaxies near the main sequence at each redshift.
We further put our results into the broader literature context, including multi-phase gas velocity dispersion and expanding the redshift range to $0<z<4$.\\

\begin{figure*}
	\centering
	\includegraphics[width=0.8\textwidth]{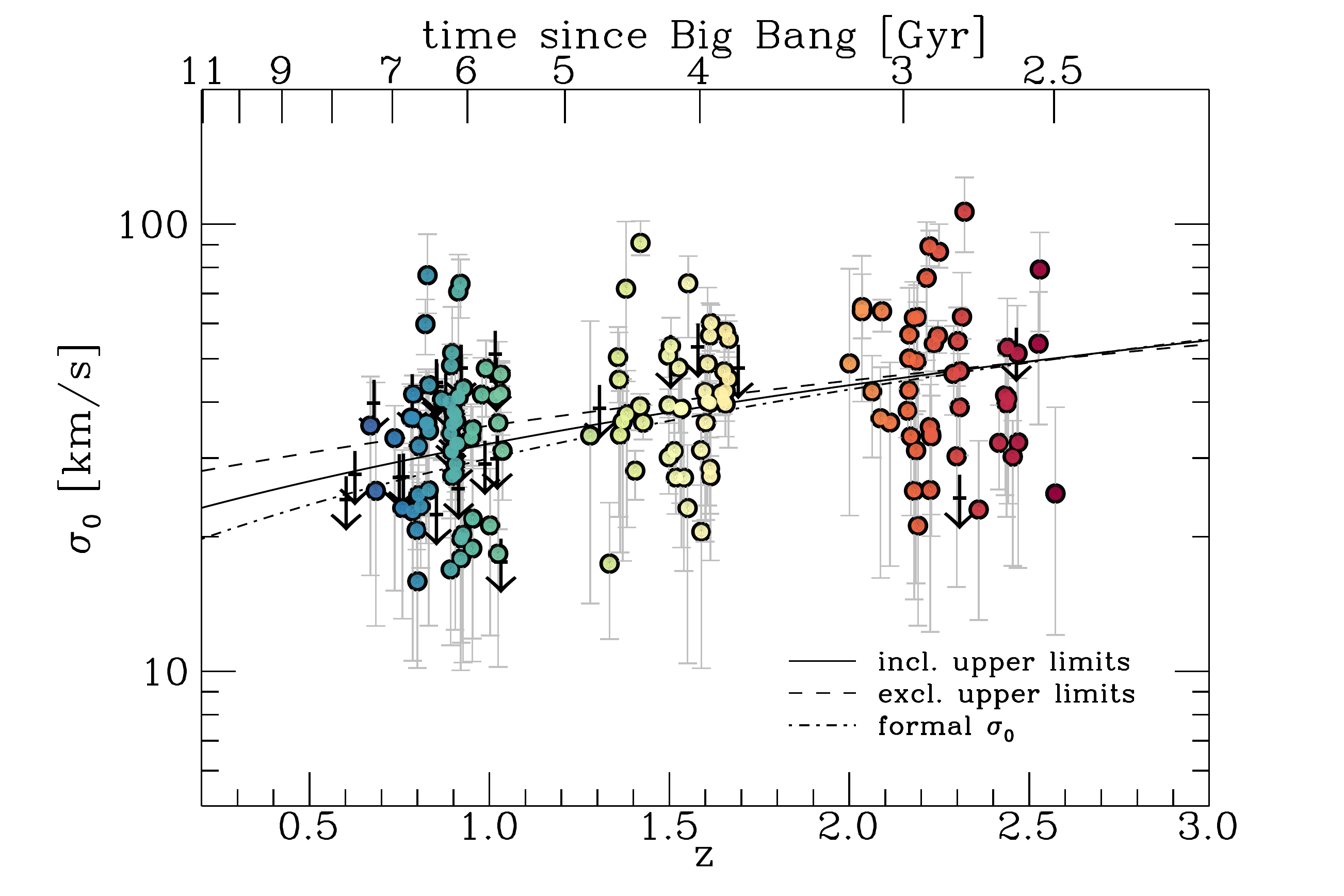}
    \caption{Intrinsic velocity dispersion $\sigma_0$ as a function of redshift and cosmic time for our kinematic sample, color-coded by redshift. Upper limits are shown as black arrows. On average, $\sigma_0$ increases with redshift, but the scatter at fixed redshift is large. The solid line shows the linear regression including the upper limits. The dashed line shows a corresponding fit for which the upper limit cases have been entirely excluded, resulting in a slightly shallower evolution. Taking the formal fit results for all galaxies at face value, we find a slightly steeper evolution (dash-dotted line).}
    \label{fig:zev}
\end{figure*}

\subsection{The \kd Velocity Dispersions from $z=2.6$ to $z=0.6$}\label{k3d}

\begin{deluxetable*}{llcccccccc}
 \tablecaption{Results from the linear regression fits of the form $\sigma_0/{\rm km\,s}=a+b\cdot z + c$ for our kinematic sample, where $a$ and $b$ are the regression coefficients, and $c$ is the intrinsic random scatter about the regression \citep[see][]{Kelly07}. For each parameter $a$, $b$, the standard deviation of $c$, and the derived linear correlation coefficient $l_{\rm corr}$ between $\sigma_0$ and $z$, we list the median together with the standard deviation of the posterior distribution. For each redshift slice we list the best-fit $\sigma_0$ value corresponding to these medians. \label{tab:bestfit} }
 \tablehead{ 
 \colhead{sample} & \colhead{N} & \colhead{$a$ } & \colhead{$b$} & \colhead{$\sigma_c$} & \colhead{$l_{\rm corr}$} & \colhead{$\sigma_0$ at $z\sim0.9$} & \colhead{$\sigma_0$ at $z\sim1.5$} & \colhead{$\sigma_0$ at $z\sim2.3$} \\
 \colhead{} & \colhead{} & \colhead{ } & \colhead{ } & \colhead{ } & \colhead{} & \colhead{[km s$^{-1}$]} & \colhead{[km s$^{-1}$]} & \colhead{[km s$^{-1}$]}
 }
 \startdata
  including upper & 175 & $21.1\pm3.0$ & $11.3\pm2.0$ & $11.3\pm1.1$ & $0.51\pm0.08$ & 31.1 & 38.3 & 46.7 \\
  \,\, limits & & & & & & & & \\
  excluding upper       & 147 & $26.2\pm3.1$ & $9.2\pm2.1$ & $10.4\pm1.1$ & $0.46\pm0.09$ & 34.3 & 40.3 & 47.1 \\
  \,\, limits (robust) & & & & & & & & \\
  using formal        & 175 & $17.2\pm3.2$ & $12.7\pm2.2$ & $13.2\pm1.1$ & $0.49\pm0.07$ & 28.4 & 36.7 & 46.1 \\
  \,\, best-fit $\sigma_0$ & & & & & & & & \\
\enddata
\end{deluxetable*}

In Figure~\ref{fig:zev} we show the intrinsic velocity dispersion of our \kd galaxies in the kinematic sample as a function of redshift, where upper limits are indicated as arrows (Section~\ref{limits}). 
Our data reflect the known trend of increasing average velocity dispersions with increasing redshift. 

To quantify the evolution, we fit a linear relation in $\sigma_0-z$ space to our best-fit data.\footnote{Our results do not depend on this particular functional form, and we list fits in $\sigma_0-\log(1+z)$ and $\log(\sigma_0)-\log(1+z)$ space in Appendix~\ref{altfits}.}
We use the Bayesian approach to linear regression by \cite{Kelly07} which allows for the inclusion of censored data (i.e., upper limits). The routine requires symmetric uncertainties, which we calculate as the mean of the asymmetric uncertainties on $\sigma_0$ from our MCMC modelling.\footnote{
We assume an uncertainty on $z$ of five times the spectral resolution in each redshift bin, translating into (negligible) uncertainties of $\delta z\sim0.001-0.002$.}
Figure~\ref{fig:zev} shows the derived fit as a solid line, with average values of $\sigma_0\sim31.1; 38.3; 46.7$~km s$^{-1}$ at $z\sim0.9; 1.5; 2.3$. The corresponding fit is described by the equation

\begin{equation}\label{eq:fit}
	\sigma_0\,[{\rm km s^{-1}}] = (21.1\pm3.0) + (11.3\pm2.0) \cdot z.
\end{equation}

We also perform a `robust' fit where the upper limit cases are not included, but entirely left out. We find a slightly shallower evolution indicated by the dashed line. If, instead, for these galaxies we do not assign upper limits but take the formal median of the posterior distribution at face value, we find a slightly steeper evolution indicated by the dash-dotted line. In Table~\ref{tab:bestfit} we list our fit parameters and uncertainties.

The $\sigma_0-$evolution we derive between $z\sim0.9$ and $z\sim2.3$ is slightly shallower than what has been reported for the first year of data from the \kd survey by \cite{Wisnioski15}. In particular, the authors cite $\sigma_0\sim24.9$~km s$^{-1}$ at $z\sim0.9$ and $\sigma_0\sim47.5$~km s$^{-1}$ at $z\sim2.3$, i.e.\ a difference of $6-7$~km s$^{-1}$ for the lowest redshift bin. We partly attribute this difference to our treatment of upper limits. Indeed, if we take the formal best-fit values of the upper limit cases at face value, we find through linear regression a value of $\sigma_0\sim28.4$~km s$^{-1}$ at $z\sim0.9$ (see Table~\ref{tab:bestfit}), reducing the difference to $\sim4$~km s$^{-1}$. 
This difference is smaller than the uncertainty on the average $\sigma_0$ value we derive through our fitting based on the standard deviation of the posterior distribution of the zero-point and slope, which is $\delta\sigma_0=4.8$~km s$^{-1}$ for the $z\sim0.9$ bin.

\subsection{Quantification of Observational Uncertainties and the Scatter in \sig}\label{scatter}

\begin{deluxetable}{lcccc}
\tablecaption{Variances of \sig around the robust best-fit relation: observed variance VAR$_{\rm obs}$, variance due to measurement uncertainties VAR$_{\delta\sigma_0}$, and intrinsic variance, VAR$_{\rm int}$. \label{tab:scatter} }
\tablehead{  \colhead{measure} & \colhead{$z\sim0.9$} & \colhead{$z\sim1.5$} & \colhead{$z\sim2.3$} & \colhead{$0.6<z<2.6$}} 
\startdata
 VAR$_{\rm obs}$ [km$^2$ s$^{-2}$] & 171 & 208 & 357 & 237 \\
 VAR$_{\delta\sigma_0}$ [km$^2$ s$^{-2}$] & 87 & 130 & 194 & 133 \\
 VAR$_{\rm int}$ [km$^2$ s$^{-2}$] & 85 & 78 & 163 & 104 \\
 VAR$_{\rm int}$/VAR$_{\rm obs}$ & 0.50 & 0.38 & 0.46 & 0.44 \\
  \enddata
\end{deluxetable}

Figure~\ref{fig:zev} shows substantial scatter in \sig at fixed redshift with values from $\sigma_0\approx20$~km s$^{-1}$ to $\sigma_0\approx100$~km s$^{-1}$. The question is whether this scatter is physical, or purely driven by observational uncertainties.

As listed in Table~\ref{tab:bestfit}, our robust best-fit relation has an intrinsic scatter around the regression with a standard deviation of 10.4~km s$^{-1}$, suggesting that part of the scatter is indeed due to real variations of the intrinsic dispersion values, and not just due to measurement uncertainties.
To quantify the intrinsic variance in each redshift slice, we first calculate the observed variance around the robust best-fit relation, i.e.\ the variance of the redshift-normalized dispersion values excluding upper limits. 
The redshift-normalized values are defined as
\begin{equation}\label{eq:norm}
	\sigma_{\rm 0, norm} = \sigma_0 - (a+b\cdot z),
\end{equation}
with coefficients $a$ and $b$ as listed in Table~\ref{tab:bestfit}. 
Then, we perform a Monte Carlo analysis of the scatter due to uncertainties: for each measurement $i$, we draw 1000 times from a normal distribution $\mathcal{N}(0, \delta\sigma_{0,i})$, where $\delta\sigma_{0,i}$ is the symmetric uncertainty of $\sigma_{0,i}$ derived from our {\sc dysmal} MCMC modelling, and calculate the corresponding sample variance per redshift slice. 

\begin{figure}
	\centering
	\includegraphics[width=\columnwidth]{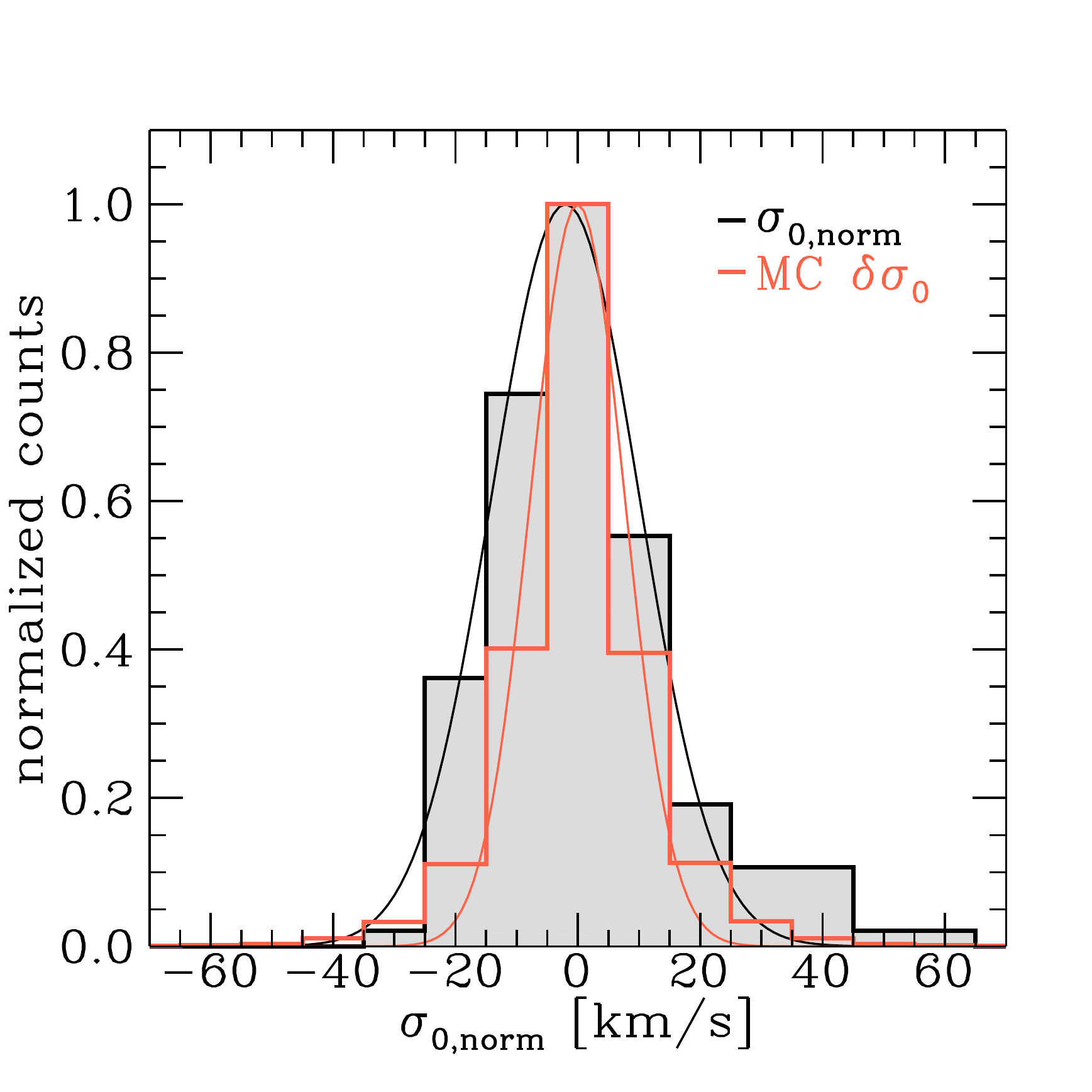}
    \caption{Histogram of redshift-normalized intrinsic dispersion values, $\sigma_{\rm 0, norm}$, in black, and histogram of the contribution to the scatter from uncertainties based on a Monte Carlo analysis in red (see Section~\ref{scatter} for details). To guide the eye we show simple Gaussian fits to the two distributions as thin curves. There is excess scatter beyond what can be accounted for by uncertainties in the distribution of $\sigma_{\rm 0, norm}$, indicating that we observe real physical variations of $\sigma_0$ at fixed redshift (see also Table~\ref{tab:scatter}).}
    \label{fig:var}
\end{figure}

We calculate this intrinsic variance as
\begin{equation}\label{eq:var}
	{\rm VAR}_{\rm int}(z) = {\rm VAR}_{\rm obs}(z) - {\rm VAR}_{\delta\sigma_0}(z),
\end{equation}
and list the corresponding values in Table~\ref{tab:scatter}. From this analysis we conclude that at least $\sim$40--50\% of the observed variance, i.e.\ $\sim$60--70\% of the observed scatter, is due to real variations of the intrinsic dispersion values, mostly independent of redshift. 
We also show a histogram of the redshift-normalized dispersion values in Figure~\ref{fig:var}, $\sigma_{\rm 0, norm}$, in black, together with a histogram of the Monte Carlo draws from the uncertainty distribution in red. Again, this clearly shows that, even though the uncertainties are substantial, there is residual scatter in our \sig distribution beyond what can be accounted for by uncertainties.
Further, if we focus on the absolute values listed in Table~\ref{tab:scatter}, the intrinsic variance increases above $z\sim1.5$ such that at $z\sim2.3$ it has doubled compare to $z\sim0.9$ and $z\sim1.5$.
This suggests that the population of galaxies in our highest redshift bin is more diverse in ISM conditions compared to the lower-redshift samples.

However, no significant residual trends with $\sigma_{0,{\rm norm}}$ and physical properties related to SFR, mass, size, or rotation velocity remain, as we show in detail in Figure~\ref{fig:normdisp_trends} in Appendix~\ref{trends}. 
That means that we cannot identify a physical source for the scatter in \sig at fixed redshift. This might be due to the limited dynamical range of our data, or it could imply that the intrinsic scatter is driven through the interplay of more than one parameter. 
Alternatively, the scatter could be due to real variations of the velocity dispersion on short timescales, for instance caused by a dynamic driver such as minor mergers or variations in gas accretion from the cosmic web. This has recently been proposed by \cite{Hung19} based on results from the FIRE simulations, where variations of intrinsic dispersion are connected to variations of the gas inflow rate on time scales $\lesssim100$~Myr.\\

\subsection{Comment on the Effect of Sample Selection}

The results presented above and in the remainder of the paper are based on our kinematic sample as defined in Section~\ref{sample}, i.e.\ 175 resolved and rotation-dominated disk galaxies that are well-fit by our dynamical model, without strong contamination from OH lines or outflows, and without close neighbours. 
If we instead consider all modelled galaxies from setup 1, about twice as many compared to the kinematic sample (see Section~\ref{dysmal}), we find a similar {\it median} evolution of $\sigma_0\approx31; 40; 49$~km s$^{-1}$ at $z\sim0.9; 1.5; 2.3$, however the {\it mean} values in the three redshift slices are systematically higher with $\sigma_0\approx34; 45; 58$~km s$^{-1}$. While at all redshifts the scatter is substantially increased due to galaxies with higher observational uncertainties or poor fits (VAR$_{\rm obs}\approx730; 850; 1560$~km$^2$ s$^{-2}$), the systematic increase of the mean values is mostly due to the inclusion of dispersion-dominated systems \citep[see e.g.][for a discussion of such galaxies]{Newman13}.

\subsection{Multi-phase Velocity Dispersions from $z=4$ to $z=0$}\label{literature}

\begin{figure*}
	\centering
	\includegraphics[width=0.8\textwidth]{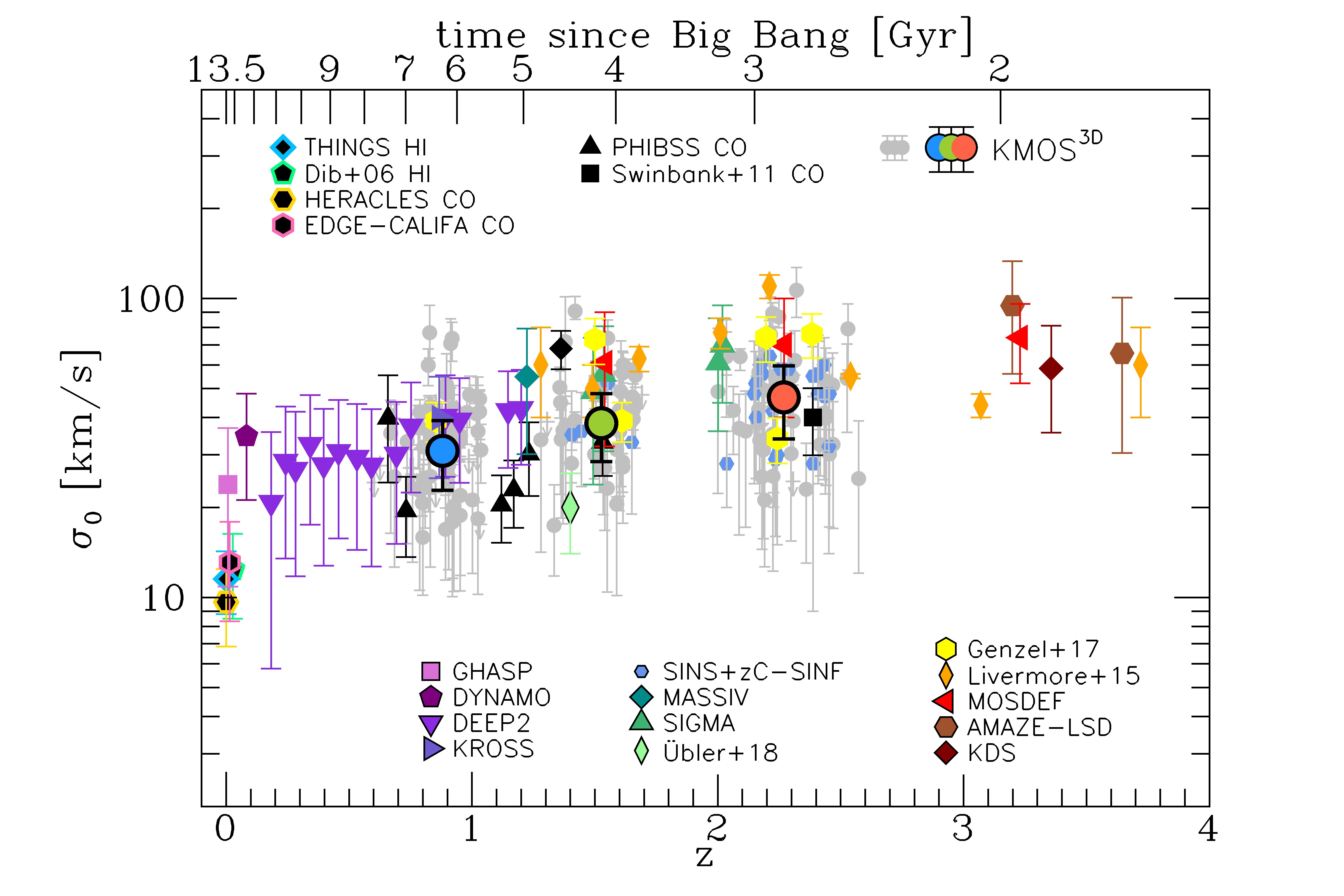}\vspace{0.4cm}
	\includegraphics[width=0.8\textwidth]{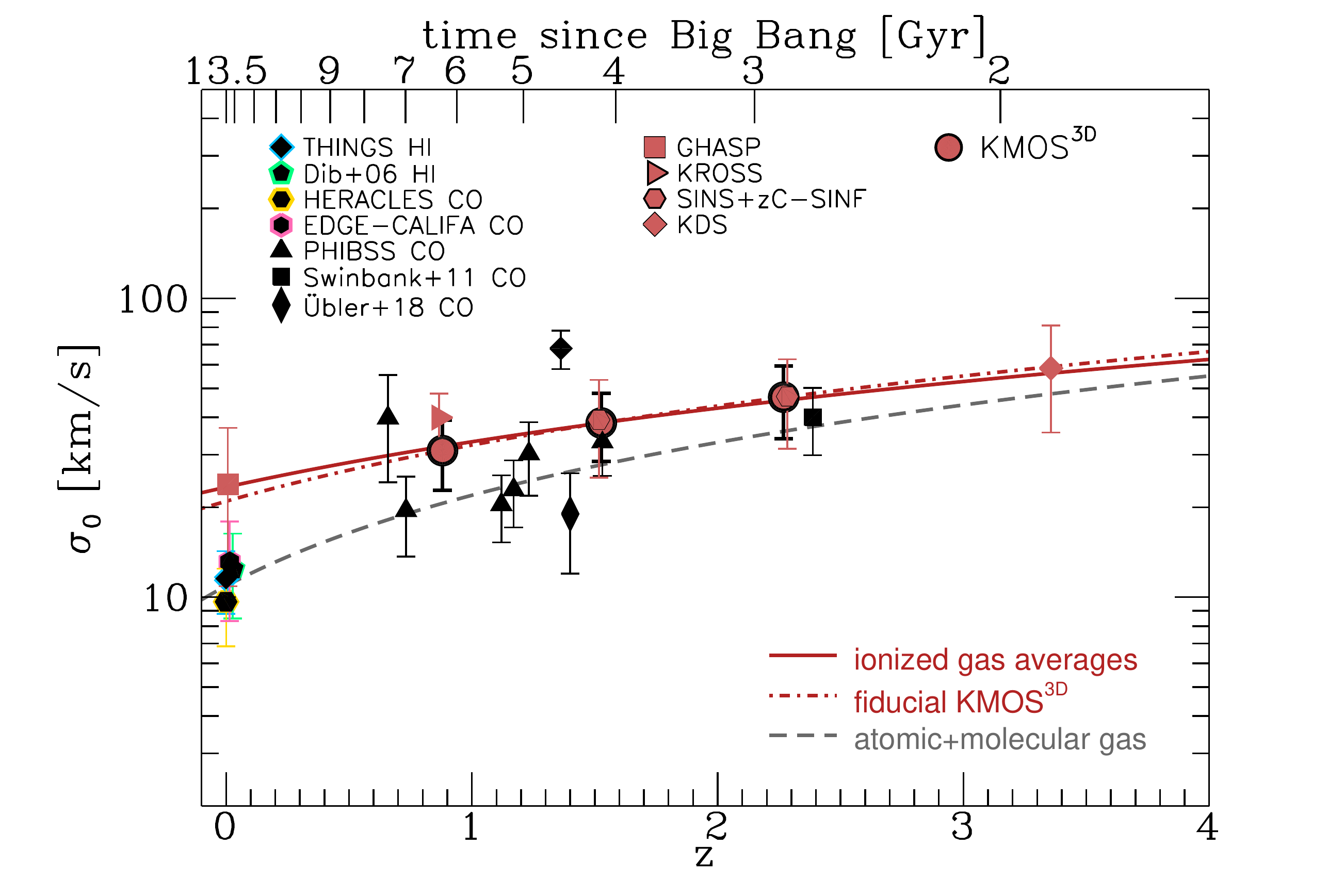}
 	 \caption{Intrinsic velocity dispersion $\sigma_0$ as a function of redshift and cosmic time for measurements from the literature at $0<z<4$ (see Table~\ref{tab:literature}). 
	 {\bf Top:} Our kinematic sample is shown in grey, with colored averages. Other individual and average ionized gas measurements are shown in color, as indicated in the legend. Molecular and atomic gas measurements are shown in black. For averages, the error bar shows the typical uncertainty of individual measurements in the sample.
	 {\bf Bottom:} Averages from selected ionized gas measurements are shown in red. Local atomic and molecular averages and individual high$-z$ molecular gas measurements are shown in black. Based on these data, we show best-fit relations (see Table~\ref{tab:bestfit_full}) for molecular gas (grey dashed) and ionized gas (red solid), as well as the best fit derived solely on our \kd data (red dash-dotted, see Section~\ref{k3d} and Table~\ref{tab:bestfit}).
	 Confirming the trend seen in our kinematic sample for the redshift range $0.6<z<2.6$, $\sigma_0$ increases with redshift over a time span of almost 12~Gyr. In the local Universe, velocity dispersions measured from molecular or atomic gas are lower than corresponding measurements from ionized gas, by ca.\ 10--15~km s$^{-1}$. The slopes derived from the molecular data and from our \kd sample are almost identical, suggesting an analogous redshift evolution of the different gas phase velocity dispersions.}
  	\label{fig:dispev_lit}
\end{figure*}

To put the evolution of velocity dispersion from $z=2.6$ to $z=0.6$ based on our \kd sample into a broader context, we collect measurements reported in the literature from $z\sim4$ to $z=0$, covering 12~Gyr of cosmic history (Table~\ref{tab:literature}).

In the top panel of Figure~\ref{fig:dispev_lit} we show again our \kd kinematic sample as clouds of grey circles, including upper limits as arrows, in the $\sigma_0-z$ space. The median values at $z\sim0.9; 1.5; 2.3$, shown as large circles in blue, green, and red, are based on the best fit plotted in Figure~\ref{fig:zev} and its uncertainties (see Table~\ref{tab:bestfit}). 
We include other individual intrinsic dispersion measurements or averages from ionized gas as colored symbols, and atomic and molecular data as black symbols, which are listed in Table~\ref{tab:literature}. 
Error bars show the mean uncertainty of individual systems in those samples. 
In our comparison, we do not apply any corrections or normalizations in mass \citep[cf.][]{Wisnioski15} which are expected to be small for main sequence galaxies \citep{Simons17}. 

In Table~\ref{tab:literature} we also list the different techniques used to correct for beam-smearing effects. As explained in Section~\ref{dysmal} and in the references listed there, we account for beam-smearing effects through a full forward-modelling of both the velocity and velocity dispersion fields with a unique PSF model for each galaxy. Techniques based on only the velocity information, or on grid-based models or look-up tables, might perform less well in their beam-smearing corrections generally resulting in overestimated intrinsic velocity dispersions. For slit surveys, systematic offsets towards higher values might be expected due to the sometimes uncertain galaxy position angle and the resulting difficulties in disentangling rotational and turbulent motions \citep[see][for a discussion and solution approach]{Price16, Price19}. Similarly, the methods chosen to calculate or model the intrinsic velocity dispersion might further introduce systematic differences. 
We note that recent work by \cite{Varidel19} on a sample of 20 local SFGs suggests that complex structure in the gas distribution may further impact the derived dispersion values.

Figure~\ref{fig:dispev_lit} shows generally good agreement of the various $\sigma_0$ measurements reported in the literature. 
Comparing slit {\it{vs.\ }}IFU techniques, the slit measurements shown here, i.e.\ data from DEEP2, SIGMA, and MOSDEF, have a tendency towards higher values as compared to the averages derived from our \kd and SINS/zC-SINF surveys, likely for the reasons discussed above, but agree within their uncertainty with the IFU data where available. 
Interestingly, the deep measurements obtained for individual targets by \cite{Genzel17}, and particularly for the lensed systems by \cite{Livermore15, Jones10} at $1.5<z<3$ also tend towards higher $\sigma_0$ values, but have moderate values at $z>3$ in agreement with the averages obtained from seeing-limited IFU and slit spectroscopy by \cite{Gnerucci11, Turner17, Price19}. 
Generally, the statistical power of these time-intensive and challenging individual measurements is still very limited.
Systematic differences in $\sigma_0$ may arise through selection effects: for instance, the nearby galaxies from the DYNAMO sample are selected to be $z\sim2$ analogues and have many physical properties, including dispersions, similar to high$-z$ SFGs \citep[see][]{Green14, White17, Fisher19}.

In contrast, the molecular and atomic data indicated by black points suggest somewhat lower values on average, particularly at $z\approx0$. \cite{Levy18} study 17 nearby, rotation-dominated SFGs in CO and ionized gas. They find consistently higher rotation velocities (<$v_{\rm CO}-v_{{\rm H}\alpha}$>$\approx14$~km s$^{-1}$) and lower velocity dispersions (<$\sigma_{\rm CO}-\sigma_{{\rm H}\gamma}$>$\approx-17$~km s$^{-1}$) for the molecular gas as compared to the ionized gas. 
At high redshift, there exist only few multi-phase measurements of the intrinsic gas velocity dispersion. Detailed observations reveal comparable values for ionized and molecular gas \citep{Genzel13, Uebler18}, however the uncertainties are larger such that differences like those found locally could be washed out.\\

\movetabledown=5cm
\begin{rotatetable*}
\begin{deluxetable*}{lcccccc}
 \tablecaption{ Literature data of the $0<z<4$ velocity dispersion measurements shown in Figure~\ref{fig:dispev_lit}.
 \label{tab:literature}
 }
  \tablehead{ 
  \colhead{Sample/References}  & \colhead{$z$} & \colhead{Primary Target Line} & \colhead{N included} & \colhead{Instrument/Method} & \colhead{BS correction} & \colhead{Comment on Selection} 
  }
\startdata
   KDS &  $3.8-3.1$ & [O{\sc iii}] &14 & KMOS/IFU & (1) & their `RD' \\
   MOSDEF  & $3.8-1.4$ & H$\alpha$, H$\beta$, [O{\sc iii}] & 108 & MOSFIRE/slit & (2) & their `resolved/aligned' \\
   AMAZE-LSD & $3.7-3.1$ & [O{\sc iii}] & 11 & SINFONI/IFU & (1) & their `rotating' \\
   \cite{Livermore15} & $3.7-1.3$ & H$\alpha$, H$\beta$ & 8 &  SINFONI+NIFS+OSIRIS/IFU+lensing & (3) & their `Disk' with $v_{\rm rot}/\sigma_0>1$ \\
   \kd & $2.6-0.6$ & H$\alpha$ & 175 & KMOS/IFU & (2) & see Section~\ref{sample} \\
   SINS/zC-SINF  & $2.5-1.4$ & H$\alpha$ & 25 & SINFONI/IFU+AO & (2) & see Section~\ref{sample} \\
   SIGMA & $2.5-1.3$ & H$\alpha$, [O{\sc iii}] & 49 & MOSFIRE/slit & (4) & \\
    \cite{Genzel17} & $2.4-0.9$ & H$\alpha$ & 6 & KMOS+SINFONI/IFU+AO & (2) & see Section~\ref{sample} \\
   MASSIV  & $1.6-0.9$ & H$\alpha$, [O{\sc iii}] & 53 & SINFONI/IFU & (1) & \\
   DEEP2 & $1.2-0.1$ & H$\alpha$, H$\beta$, [O{\sc ii}], [O{\sc iii}] & 544 & DEIMOS/slit & (4) & \\
   KROSS & $1.0-0.8$ & H$\alpha$ & 171 & KMOS/IFU & (4) & their `sigma0\textunderscore flag=O' with $v_{2.2}/\sigma_0>1$ \\
   DYNAMO & $\sim0.1$ & H$\alpha$ & 25 & SPIRAL+WiFeS/IFU & (1) & their `RD' \\
   GHASP & local & H$\alpha$ &153 & scanning Fabry-Perot & (1) & \\    
   \cite{Uebler18} & 1.4 & H$\alpha$, CO(3--2) & 1 & LUCI/slit+NOEMA/interferometry & (2) & see Section~\ref{sample}\\    
  \cite{Swinbank11}  & 2.4 & CO(6--5), CO(1--0) & 1 & IRAM+EVLA/interferometry  & (3) & \\
   PHIBSS & $1.5-0.7$ & CO(3--2) & 7 & IRAM+NOEMA/interferometry & (4) & see Section~\ref{sample} \\
   HERACLES & local & CO(2--1) & 13 & IRAM/single dish & (5) & \\
   EDGE-CALIFA & local & CO(1--0) & 17 & CARMA/interferometry & (1) & \\
   THINGS & local & H{\sc i} & 35 & VLA/interferometry & (5) & \\
   \cite{Dib06} & local & H{\sc i} & 13 & literature compilation &  & \\  
  \enddata
  \tablecomments{Beam-smearing (BS) correction methods: (1) correction map derived from the model velocity field including two-dimensional PSF; (2)  simultaneous forward-modelling of both the velocity and dispersion including two-dimensional PSF, cf.\ Section~\ref{dysmal}; (3) subtraction in quadrature of the velocity gradient across each spaxel; (4) model-based look-up grid; (5) exclusion of regions most strongly affected based on a model grid. 
  References: KDS \cite{Turner17}, MOSDEF \cite{Kriek15, Price19}, AMAZE-LSD \cite{Gnerucci11}, \kd \cite{Wisnioski15}; Wisnioski et al.\ (in prep.), SINS/zC-SINF \cite{FS06b, FS09, FS18}, SIGMA \cite{Simons16, Simons17}, MASSIV \cite{Contini12, Epinat12}, DEEP2  \cite{Davis03, Kassin07, Kassin12}, KROSS \cite{Stott16, Johnson18}, DYNAMO \cite{Green14}, GHASP \cite{Epinat08,Epinat10}, PHIBSS \cite{Tacconi13, Freundlich19}, HERACLES \cite{Leroy09, Tamburro09, CalduPrimo13, Mogotsi16}, EDGE-CALIFA \cite{Bolatto17, Levy18}, THINGS \cite{Walter08, Tamburro09, Ianjamasimanana12, CalduPrimo13, Mogotsi16}.}
\end{deluxetable*}
\end{rotatetable*}

\subsection{Multi-phase Gas Velocity Dispersions Evolve Similarly with Redshift}\label{multiphase}

We quantify the difference between the atomic+molecular and the ionized gas velocity dispersions over cosmic time in the bottom panel of Figure~\ref{fig:dispev_lit}. Fitting a robust linear relation\footnote{
We use the least trimmed squares method by \cite{Cappellari13}.} 
to the average local and individual high$-z$ measurements of atomic+molecular gas, we find a zero point of $a=10.9\pm0.6$~km s$^{-1}$ and a slope of $b=11.0\pm2.0$~km s$^{-1}$ (grey dashed line). 
For the ionized velocity dispersion, we choose in addition to our own averages from the \kd and SINS/zC-SINF surveys the other large KMOS surveys, KROSS and KDS, and the local average from the GHASP survey. This choice maximizes the redshift range and avoids systematic effects at $z>0$ through different instrumentation. We find a higher zero-point offset of $a=23.3\pm4.9$ and a somewhat shallower slope of $b=9.8\pm3.5$, while the extrapolation of our best fit to the \kd data only gives $a=21.1\pm3.0$ and $b=11.3\pm2.0$ (Table~\ref{tab:bestfit}). 
Fixing the slope to that of the atomic+molecular fit, the zero point shifts in between these measurements, with $a=22.8$. In Table~\ref{tab:bestfit_full} we list our fit parameters and uncertainties.

\begin{deluxetable}{lcc}
 \tablecaption{Results and standard deviations from the robust least-squares linear regression fits of the form $\sigma_0/{\rm km\,s}=a+b\cdot z$ to the data sets shown in the bottom panel of Figure~\ref{fig:dispev_lit}.
 \label{tab:bestfit_full} }
 \tablehead{ \colhead{sample} & \colhead{$a$ [km s$^{-1}$]} & \colhead{$b$ [km s$^{-1}$]} }
 \startdata
	ionized gas (best averages) & 23.3$\pm$4.9 & 9.8$\pm$3.5 \\
	... fixing slope to atomic+molecular & 22.8 & 11.0 (fixed) \\
	\kd incl.\ upper limits (Table~\ref{tab:bestfit}) & 21.1$\pm3.0$ & 11.3$\pm$2.0 \\
	atomic+molecular gas & 10.9$\pm$0.6 & 11.0$\pm$2.0 \\
 \enddata
\end{deluxetable}

This suggests that the redshift evolution of the intrinsic velocity dispersion in all gas phases is quite comparable, but their normalization differs. 
Typical thermal broadening of the atomic/molecular and the ionized gas due to their characteristic temperatures are $\sim5$~km s$^{-1}$ and $\sim10$~km s$^{-1}$, respectively, meaning the measured velocity dispersions are super-thermal even in the local Universe. 
Part of the difference between atomic+molecular and ionized gas velocity dispersions can be explained through the expansion of H{\sc{ii}} regions from which the ionized emission originates, with typical values of $10-25$~km s$^{-1}$ \citep{Shields90}, accounting for another $\sim5-15$~km s$^{-1}$ when added in quadrature. In combination, these effects can explain the difference in the local normalizations of the gas phase velocity dispersions, as well as their average offset of $\sim10-15$~km s$^{-1}$ at fixed redshift.

Clearly, more studies of high$-z$ molecular kinematics are warranted to corroborate our result, which potentially has important implications for work on ionized gas kinematics.\\

\subsection{Comments on Thin {\it vs.\ }Thick Disk Evolution}\label{thickthin}

Figure~\ref{fig:dispev_lit} shows a smooth evolution of velocity dispersion with redshift over the past $\sim12$~Gyr, likely connected to decreasing accretion rates and gas fractions with cosmic time (see Sections~\ref{gravity} and \ref{feedbackgravity}). 
This evolution suggests that also the typical thickness of the {\it young, star-forming gas disk} is lower for lower redshift SFGs, as has also been found in state-of-the-art cosmological simulations \citep{Pillepich19}. 

This is potentially interesting in the context of Galactic archeology: 
early research of the vertical structure of our Milky Way found evidence for two main, distinct exponential disks with scale heights of $\sim300$~pc and $\sim1450$~pc \citep{Gilmore83}. This was confirmed through later work on the Milky Way as well as nearby edge-on galaxies \citep[e.g.][]{Dalcanton02, Yoachim06, Juric08}. The thick disk components have been found to be generally older ($>6$~Gyr) than the thin disks, raising the question of distinct formation periods. 
Naturally, observations of the typically thick high$-z$ disks also prompted the question of the connection between these early thick disks and modern disk structure \citep[e.g.][]{ElmegreenB06}.

To explicitly address the question of distinct formation periods of thin {\it vs.\ }thick disks, we make the simple assumption of a step function describing \sig of the ionized gas below and above $z=1$. Unsurprisingly, the resulting fit with $\sigma_0=28$~km s$^{-1}$ at $z<1$ and $\sigma_0=42$~km s$^{-1}$ at $z>1$ is not a good description of the compiled data, with a goodness of fit that is a factor $\sim20$ worse compared to the linear fit shown in the bottom panel of Figure~\ref{fig:dispev_lit}. 

Our results suggest that in the absence of recent major mergers it should depend primarily on the star-formation history (connected to gas accretion) if present-day galaxies have distinct disks of different age and scale height, or if there is rather one component with a vertical age gradient \citep[see also][]{Leaman17}. 
This interpretation is in agreement with the recent work by \cite{Bovy12b, Bovy16, Rix13} who argue based on elemental abundances that the Milky Way has a continuous range of different scale heights, with no sign of a thin-thick disk bimodality. Simulations by e.g.\ \cite{Burkert92, Aumer16, Aumer17, Grand16} support this picture.

However, in this context it is important to remember that based on stellar and gas masses of our galaxies and results from co-moving number density studies \citep[e.g.][]{Brammer11}, only the lower mass, lower redshift systems in our sample may evolve into present-day disk galaxies, while the galaxies that have high baryonic masses already at high redshift will most likely evolve into present-day's early-type galaxies. With our data, we do therefore not necessarily track the change in star-forming scale height over time for progenitor-descendant populations, but rather the change in average star-forming scale height of main sequence galaxies at different epochs.\\

\section{What Drives the Gas Velocity Dispersion?}\label{drivers}

\subsection{Galaxy-scale Velocity Dispersion Correlates with Gas Mass and SFR Properties}\label{correlations}

The redshift dependence of \sig suggests that one or more physical galaxy properties that are themselves redshift-dependent drive velocity dispersion. Consistent with previous findings in the literature \citep[e.g.][]{Johnson18}, we find several properties positively correlating with $\sigma_0$, particularly, SFR, SFR surface density $\Sigma_{\rm SFR}$, gas and stellar mass, and their surface densities. 
We list direct and residual (after correcting for redshift dependence) Spearman rank correlations in Table~\ref{tab:physpropcorr} and show plots for several quantities in Figure~\ref{fig:sig0_trends} in Appendix~\ref{trends}. In Table~\ref{tab:physpropcorr}, we also list SFR$_{\rm H\alpha}$ and $\Sigma_{\rm SFR, H\alpha}$ derived from the H$\alpha$ fluxes (see Wisnioski et al., in prep.), tracing the more recent star formation history, but find no appreciable difference in correlations compared to our fiducial SFR properties (see Section~\ref{data}).

We emphasize that due to the limited dynamical range covered by the individual redshift slices, we do not find significant correlations of \sig within one redshift slice with any of the above properties, such that we cannot readily connect the scatter in \sig at fixed redshift to a physical driving source.
Similarly, if we remove the redshift dependence of \sig by normalizing with our best-fit relation, we do not find any significant correlations of the normalized \sig with physical properties (see Section~\ref{scatter} and Figure~\ref{fig:normdisp_trends}).

\begin{deluxetable}{lcccc}
 \tablecaption{Spearman rank correlation coefficients $\rho_{\rm S}$, and their significance $\sigma_\rho$, between $\sigma_0$ and different galaxy properties for our robust sample before and after accounting for the redshift dependence of \sig.
 \label{tab:physpropcorr}}
 \tablehead{ 
  & \multicolumn{2}{c}{$\sigma_0(z)$} & \multicolumn{2}{c}{$\sigma_{\rm 0, norm}$} \\
  \colhead{quantity} & \colhead{$\rho_{\rm S}$} & \colhead{$\sigma_\rho$} & \colhead{$\rho_{\rm S}$} &\colhead{ $\sigma_\rho$} }
 \startdata   
 	z & 0.33 & 4.0 & -- & -- \\
	SFR & 0.38 & 4.6 & 0.18 & 2.1 \\
	SFR$_{\rm H\alpha}$ & 0.36 & 4.4 & 0.14 & 1.7 \\
	$\Sigma_{\rm SFR}$ & 0.32 & 3.9 	& 0.06 & 1.0\\
	$\Sigma_{\rm SFR, H\alpha}$ & 0.30 & 3.7  & 0.08 & 0.9\\
	$M_{\rm gas}$ & 0.38 & 4.6 & 0.19 & 2.3\\
	$\Sigma_{{\rm gas}}$ & 0.31 & 3.8 	& 0.07 & 0.9 \\
	$M_{*}$ & 0.26 & 3.1 & 0.20 & 2.4 \\
	$\Sigma_{*}$ & 0.26 & 3.1 & 0.14 & 1.6 \\
	$M_{\rm bar}$ & 0.32 & 3.9 & 0.20 & 2.4\\
	$\Sigma_{{\rm bar}}$ & 0.30 & 3.6 & 0.12 & 1.5 \\	
	$\Delta$MS & -- & -- 	& 0.15 & 1.8 \\
	$\Delta$MR & -- & -- 	& -0.05 & 0.6 \\
 \enddata
\end{deluxetable}

Over the full redshift range covered by our \kd survey, SFR and gas mass correlate most strongly and significantly with intrinsic velocity dispersion. 
In order to identify which of these two physical quantities is most directly tied to the elevated velocity dispersions at high redshift, we discuss in the following sections the physical mechanisms through which quantities such as SFR and gas mass may affect velocity dispersion, namely stellar feedback and gravitational instabilities, and we comment on the tentative connection to AGN feedback for individual galaxies.\\

\subsection{Stellar Feedback}\label{feedback}

Turbulence-driving can be provided through thermal and momentum feedback from massive stars. 
Correlations between intrinsic velocity dispersion and SFR properties have previously been reported in the literature \citep[e.g.][]{Dib06, Lehnert09, Lehnert13, Green10, Green14, Moiseev15, Johnson18}, and often invoked the argument for stellar feedback-driven turbulence.

From a theoretical point of view, feedback-driven turbulence is mainly generated through momentum injection from supernovae into the ISM (contributions to the momentum injection from e.g.\ expanding H{\sc{ii}} regions or stellar winds are minor, see \citealp{MacLow04, OstrikerE11}). 
Feedback-driven turbulence should therefore primarily depend on the decay rate of turbulence, the momentum injected per supernova, and the supernova rate, where the latter is the quantity connecting turbulence to SFR and $\Sigma_{\rm SFR}$. 
\cite{OstrikerE11} and \cite{Shetty12} derive a weak dependence of $\sigma_0$ on star formation rate surface density. 
Even considering the case where feedback-driven turbulence vertically stabilizes the disk, the resulting velocity dispersions are low \citep[Equation (22) by][]{OstrikerE11}: 
\begin{equation}
	\sigma_z = 5.5~{\rm km s^{-1}} \cdot \frac{f_p}{(1+\chi)^{1/2}} \bigg(\frac{\epsilon_{\rm ff}(\rho_0)}{0.005}\bigg) \bigg(\frac{p_*/m_*}{3000~{\rm km s^{-1}}}\bigg).
\end{equation}
Here, $f_p$ is a factor characterizing the evolution of turbulence, with $f_p=1$ for strong dissipation, and $f_p=2$ for weak dissipation. 
$\chi$ is a measure of the importance of the gas disk's self-gravitational weight, and is below 0.5 for marginally stable disks, such that the first factor is in the range $\sim0.8-2$.
The mean star formation efficiency $\epsilon_{\rm ff}(\rho_0)$ is assumed to be approximately constant with a fiducial value of $\epsilon_{\rm ff}(\rho_0)=0.005$. 
$p_*/m_*=3000$~km s$^{-1}$ is the fiducial value of momentum injection per supernova \citep[but see e.g.][for arguments for a $z$-dependent $p_*/m_*$]{Fisher19}.
As a result, the gas velocity dispersion is expected to vary only mildly due to supernova explosions. 

\begin{figure}
	\centering
	\includegraphics[width=\columnwidth]{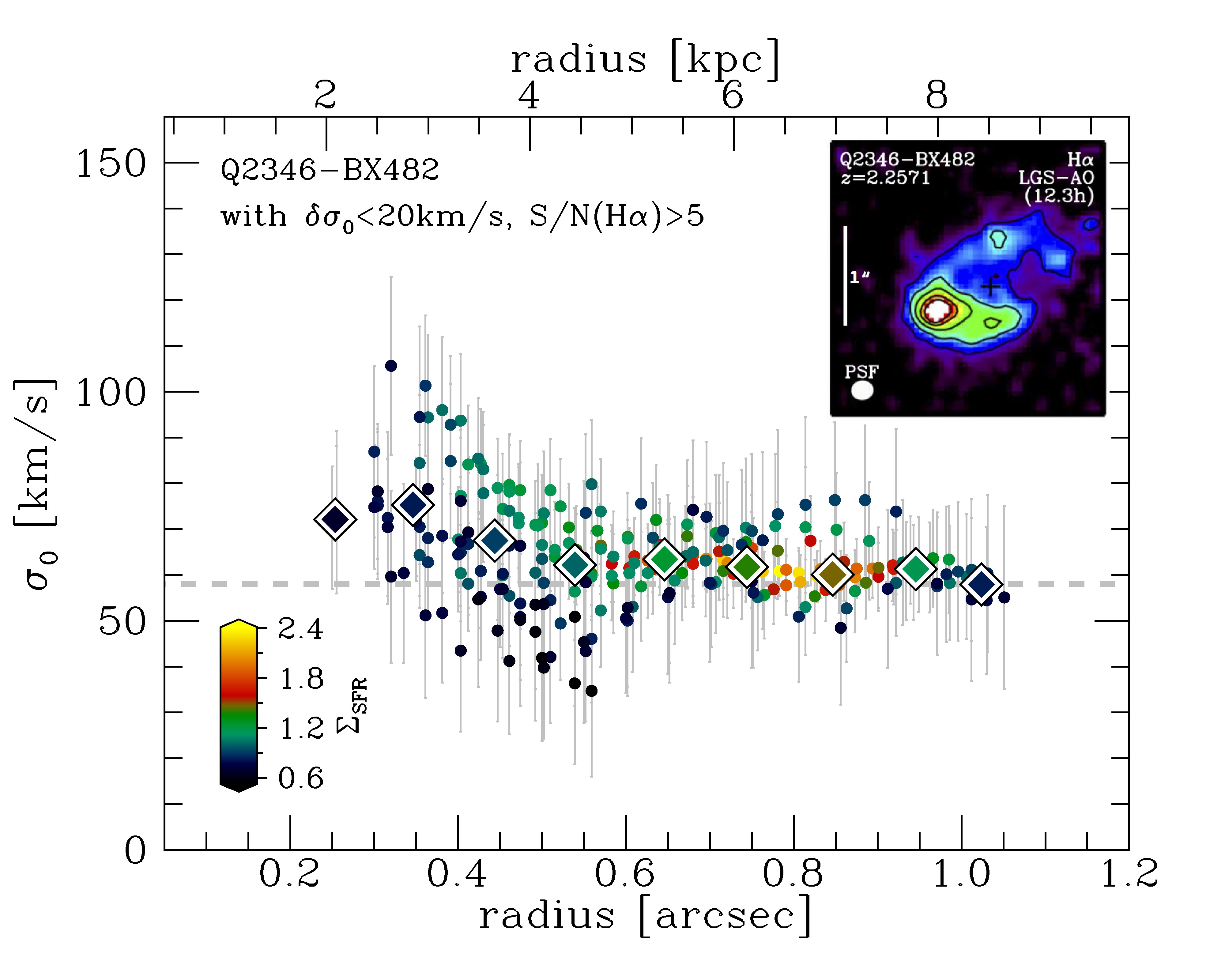}
    \caption{Intrinsic local velocity dispersion \sig as a function of radius for galaxy Q2346-BX482, measured from individual spaxels (circles) and color-coded by local $\Sigma_{\rm SFR}$ \citep[adopted from Figure~A1 by][]{Genzel11}. Larger diamonds show the running median. The grey dashed line shows the best-fit intrinsic velocity dispersion from kinematic modelling where $\sigma_0$ is assumed to be constant. The inset in the top right corner shows the projected map of H$\alpha$ flux, featuring the bright star-forming clump to the South-East, adopted from Figure~16 by \cite{FS18}. There is no correlation between local $\Sigma_{\rm SFR}$ and local velocity dispersion.}
    \label{fig:bx482}
\end{figure}

\begin{figure}
	\centering
	\includegraphics[width=\columnwidth]{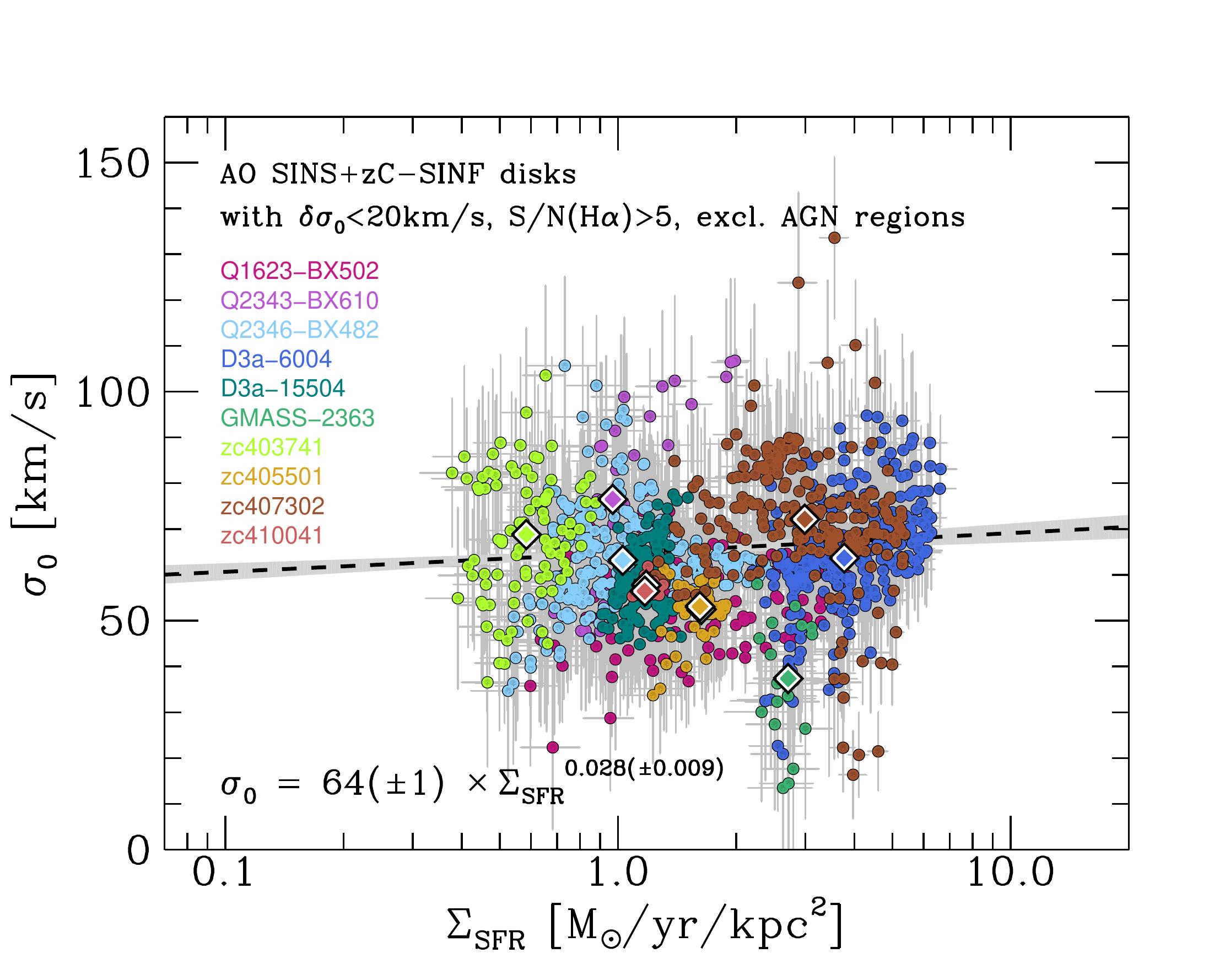}
    \caption{Intrinsic local velocity dispersion \sig as a function of star formation rate surface density $\Sigma_{\rm SFR}$, measured from individual spaxels in ten galaxies from the SINS/zC-SINF survey adaptive optics follow-up. We select spaxels with $\delta\sigma_0<20$~km s$^{-1}$, S/N(H$\alpha$)>5, and exclude the regions of three galaxies that are affected by AGN feedback. Colored circles correspond to data from the different galaxies as listed in the legend, and larger diamonds show the median values. The black dashed line shows the linear regression to the individual spaxel data, with fit uncertainties shown as grey shading, as given in the bottom of the figure}
    \label{fig:aosins}
\end{figure}

Similar results are obtained by other theoretical models investigating stellar feedback as the sole driver of the turbulence in the ISM, for instance the models discussed by \cite{Dib06, Joung09, Kim13}.
In fact, the resulting velocity dispersions in the ISM do not even seem to depend much on the supernova rate. 
Rather, very high supernova rates might create super-bubble structures that, instead of stirring the ambient medium, will eventually blow out of the galactic disk, thus transferring energy and metals into the circum-galactic medium \citep{MacLow89, Joung09, Baumgartner13, Kim18}. 
This is an important result because at higher redshift also the supernova rate is higher. However, other work indicates that not only the rate but also the {\it location} of supernovae is crucial for the efficiency of stellar feedback turbulence driving: considering peak driving, where supernovae go off in the densest ISM regions (e.g.\ their birth clouds), \cite{Gatto15} find local H$\alpha$ velocity dispersions of up to 60~km s$^{-1}$ for gas mass surface densities of $\Sigma_{\rm gas}\sim100~M_{\odot}~$pc$^{-2}$. This is similar to high$-z$ conditions and therefore suggests that stellar feedback can more easily maintain elevated velocity dispersions at higher redshift. Also, some idealized simulations of isolated galaxies are able to produce velocity dispersion of $\sim50$~km s$^{-1}$ from strong stellar feedback \citep{Hopkins11}.

If stellar feedback is an important factor in powering turbulence, then not only would the (observed) global scaling of velocity dispersion with SFR or $\Sigma_{\rm SFR}$ be expected, but particularly locally elevated velocity dispersion in regions of high star formation rate density \citep[cf.][]{Gatto15}. 
We exploit the high-resolution data from the SINS-zC/SINF AO survey \citep{FS18} to study local correlations between $\Sigma_{\rm SFR}$ and $\sigma_0$. 
In Figure~\ref{fig:bx482} we show the local intrinsic velocity dispersion per spaxel of galaxy Q2346-BX482 as a function of radius, color-coded by $\Sigma_{\rm SFR}$ \citep[adopted from Figure~A1 by][]{Genzel11}. The local intrinsic velocity dispersion is derived from the observed dispersion map, after correcting all instrumental and beam-smearing effects through modelling. 
In the vicinity of the giant star-forming clump $\sim6.5$~kpc South-East from the center (inset), no elevated velocity dispersion can be registered. 
 
In Figure~\ref{fig:aosins} we show the local intrinsic velocity dispersion per spaxel as a function of local $\Sigma_{\rm SFR}$ for ten SINS/zC-SINF galaxies. The velocity dispersions of these galaxies with a mean redshift of $z\sim2.2$ have somewhat higher values compared to our \kd sample, consistent with their higher average SFR and $\Sigma_{\rm SFR}$. Only two of these galaxies show an intrinsic scaling of $\sigma_0$ with $\Sigma_{\rm SFR}$. The best-fit power-law relation derived from this sub-galactic, high-quality data shows a very weak dependence of local \sig on $\Sigma_{\rm SFR}$,\footnote{This finding does not extend to nuclear regions, since more complex circum-nuclear kinematic structure caused by a combination of nuclear outflows, radial inflow and bulge induced rotation in a number of cases generates unresolved velocity fields that appear as an increased velocity dispersion. To explore its true nature will require $<0.1\arcsec$ IFU spectroscopy on 30~m class telescopes.} confirming the earlier findings by \cite[][see also \citealp{Patricio18, Tadaki18}; but \citealp{Swinbank12b}]{Genzel11}. 
Similar results are found for both ionized gas \citep{Varidel16, Zhou17} and molecular gas \citep{CalduPrimo16} in local galaxies. For atomic gas, several studies of local galaxies find correlations with SFR or $\Sigma_{\rm SFR}$ that are too weak to explain the turbulent velocities in the galaxy outskirts \citep[e.g.][]{Tamburro09, Ianjamasimanana15, Utomo19}.

In summary, while global $\sigma_0$ correlates with SFR properties, we do not find a direct connection between high, {\it local} star-formation activity and elevated $\sigma_0$, as suggested by some simulations. Generally, however, simulations and models agree that stellar feedback is able to maintain galaxy-wide turbulence on scales of 10-20~km s$^{-1}$.\\

\subsection{Marginally Toomre-stable Disks}\label{gravity}

Turning to gravity-driven turbulence, an empirical model to describe the redshift evolution of velocity dispersion is that of marginally stable disks, where (non-interacting) galaxies are subject to gas replenishment from the halo or the cosmic web, and to gas loss through either outflows or star formation \citep[][]{Noguchi99, Silk01, Immeli04a, Immeli04b, FS06b, ElmegreenD07, Genzel08, Dekel09, Dekel14, Bouche10, Krumholz10, Krumholz16, Cacciato12, Dave12, Lilly13, Saintonge13, Rathaus16, Leaman17}. In this framework, the (in)stability of the disk directly corresponds to the level of turbulence in the interstellar medium, where turbulence is fed through external (accretion) and internal (radial flows, clump formation) events via the release of gravitational energy, creating a self-regulation cycle to maintain marginal stability (\citealp{Dekel09, Genel12b}; but see \citealp{ElmegreenB10}).

For a snapshot in time that represents the observation of a high$-z$ galaxy, this equilibrium situation is captured through the Toomre $Q$ parameter \citep{Toomre64}, where generally $Q<Q_{\rm crit}\approx1$ indicates gravitational instability. Considering the one-component approximation for a gas disk, we can write \citep{BT08, Escala08, Dekel09b}
\begin{eqnarray}\label{eq:tq}
	Q_{\rm gas} &=& \frac{\sigma_0 \kappa}{\pi G \Sigma_{\rm gas}} = \frac{\sigma_0}{\pi G \Sigma_{\rm gas}} \frac{a v_c(r)}{r}.
\end{eqnarray}
Here, $\kappa$ is the epicyclic frequency, $a$ is a constant taking values of $1$ and $\sqrt{2}$ for Keplerian and constant rotation velocity, respectively, and $v_c$ is the circular velocity tracing the dynamical mass.

\begin{figure}
	\centering
	\includegraphics[width=\columnwidth]{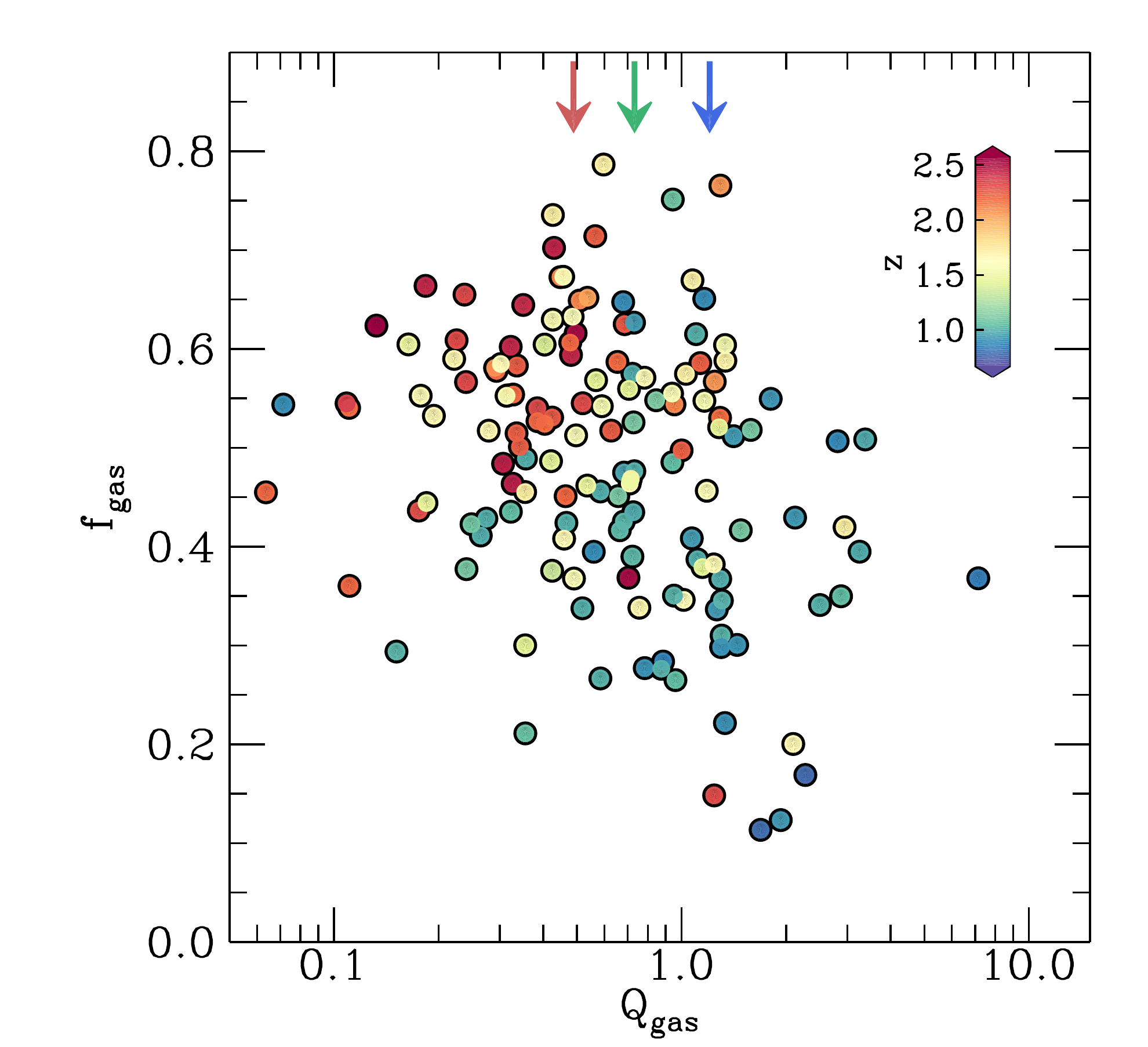}
 \caption{Gas-to-baryonic mass fraction $f_{\rm gas}$ as a function of $Q_{\rm gas}$, color-coded by redshift. The arrows indicate the average value of $Q_{\rm gas}$ at $z\sim0.9$ (blue), $z\sim1.5$ (green), and $z\sim2.3$ (red). $f_{\rm gas}$ and $Q_{\rm gas}$ are weakly anti-correlated with Spearman rank correlation coefficient $\rho_S=-0.30$ and significance $\sigma_\rho=3.6$. Higher$-z$ galaxies with higher gas fractions reach values below $Q_{\rm gas}=1$.}
  \label{fig:tqfgas}
\end{figure}

The framework of Toomre-(in)stability generally refers to the linear regime, where perturbations are assumed to be axisymmetric. The galaxies studied here, however, are in the non-linear limit where the ISM is turbulent and many stars have formed \citep{Mandelker14}. \cite{Inoue16} investigated the stability of simulated high$-z$ disks, finding that large parts of the disks are in the non-linear regime with $Q>1-3$. This result however depends on gas fraction, which is generally too low in the simulations. Indeed, for those simulated galaxies with the highest gas fractions ($f_{\rm gas}\sim0.4$, still lower than for typical $z\sim2$ SFGs), \cite{Inoue16} find values more compatible with observational findings. \cite{Meng18} argue in recent work that the Toomre$-Q$ linear stability analysis is still applicable to simulated high$-z$ galaxies, with values of $Q\sim0.5-1$ in gas-rich regions \citep[see also][for simulations of isolated gas-rich disks]{Behrendt15}.

Generally, for a multi-component system an effective $Q$ parameter has to be computed, $Q_{\rm eff}^{-1}=\sum\nolimits_{i}Q_i^{-1}$, where $i$ refers to e.g.\ stars or different gas phases \citep[e.g.][and references therein]{Wang94, Escala08, Genzel11, Obreschkow15}.
Simulations of galaxy formation support a picture where $Q_{\rm eff}\sim1$ for high$-z$ galaxies, and $Q_{\rm eff}\sim2-3$ for low$-z$ galaxies where the increasing impact of a stellar disk increases $Q_{\rm crit}$ \citep{Hohl71, Athanassoula86, Bottema03, Immeli04a, Kim07, Agertz09a, Agertz09b, Aumer10, Ceverino10, Hopkins11, Genel12a, Danovich15}.
For gas-rich, thick disks instead $Q_{\rm crit}$ decreases, such that for $z\gtrsim1$ galaxies values $Q_{\rm crit}\approx0.7$ are expected \citep[e.g.][]{Goldreich65, Kim07, Wang10, Behrendt15}.

It has been shown that the gas-rich, star-forming disks observed at high redshift are at most marginally stable to gravitational fragmentation (\citealp{Genzel11}; see also \citealp{Swinbank17, Johnson18, Tadaki18}; and \citealp{Fisher17} for local high$-z$ analogues), and \cite{Wisnioski15} have shown that the redshift evolution predicted by Equation~\eqref{eq:tq} for $Q\sim1$ gas disks is in remarkable agreement with observations \citep[see also e.g.][]{Green14, Turner17, White17, Johnson18}.
In addition, \cite{Genzel11} have shown that on sub-kpc spatially resolved scales, values of $Q\sim0.2$ can be reached in regions of star-forming clumps, possibly demonstrating gravitational fragmentation at work. 

We calculate $Q_{\rm gas}$ for our galaxies following Eq.~\eqref{eq:tq} by evaluating the circular velocity at $v_c(r=1.38 R_e)$. As mentioned in Section~\ref{sample}, this radius corresponds to the theoretical peak of a Noordermeer disk with $n_{\rm S}=1$, such that the local gradient of the rotation curve is flat, leading to $a=\sqrt{2}$. The circular velocity $v_c$ is computed from the model rotation velocity corrected for pressure support from the turbulent motions \citep{Burkert10, Burkert16, WuytsS16}. 
In Figure~\ref{fig:tqfgas} we show $f_{\rm gas}=M_{\rm gas}/M_{\rm bar}$, with $M_{\rm bar}=M_*+M_{\rm gas}$, as a function of $Q_{\rm gas}$, color-coded by redshift as in Figure~\ref{fig:zev}. Albeit large scatter, an anti-correlation between $f_{\rm gas}$ and $Q_{\rm gas}$ is evident, such that galaxies with higher gas fractions have lower $Q$ (Spearman rank correlation coefficient $\rho_S=-0.30$ with significance $\sigma_\rho=3.6$). This is in agreement with the theoretical prediction that SFGs that are more gas rich have lower $Q$ values.
The average $Q_{\rm gas}$ for our galaxies in the redshift bins $z\sim0.9; 1.5; 2.3$ is $Q_{\rm gas}=1.2; 0.7; 0.5$ (arrows in Figure~\ref{fig:tqfgas}). 
Our results on the average offset of ionized {\it vs.\ }atomic+molecular gas from Section~\ref{multiphase} suggest that the cold gas tracing the bulk of the gas mass might have lower velocity dispersion by $10-15$~km $^{-1}$. This would lower the $Q_{\rm gas}$ values by a factor $\sim1.2-2$. 
While our calculation of the Toomre$-Q$ parameter is simplified through the omission of the stellar component, this suggests that thick high$-z$ disks with high gas fractions of $\gtrsim50$~\% can be marginally stable even down to $Q_{\rm gas}<0.7$.

\subsection{Combining Feedback and Gravity}\label{feedbackgravity}

While gravitational instabilities are likely important drivers of the elevated velocity dispersions at $z>1$, the contribution from stellar feedback-driven turbulence of the order of $10-20$~km s$^{-1}$ could become comparable or even dominant for lower$-z$, low$-\sigma_0$ galaxies. Therefore, one must consider both processes to get a complete picture.

The combination of stellar feedback and gravitational processes for turbulence driving has recently been investigated through the analytic model for structure and evolution of gas in galactic disks by \cite{Krumholz18}, who combine prescriptions for star formation, stellar feedback, and gravitational instabilities into a unified `transport+feedback' model to explain the range of observed dispersions from $z=3$ to the present day. In their model, gas is in vertical hydrostatic equilibrium and energy equilibrium. 
This model assumes (isolated) rotating galactic disks built of gas and stars within a quasi-spherical dark matter halo over a wide redshift range. Disks are stable or marginally stable to gravitational collapse, regulated by mass transport through the disk. The gas is in vertical hydrostatic equilibrium, and in energy equilibrium such that losses through the decay of turbulence are balanced by energy input into the system via stellar feedback and the release of gravitational energy via mass transport through the disk.

Consistent with the discussion above, \cite{Krumholz18} show in their model that stellar feedback may maintain velocity dispersions of $\sim10$~km s$^{-1}$, creating a dispersion floor, while gravitational instabilities, for instance created through radial mass transport through the disk, are necessary to constantly drive velocity dispersions beyond $\sigma_0\sim20$~km s$^{-1}$ for moderate star-formation rates \citep[cf.\ also Figure~4 by][]{Krumholz18}. 
They make a prediction for galactic gas velocity dispersion and its correlation with SFR. Particularly, they show that (see their Equation~(60))
\begin{eqnarray}\label{eq:60k18}
	{\rm SFR} &=& \frac{0.42}{\pi G}\frac{1}{Q} \cdot f_{\rm gas} v_{\rm circ}^2 \sigma_0,
\end{eqnarray}
where we have substituted appropriate constants for high$-z$ galaxies following \cite{Krumholz18}. Specifically, we adopt a rotation curve slope of $\beta=0$, an offset between resolved and unresolved star formation law normalizations of $\phi_a=3$, a fraction of ISM in the star-forming phase $f_{\rm sf}=1$, a ratio of total pressure to turbulent pressure at the midplane of $\phi_{\rm mp}=1.4$, a star-formation efficiency per free-fall time of $\epsilon_{\rm ff}=0.0015$, an orbital period of $t_{\rm orb,out}=200$~Myr, and a maximum star-formation time-scale of $t_{\rm sf,max}=2$~Gyr.

We make two adjustments to our data to properly compare to the model: 
here, and for all of Section~\ref{feedbackgravity}, we subtracted 15~km s$^{-1}$ in quadrature from our intrinsic dispersion values, denoted by $\sigma_{0,15}$, to ensure consistency with the theoretical model (see \citealp{Krumholz16} and \citealp{Krumholz18}, Appendix~B). These 15~km s$^{-1}$ represent the average combination of thermal motions and expansion of H{\sc ii} regions that enter our ionized gas velocity dispersion measurement (see also Sections~\ref{limits} and \ref{multiphase}).
We also modify our gas mass fractions: the corresponding parameter used by \cite{Krumholz18} describes an effective gas fraction at the mid plane. This has typically higher values than our gas fraction $f_{\rm gas}$ because of the larger stellar scale heights compared to the gas scale heights. For the comparison here we adopt a scaling factor of 1.5 for our gas mass fractions, motivated by measurements in the Solar neighbourhood (\citealp{McKee15, Krumholz18}; M.\ Krumholz, private communication).

\begin{figure}
	\centering
	\includegraphics[width=\columnwidth]{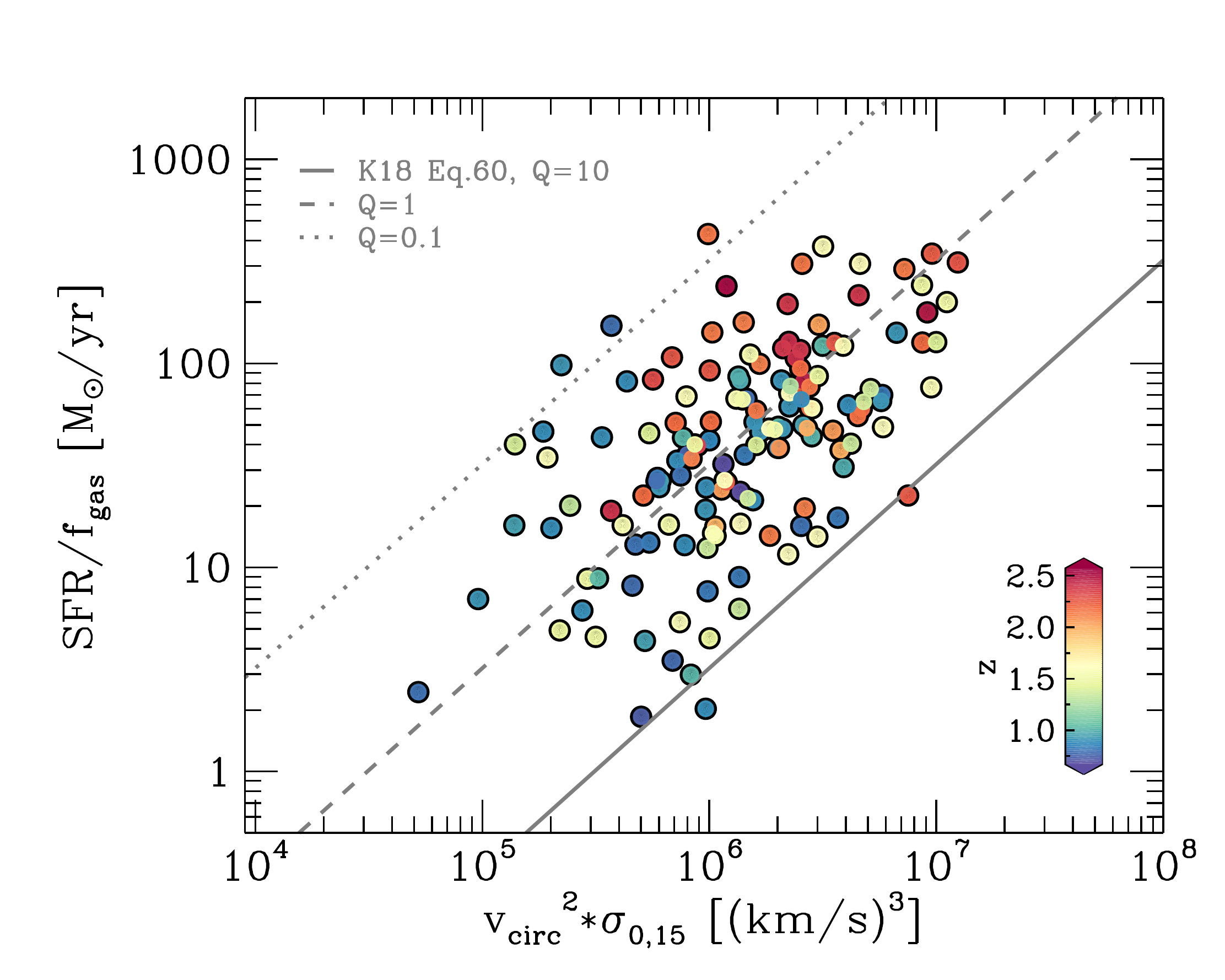}
 \caption{SFR divided by gas fraction as a function of circular velocity squared times intrinsic velocity dispersion for our kinematic sample, color-coded by redshift. The lines show predictions from the `transport+feedback' model by \cite{Krumholz18} for different values of $Q$ (Equation~\eqref{eq:60k18}). We find a strong correlation between the displayed quantities ($\rho_{\rm S}=0.57$; $\sigma_\rho=6.8$), where galaxies scatter around constant $Q$, suggesting dominant self-regulation processes in our galaxies at all redshifts.}
  \label{fig:k18eq60}
\end{figure}

To compare the model prediction from Equation~\eqref{eq:60k18} to our data, we group correlated quantities and separate the star formation properties SFR and $f_{\rm gas}$ from the kinematic tracers $v_{\rm circ}$ and $\sigma_0$. We show the result for our kinematic sample in Figure~\ref{fig:k18eq60}, specifically SFR divided by gas fraction as a function of circular velocity squared times intrinsic velocity dispersion. 
Figure~\ref{fig:k18eq60} reveals a clear trend between the displayed quantities, with a Spearman rank correlation of $\rho_{\rm S}=0.57$ with significance $\sigma_\rho=6.8$. We also show model predictions from \cite{Krumholz18} as quoted in Equation~\eqref{eq:60k18} for three values of $Q$. There is a tendency for higher$-z$ galaxies to have a predicted $Q\lesssim1$, consistent with our results presented in Figure~\ref{fig:tqfgas}. Generally, however, our galaxies scatter around $Q=1$ at all redshifts. This suggests that SFGs self-regulate at all times such that the population of SFGs evolves roughly along lines of constant $Q$. This result is largely independent from the specific choices of parameters such as $\phi_a$ or $f_{\sf}$, which will only affect the average $Q$ value.

It is important to realize that the above correlation is predicted for both the combined `transport+feedback' model and a model without feedback, but not for models lacking the `transport' component accounting for gravitational instabilities. Therefore the correlation between SFR/$f_{\rm gas}$ and $v_{\rm circ}^2\sigma_0$ illustrates the importance of gravitational instabilities as drivers of turbulence for our kinematic sample.

\begin{figure}
	\centering
	\includegraphics[width=\columnwidth]{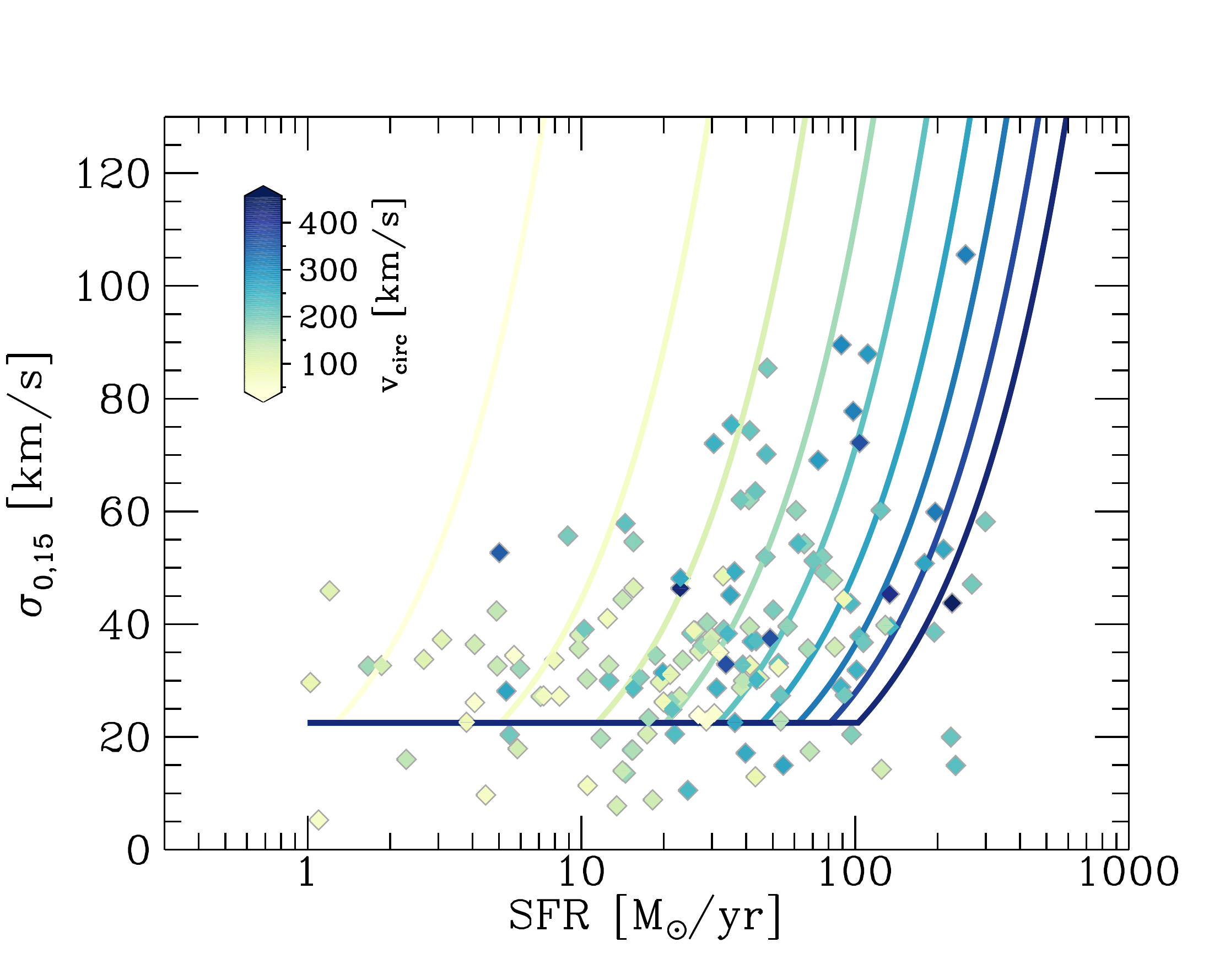}
 \caption{Intrinsic velocity dispersion $\sigma_{0,15}$ as a function of star formation rate SFR, color coded by circular velocity. The data points show our kinematic sample. The lines are predictions from the `transport+feedback high$-z$' model by \cite{Krumholz18}, where we additionally vary the galaxy circular velocity $v_{\rm circ}$ between 50~km s$^{-1}$ and 450~km s$^{-1}$ in steps of 50~km s$^{-1}$. 
For 60~\% of our galaxies in the $\sigma_{0,15}-$SFR parameter space, the model predicts the correct rotation velocity, with all other parameters being fixed as specified in the main text.}
  \label{fig:k18_vrot}
\end{figure}

In the following, we now investigate separately changes of circular velocity and gas fraction in the $\sigma_0-$SFR parameter space. In Figure~\ref{fig:k18_vrot} we show for our kinematic sample the intrinsic velocity dispersion as a function of SFR, color-coded by circular velocity. 
As expected from the main sequence and Tully Fisher relation \citep{Tully77}, which is in place for our data set at all redshifts \citep{Uebler17}, our data display a gradient such that circular velocity on average increases with increasing SFR. 
As lines we plot the high$-z$ model by \cite{Krumholz18}, but we modify it such that we vary the galaxy circular velocity from $v_{\rm circ}=50$~km s$^{-1}$ to $v_{\rm circ}=450$~km s$^{-1}$ in order to appropriately cover the range of observed velocities in our kinematic sample. 
In the model framework, stellar feedback creates sustains a dispersion floor, represented through the horizontal regime of the model lines. The predicted rapid increase of velocity dispersion with SFRs, the exact location here dependent on circular velocity, requires the release of gravitational energy through radial transport through the disk \citep[see][for details]{Krumholz18}.  
The agreement between the theoretical model and our data is remarkably good: $\sim60$\% of our data are matched by the model for this simple variation of only the circular velocity, with all other parameters being fixed to the fiducial `transport+feedback high-z' parameters.

\begin{figure}
	\centering
	\includegraphics[width=\columnwidth]{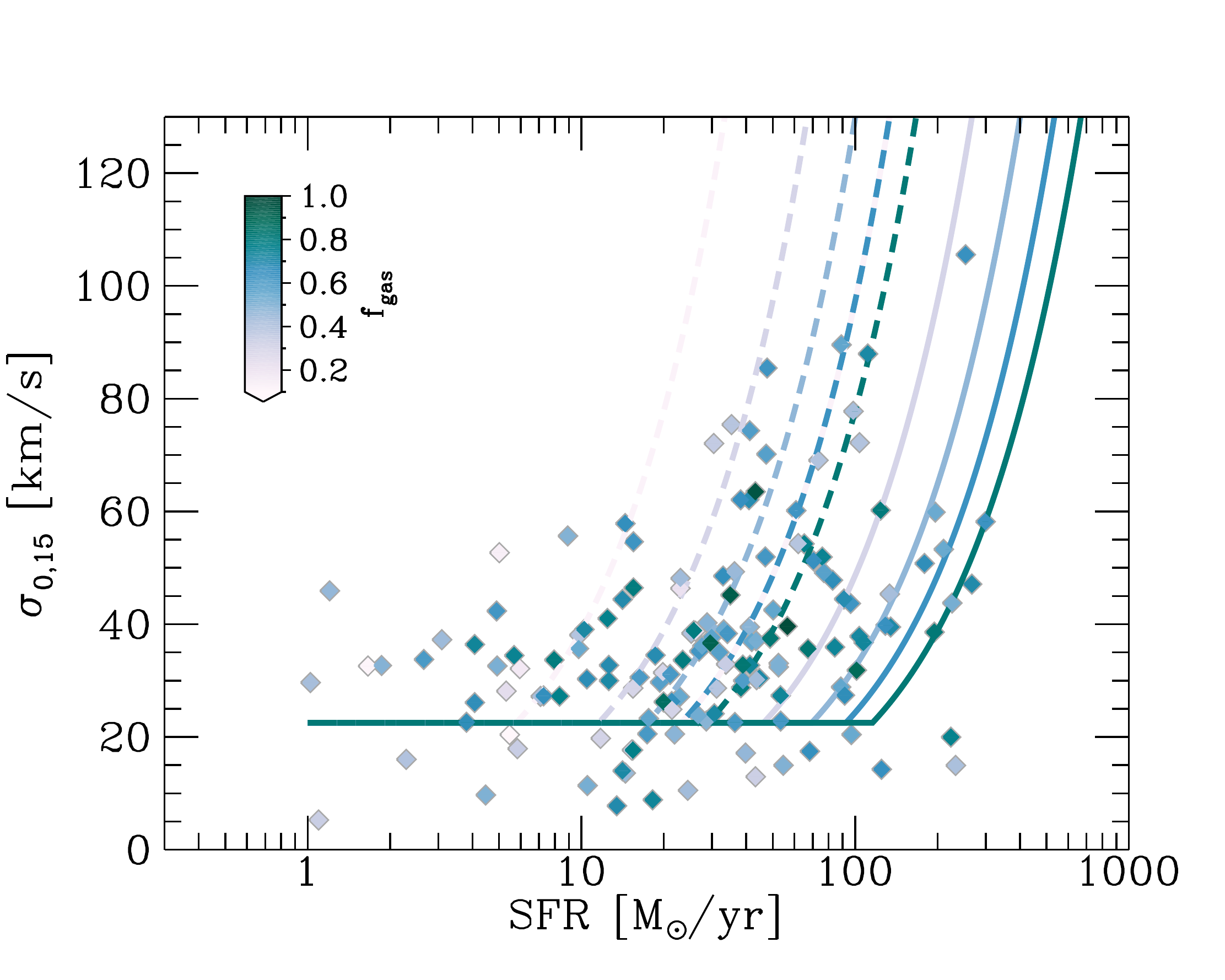}
 \caption{Intrinsic velocity dispersion $\sigma_{0,15}$ as a function of star formation rate SFR, color coded by (scaled) gas fraction (see main text for details). The data points show our kinematic sample. The lines are predictions from the `transport+feedback high$-z$' model by \cite{Krumholz18}, where we additionally vary $f_{g,Q}$ in lock-step with $f_{g,P}$ between 0.2 and 1 in steps of 0.2, and the galaxy rotation velocity $v_\phi$ from 200~km s$^{-1}$ (dashed lines) to 400~km s$^{-1}$ (solid lines). The location of the model predictions illustrate how the observed scatter in gas fractions at fixed SFR and $\sigma_0$ may be caused by different rotation velocities.}
  \label{fig:k18_fgas}
\end{figure}

For the gas fraction we can make only an approximate comparison. As mentioned in Section~\ref{data}, gas masses for our galaxies are calculated applying the scaling relation by \cite{Tacconi18}, since direct gas mass measurements are not available for most of our galaxies. With this, we get the total gas mass over the total baryonic mass per galaxy. Again, we use a scaling factor of 1.5 for our gas mass fractions. 
In Figure~\ref{fig:k18_fgas} we show the same parameter space as in Figure~\ref{fig:k18_vrot} but now color coded by gas fraction. 
While galaxies with SFR~$\lesssim10~M_{\odot}$ yr$^{-1}$ have on average lower gas fractions, no strong trend is apparent at higher SFRs. 
We show again lines based on the `transport+feedback high$-z$' model by \cite{Krumholz18}, but now we vary the gas fraction (and with it $f_{g,P}$) from $f_{g,Q}=0.2$ to $f_{g,Q}=1.0$ in order to explore the range of scaled gas fractions of galaxies in our kinematic sample.
With solid lines we show models with $v_\phi=400$~km s$^{-1}$, and dashed lines show $v_\phi=200$~km s$^{-1}$. 
It becomes clear that in the model framework galaxies at fixed SFR and $\sigma_0$ can have higher $f_{g,Q}$ and lower $v_\phi$, or lower $f_{g,Q}$ and higher $v_\phi$, but rotation velocity has to be varied to cover the full range of SFRs in our observations.

Horizontal variations of the model predictions can be reached through changing the fraction of gas assumed to be in the star-forming ISM, and through changes in the outer rotation curve slope. For instance, assuming only 20$\%$ of the gas to be in the star-forming phase pushes the horizontal floor of the model below 10~km s$^{-1}$, and lowers the predicted SFR by almost an order of magnitude.
Assuming a dropping rotation curve, on the other hand, lifts the horizontal floor and increases the predicted SFR. Assuming an outer rotation curve slope of $\beta=-0.5$ increases the horizontal saturation of the model to $\sim32$~km s$^{-1}$, whilst increasing the star-formation rate only marginally.
\cite{Lang17} have shown that the typical outer rotation curve slope of galaxies in our sample is negative. This is more pronounced at higher redshift, possibly offering an additional reason for the elevated velocity dispersions at $z\gtrsim2$ in this model framework.

Considering these analytic model prescriptions, and the typical uncertainty of the intrinsic dispersion measurements of $\delta\sigma_0\sim10$~km s$^{-1}$ in our kinematic sample, we conclude that galaxies with $\sigma_0\gtrsim35$~km s$^{-1}$ are dominated by gravitational instability-driven turbulence. This encompasses more than 60$\%$ of galaxies in our sample, underlining the importance of gravity-driven turbulence in star forming galaxies at $z\sim1-3$.

\subsection{AGN feedback}\label{agn}

As a final remark, we briefly want to comment on AGN feedback as a potential additional source for elevated velocity dispersions in the SFGs in our kinematic sample. While we excluded galaxies, or regions of galaxies, that are so strongly affected by the AGN and associated outflows that the disk kinematics cannot be recovered, we do not entirely exclude AGN. This ensures that we can explore the full mass range covered by the \kd survey, including the high-mass end where at log$(M_*/M_{\odot})>11$, above the Schechter mass, the fraction of AGN increases rapidly \citep{FS14, FS18, Genzel14b}.

\begin{figure}
	\centering
	\includegraphics[width=\columnwidth]{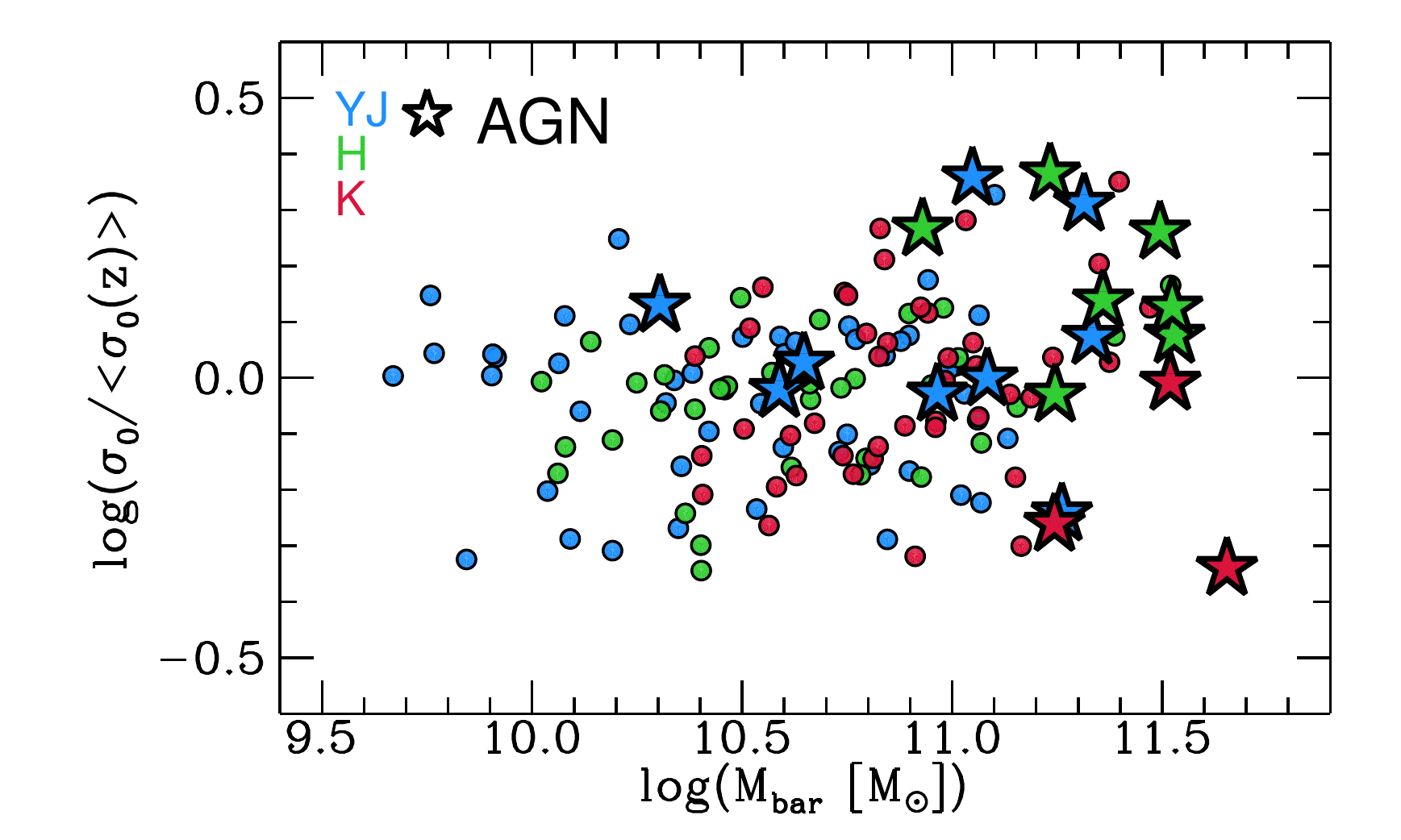}
    \caption{Redshift-normalized intrinsic velocity dispersion as a function of baryonic mass. Blue, green, and red colors indicate $z\sim0.9$, $z\sim1.5$, and $z\sim2.3$ SFGs, respectively. Galaxies that host an AGN are shown as stars. Most galaxies with log$(M_{\rm bar}/M_{\odot})\gtrsim11.3$ have above-average velocity dispersions and about half of them host an identified AGN.}
    \label{fig:agn}
\end{figure}

While we do not find significant correlations between $z-$normalized $\sigma_0$ and mass properties (Table~\ref{tab:physpropcorr}), we do note a cloud of galaxies from {\it all redshifts} with dispersions above average for the highest stellar (log$(M_*/M_{\odot})>11$) and baryonic masses (log($M_{\rm bar}/M_{\odot})\gtrsim11.3$) as shown in Figure~\ref{fig:agn}.
About half of the log($M_{\rm bar}/M_{\odot})\gtrsim11.3$ above-average dispersion galaxies are known to host an AGN (stars in Figure~\ref{fig:agn}). 
We speculate that the energy deposited by strong AGN feedback in the form of nuclear outflows could induce turbulence in the disk via the re-accretion of material at larger radii.

It is important to keep in mind that outflow components with velocities similar to the galaxy rotation velocity can broaden the line width but may not be distinguishable from the star-forming regions due to S/N limitations. Comparing to the deep AO data from the SINS/zC-SINF survey that we show in Figure~\ref{fig:aosins}, one of the three identified, log($M_*/M_{\odot})\gtrsim11$ AGN (Q2343-BX610) shows above-average velocity dispersions (after excluding the regions clearly affected by the nuclear outflow), while the other two (D3a-6004, D3a-15504) have average dispersions.\\

\section{Conclusions}\label{conclusion}

We have investigated the evolution of the ionized gas intrinsic velocity dispersions, $\sigma_0$, between $0.6<z<2.6$ based on data from our full \kd survey. We have selected a high-quality, rotation-dominated ($v_{\rm rot}/\sigma_0\geq1$) sample for which we forward-modelled in a Bayesian framework the one-dimensional galaxy kinematics extracted from the H$\alpha$ velocity and velocity dispersion maps, taking into account instrumental effects, beam smearing, and pressure support. 
Our main conclusions are as follows:

\begin{itemize}

	\item Assuming an isotropic and radially constant intrinsic velocity dispersion, we find an average decrease of the H$\alpha$ intrinsic dispersion for our kinematic sample from $\sigma_0\sim46$~km s$^{-1}$ at $z\sim2.3$ to $\sigma_0\sim31$~km s$^{-1}$ at $z\sim0.9$, solidifying trends previously reported in the literature (Section~\ref{k3d}). Putting our sample into the broader context of literature measurements from $z=4$ to $z=0$, tracing ionized, molecular, and atomic gas phases, confirms the general increase of intrinsic galaxy velocity dispersion with redshift (Section~\ref{literature}).
	
	\item Comparing the redshift evolution of ionized and molecular plus atomic gas velocity dispersion, we find that the ionized gas dispersion is on average higher by $\sim12$~km s$^{-1}$ (Section~\ref{multiphase}). This offset can in principle be accounted for through the different gas temperatures together with the line broadening through expansion of H{\sc ii} regions the ionized gas emission typically originates from.
	
	\item For our \kd kinematic sample, we find that there is intrinsic scatter in the $\sigma_0$ distribution at fixed redshift after accounting for measurement and modelling uncertainties, and it increases for our highest redshift slice (Section~\ref{scatter}). However, we cannot single out a physical mechanism behind this scatter. This could imply that the velocity dispersion is highly variable in time, due to a dynamic mechanism such as minor mergers or variation in accretion \citep[see][for evidence from simulations]{Hung19}. 
	Alternatively, the scatter could be caused by the interplay of different physical properties responsible to maintain marginal stability (see Section~\ref{feedbackgravity}).
		
	\item Investigating the physical driver of the elevated velocity dispersions at higher redshift, we find that  galaxies in our kinematic sample are at most marginally Toomre-stable, i.e.\ they are consistent with their turbulence being powered through gravitational instabilities in a self-regulated environment (Section~\ref{gravity}).
	
	\item We find no evidence from our high-resolution SINS/zC-SINF AO data that stellar feedback as traced through $\Sigma_{\rm SFR}$ typically increases the velocity dispersion on sub-galactic scales beyond the average level, or that the {\it local} velocity dispersion correlates strongly with $\Sigma_{\rm SFR}$, suggesting that contributions from stellar feedback to turbulence driving are minor for our $z>1$ SFGs (Section~\ref{feedback}).
	
	\item We find good agreement between data from our \kd kinematic sample and predictions from the state-of-the-art analytical model of galaxy formation and evolution by \cite{Krumholz18}, further strengthening the evidence that the majority of our galaxies ($\gtrsim60$\%) are dominated by gravity-driven turbulence (Section~\ref{feedbackgravity}).
	
\end{itemize}

The measurement of intrinsic gas velocity dispersion at $z>0$ is challenging. Next-generation instruments such as ERIS+AO at the VLT or HARMONI at the ELT will expand current samples on spatial scales that are currently only achievable for strongly lensed objects, and push spectral scales down to $\sim15$~km s$^{-1}$.
The statistics from these observations will facilitate further investigation of the scatter of the intrinsic velocity dispersion at fixed redshift, and tests of theoretical predictions such as the transition regime from gravity-driven turbulence to feedback-driven turbulence as a function of redshift and mass \citep{Krumholz18}.
Deep, high-S/N observations of particularly molecular gas reaching 1--2~kpc resolution at $z>1$ with NOEMA or ALMA are necessary to test if the redshift evolution of molecular and ionized gas velocity dispersion is indeed comparable.\\


\acknowledgements 
We are grateful to the anonymous referee for a very constructive report that helped to improve the quality of this manuscript. 
We thank the ESO Paranal staff for their helpful support with the KMOS observations for this work. 
We are grateful to Raymond Simons and Susan Kassin for providing us with data from the DEEP2 and SIGMA surveys.
We thank Amiel Sternberg, Vadim Semenov, and Chia-Yu Hu for fruitful discussions, and we are particularly grateful to Blakesley Burkhart, Mark Krumholz, and Andreas Schruba for valuable comments on various aspects of this work. 
DW acknowledges the support of the Deutsche Forschungsgemeinschaft via projects WI 3871/1-1 and WI 3871/1-2. 
EW acknowledges support by the Australian Research Council Centre of Excellence for All Sky Astrophysics in 3 Dimensions (ASTRO 3D), through project number CE170100013. 
MF acknowledges the support from the European Research Council (ERC) under the European Union's Horizon 2020 research and innovation programme (grant agreement No.\ 757535). 
PL acknowledges funding from the ERC under the European Union's Horizon 2020 research and innovation programme (grant agreement No.\ 694343).\\




\appendix

\section{Example Galaxies and Fits}\label{append}

\begin{figure*}
	\centering
	\includegraphics[width=\textwidth]{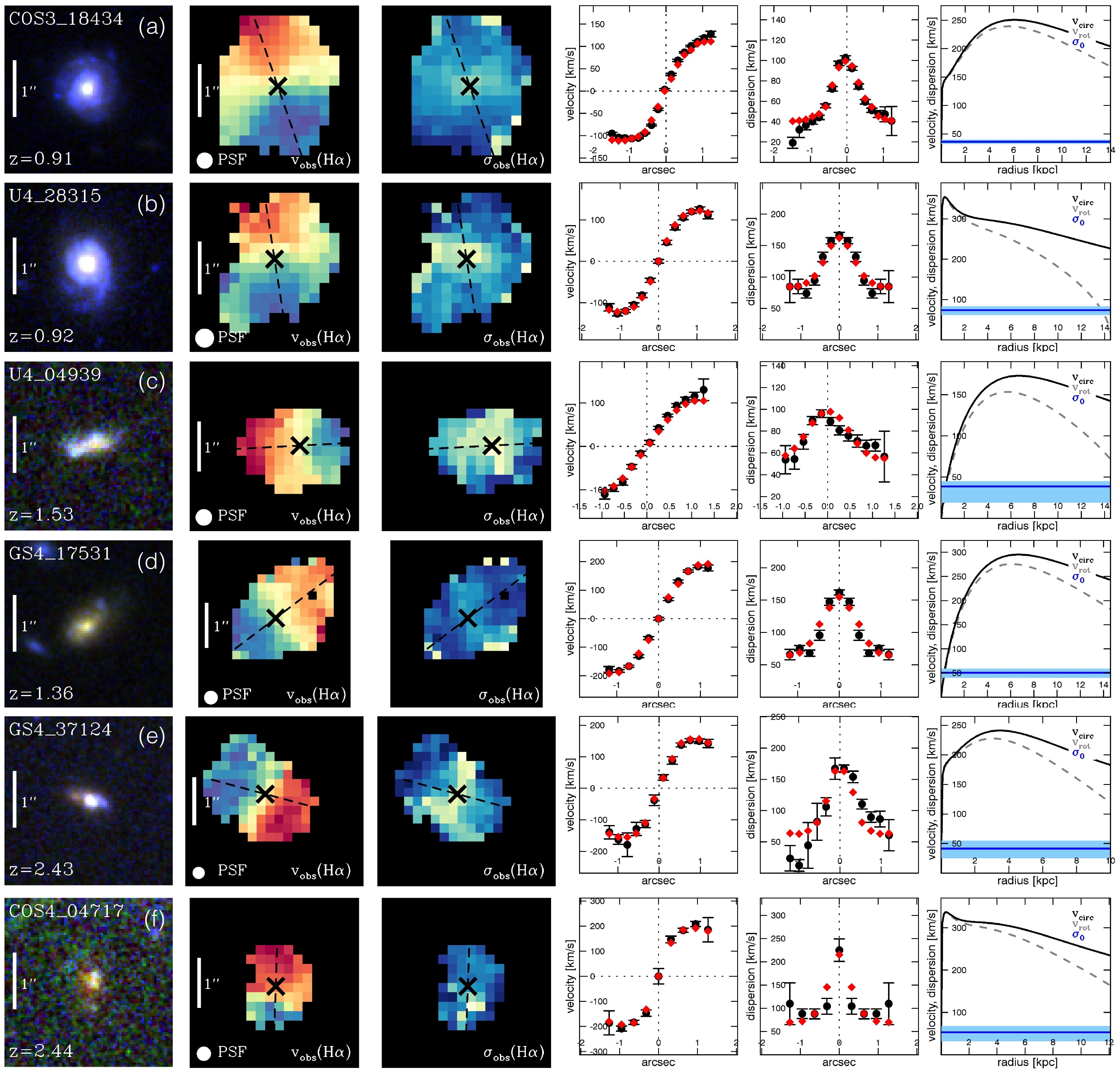}
    \caption{Example cases of galaxies in our kinematic sample. From top to bottom, we show for each redshift slice a galaxy modelled with setup 1 and with setup 2 (see Section~\ref{dysmal}). From left to right, we show an $IJH$ HST color-composite image; the projected H$\alpha$ velocity map; the projected H$\alpha$ velocity dispersion map; the observed velocity $v_{\rm rot}(r)\cdot\sin(i)$ along the kinematic major axis (black) and the best-fit model (red); the observed velocity dispersion $\sigma(r)$ correspondingly; and the intrinsic model circular velocity $v_{\rm circ}$ (black), rotation velocity $v_{\rm rot}$ (grey dashed), and intrinsic velocity dispersion (blue) together with its uncertainties derived from the MCMC posterior distribution (blue shading). The kinematic maps and profiles are corrected for the instrument line-spread function, but not for beam-smearing. The kinematic major axis is indicated by the black dashed line on top of the velocity and dispersion maps, and the black crosses indicate the midpoint between the observed minimum and maximum velocities (not necessarily the kinematic center). Note that the intrinsic rotation curves are falling by construction because we do not include a dark matter halo (but see Section~\ref{dysmal}). Rows (a), (c), and (e) show examples from setup 1, and rows (b), (d), and (f) show examples from setup 2.}
    \label{fig:ex}
\end{figure*}

We show examples of galaxies in our kinematic sample together with their best-fit kinematic models in Figure~\ref{fig:ex}. See the figure caption for details.\\

\section{Alternative Fits to our \kd Velocity Dispersions}\label{altfits}

We list fits to our \kd velocity dispersion data from $z=2.6$ to $z=0.6$ in $\sigma_0-\log(1+z)$ space and $\log(\sigma_0)-\log(1+z)$ space in Tables~\ref{tab:altfit1} and \ref{tab:altfit2}, respectively. These results agree with our fiducial fits in $\sigma_0-z$ space listed in Table~\ref{tab:bestfit}, and do not change our conclusions.\\

\clearpage 

\begin{deluxetable*}{llcccccccc}
 \tablecaption{Results from the linear regression fits of the form $\sigma_0/{\rm km\,s}=a+b\cdot\log(1+z) + c$ for our kinematic sample, where $a$ and $b$ are the regression coefficients, and $c$ is the intrinsic random scatter about the regression \citep[see][]{Kelly07}. For each parameter $a$, $b$, the standard deviation of $c$, and the derived linear correlation coefficient $l_{\rm corr}$ between $\sigma_0$ and $z$, we list the median together with the standard deviation of the posterior distribution. For each redshift slice we list the best-fit $\sigma_0$ value corresponding to these medians. \label{tab:altfit1}}
 \tablehead{ 
 \colhead{sample} & \colhead{N} & \colhead{$a$ } & \colhead{$b$} & \colhead{$\sigma_c$} & \colhead{$l_{\rm corr}$} & \colhead{$\sigma_0$ at $z\sim0.9$} & \colhead{$\sigma_0$ at $z\sim1.5$} & \colhead{$\sigma_0$ at $z\sim2.3$} \\
 \colhead{} & \colhead{} & \colhead{ } & \colhead{ } & \colhead{ } & \colhead{} & \colhead{[km s$^{-1}$]} & \colhead{[km s$^{-1}$]} & \colhead{[km s$^{-1}$]}
 }
 \startdata
  including upper & 175 & $19.4\pm3.8$ & $52.6\pm10.1$ & $9.8\pm1.1$ & $0.48\pm0.08$ & 33.8 & 40.6 & 46.5 \\
  \,\, limits & & & & & & & & \\
  excluding upper       & 147 & $19.5\pm4.5$ & $53.1\pm11.4$ & $10.4\pm1.1$ & $0.46\pm0.09$ & 34.1 & 40.9 & 46.9 \\
  \,\, limits (robust) & & & & & & & & \\
  using formal        & 175 & $19.5\pm4.5$ & $53.1\pm11.3$ & $10.4\pm1.1$ & $0.46\pm0.09$ & 34.1 & 40.9 & 46.8 \\
  \,\, best-fit $\sigma_0$ & & & & & & & & \\
\enddata
\end{deluxetable*}
\begin{deluxetable*}{llcccccccc}
 \tablecaption{Results from the linear regression fits of the form $\log(\sigma_0/{\rm km\,s})=a+b\cdot\log(1+z) + c$ for our kinematic sample, where $a$ and $b$ are the regression coefficients, and $c$ is the intrinsic random scatter about the regression \citep[see][]{Kelly07}. For each parameter $a$, $b$, the standard deviation of $c$, and the derived linear correlation coefficient $l_{\rm corr}$ between $\sigma_0$ and $z$, we list the median together with the standard deviation of the posterior distribution. For each redshift slice we list the best-fit $\sigma_0$ value corresponding to these medians. \label{tab:altfit2} }
 \tablehead{ 
 \colhead{sample} & \colhead{N} & \colhead{$a$ } & \colhead{$b$} & \colhead{$\sigma_c$} & \colhead{$l_{\rm corr}$} & \colhead{$\sigma_0$ at $z\sim0.9$} & \colhead{$\sigma_0$ at $z\sim1.5$} & \colhead{$\sigma_0$ at $z\sim2.3$} \\
 \colhead{} & \colhead{} & \colhead{} & \colhead{} & \colhead{ } & \colhead{} & \colhead{[km s$^{-1}$]} & \colhead{[km s$^{-1}$]} & \colhead{[km s$^{-1}$]}
 }
 \startdata
  including upper & 175 & $1.29\pm0.05$ & $0.77\pm0.13$ & $0.12\pm0.01$ & $0.54\pm0.08$ & 31.8 & 39.9 & 48.6 \\
  \,\, limits & & & & & & & & \\
  excluding upper       & 147 & $1.38\pm0.05$ & $0.61\pm0.13$ & $0.11\pm0.01$ & $0.49\pm0.09$ & 35.3 & 42.2 & 49.4 \\
  \,\, limits (robust) & & & & & & & & \\
  using formal        & 175 & $1.30\pm0.05$ & $0.75\pm0.13$ & $0.12\pm0.01$ & $0.53\pm0.07$ & 32.3 & 40.3 & 49.0 \\
  \,\, best-fit $\sigma_0$ & & & & & & & & \\
\enddata
\end{deluxetable*}

\section{Correlations of Physical Properties with Velocity Dispersion and Redshift-normalized Velocity Dispersion}\label{trends}

\begin{figure*}
	\centering
	\includegraphics[width=0.9\textwidth]{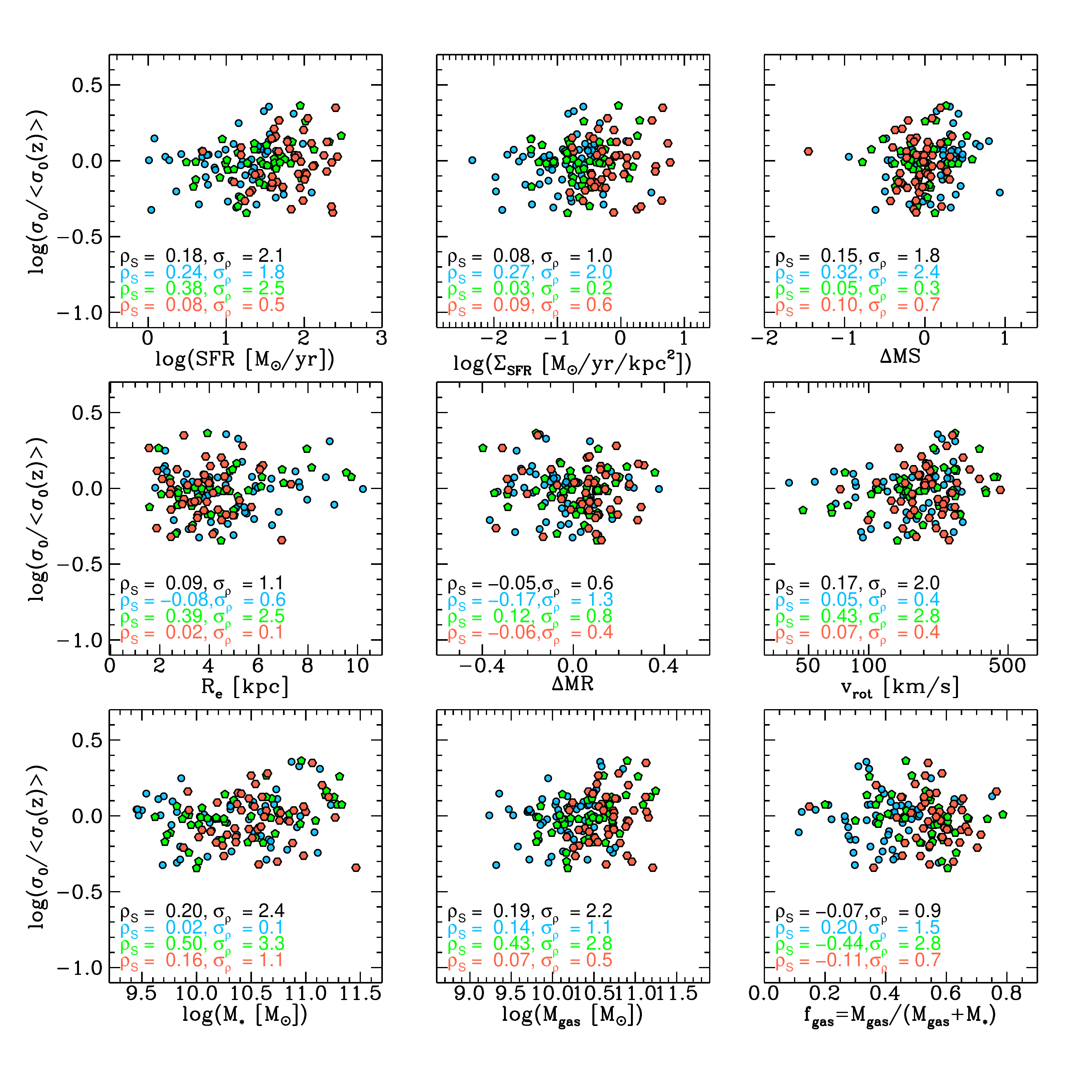}
    \caption{Redshift-normalized velocity dispersion (see Equation~\eqref{eq:norm}) as a function of several physical properties. Colors show our redshift subsamples at $z\sim0.9$ (blue), $z\sim1.5$ (green), and $z\sim2.3$ (red). Spearman rank correlation coefficients $\rho_{\rm S}$ and their significance $\sigma_\rho$ are listed in each panel for the full sample (black) and the individual redshift bins (colors). We do not find any significant correlations between redshift-normalized velocity dispersion and the considered physical properties (see Table~\ref{tab:physpropcorr} for additional quantities) for our kinematic \kd sample, meaning that we cannot identify a single physical driving source behind the intrinsic scatter in velocity dispersion.}
    \label{fig:normdisp_trends}
\end{figure*}

\begin{figure*}
	\centering
	\includegraphics[width=0.9\textwidth]{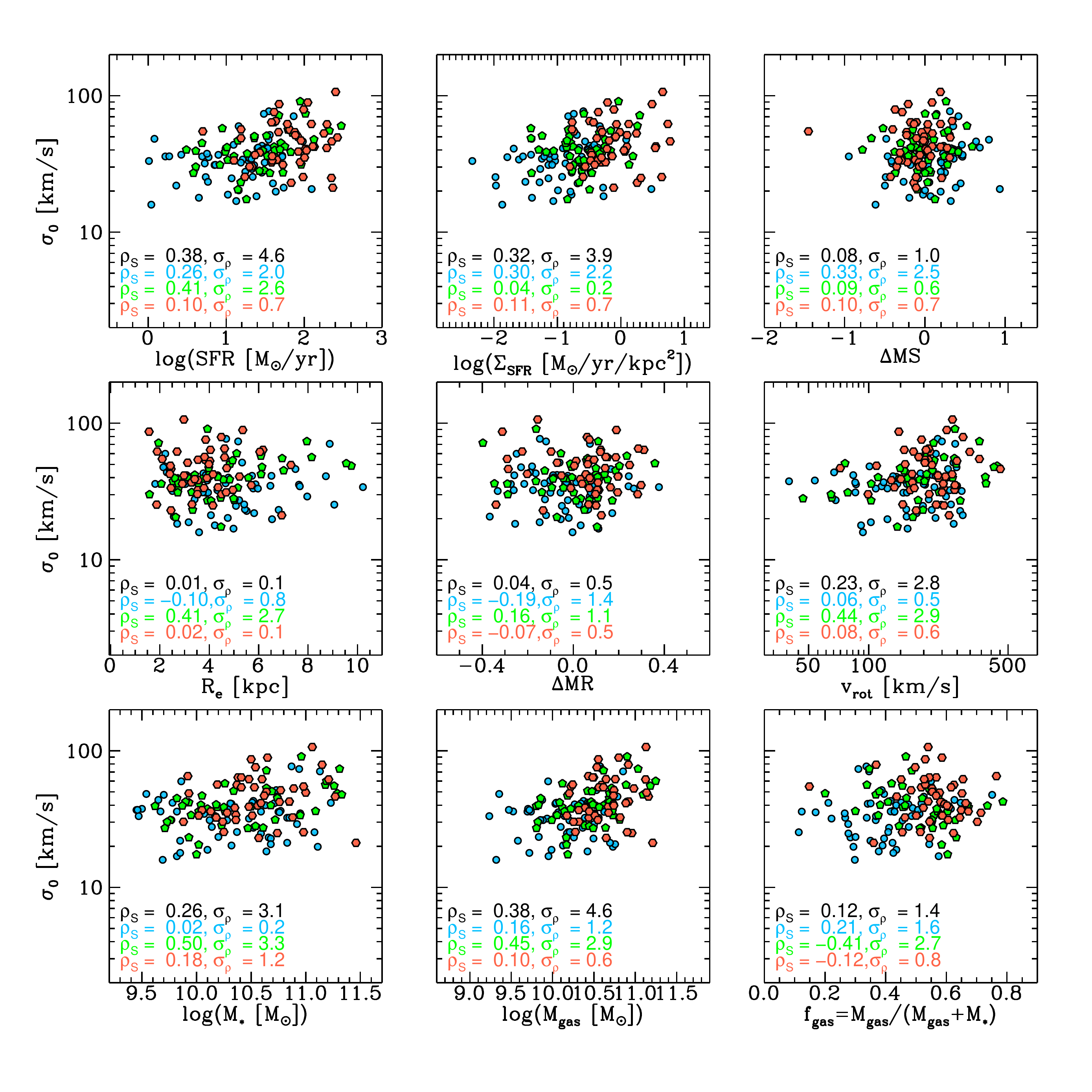}
    \caption{Velocity dispersion as a function of several physical properties. Colors show our redshift subsamples at $z\sim0.9$ (blue), $z\sim1.5$ (green), and $z\sim2.3$ (red). Spearman rank correlation coefficients $\rho_{\rm S}$ and their significance $\sigma_\rho$ are listed in each panel for the full sample (black) and the individual redshift bins (colors). Velocity dispersion positively correlates with several physical properties, some of which correlate themselves with redshift. For our kinematic \kd sample, we find the strongest and most significant correlations between $\sigma_0$ and SFR, as well as $M_{\rm gas}$, which we further investigate in Section~\ref{drivers}.}
    \label{fig:sig0_trends}
\end{figure*}

We show correlations of various physical properties with velocity dispersion after (see Equation~\eqref{eq:norm}) and before correcting for the redshift dependence of $\sigma_0$ in Figures~\ref{fig:normdisp_trends} and \ref{fig:sig0_trends} (see also Table~\ref{tab:physpropcorr}). While several properties positively correlate with $\sigma_0$, particularly SFR and $M_{\rm gas}$, we do not find any significant correlation after correcting for the redshift-dependence of $\sigma_0$. This means that we cannot readily identify a single physical driving source behind the intrinsic scatter in $\sigma_0$ (see discussions in Sections~\ref{scatter} and \ref{drivers}).

\clearpage

\bibliographystyle{aasjournal.bst}
\bibliography{literature}

\begin{thebibliography}{}
\expandafter\ifx\csname natexlab\endcsname\relax\def\natexlab#1{#1}\fi

\bibitem[{{Agertz} {et~al.}(2009{\natexlab{a}}){Agertz}, {Lake}, {Teyssier},
  {Moore}, {Mayer}, \& {Romeo}}]{Agertz09a}
{Agertz}, O., {Lake}, G., {Teyssier}, R., {et~al.} 2009{\natexlab{a}}, \mnras,
  392, 294

\bibitem[{{Agertz} {et~al.}(2009{\natexlab{b}}){Agertz}, {Teyssier}, \&
  {Moore}}]{Agertz09b}
{Agertz}, O., {Teyssier}, R., \& {Moore}, B. 2009{\natexlab{b}}, \mnras, 397,
  L64

\bibitem[{{Athanassoula} \& {Sellwood}(1986)}]{Athanassoula86}
{Athanassoula}, E., \& {Sellwood}, J.~A. 1986, \mnras, 221, 213

\bibitem[{{Aumer} \& {Binney}(2017)}]{Aumer17}
{Aumer}, M., \& {Binney}, J. 2017, \mnras, 470, 2113

\bibitem[{{Aumer} {et~al.}(2016){Aumer}, {Binney}, \&
  {Sch{\"o}nrich}}]{Aumer16}
{Aumer}, M., {Binney}, J., \& {Sch{\"o}nrich}, R. 2016, \mnras, 462, 1697

\bibitem[{{Aumer} {et~al.}(2010){Aumer}, {Burkert}, {Johansson}, \&
  {Genzel}}]{Aumer10}
{Aumer}, M., {Burkert}, A., {Johansson}, P.~H., \& {Genzel}, R. 2010, \apj,
  719, 1230

\bibitem[{{Baumgartner} \& {Breitschwerdt}(2013)}]{Baumgartner13}
{Baumgartner}, V., \& {Breitschwerdt}, D. 2013, \aap, 557, A140

\bibitem[{{Behrendt} {et~al.}(2015){Behrendt}, {Burkert}, \&
  {Schartmann}}]{Behrendt15}
{Behrendt}, M., {Burkert}, A., \& {Schartmann}, M. 2015, \mnras, 448, 1007

\bibitem[{{Binney} \& {Tremaine}(2008)}]{BT08}
{Binney}, J., \& {Tremaine}, S. 2008, {Galactic Dynamics: Second Edition}
  (Princeton University Press)

\bibitem[{{Bolatto} {et~al.}(2017){Bolatto}, {Wong}, {Utomo}, {Blitz}, {Vogel},
  {S{\'a}nchez}, {Barrera-Ballesteros}, {Cao}, {Colombo}, {Dannerbauer},
  {Garc{\'\i}a-Benito}, {Herrera-Camus}, {Husemann}, {Kalinova}, {Leroy},
  {Leung}, {Levy}, {Mast}, {Ostriker}, {Rosolowsky}, {Sandstrom}, {Teuben},
  {van de Ven}, \& {Walter}}]{Bolatto17}
{Bolatto}, A.~D., {Wong}, T., {Utomo}, D., {et~al.} 2017, \apj, 846, 159

\bibitem[{{Bottema}(2003)}]{Bottema03}
{Bottema}, R. 2003, \mnras, 344, 358

\bibitem[{{Bouch{\'e}} {et~al.}(2010){Bouch{\'e}}, {Dekel}, {Genzel}, {Genel},
  {Cresci}, {F{\"o}rster Schreiber}, {Shapiro}, {Davies}, \&
  {Tacconi}}]{Bouche10}
{Bouch{\'e}}, N., {Dekel}, A., {Genzel}, R., {et~al.} 2010, \apj, 718, 1001

\bibitem[{{Boulanger} \& {Viallefond}(1992)}]{Boulanger92}
{Boulanger}, F., \& {Viallefond}, F. 1992, \aap, 266, 37

\bibitem[{{Bovy} {et~al.}(2012){Bovy}, {Rix}, {Liu}, {Hogg}, {Beers}, \&
  {Lee}}]{Bovy12b}
{Bovy}, J., {Rix}, H.-W., {Liu}, C., {et~al.} 2012, \apj, 753, 148

\bibitem[{{Bovy} {et~al.}(2016){Bovy}, {Rix}, {Schlafly}, {Nidever},
  {Holtzman}, {Shetrone}, \& {Beers}}]{Bovy16}
{Bovy}, J., {Rix}, H.-W., {Schlafly}, E.~F., {et~al.} 2016, \apj, 823, 30

\bibitem[{{Brammer} {et~al.}(2011){Brammer}, {Whitaker}, {van Dokkum},
  {Marchesini}, {Franx}, {Kriek}, {Labb{\'e}}, {Lee}, {Muzzin}, {Quadri},
  {Rudnick}, \& {Williams}}]{Brammer11}
{Brammer}, G.~B., {Whitaker}, K.~E., {van Dokkum}, P.~G., {et~al.} 2011, \apj,
  739, 24

\bibitem[{{Brammer} {et~al.}(2012){Brammer}, {van Dokkum}, {Franx},
  {Fumagalli}, {Patel}, {Rix}, {Skelton}, {Kriek}, {Nelson}, {Schmidt},
  {Bezanson}, {da Cunha}, {Erb}, {Fan}, {F{\"o}rster Schreiber}, {Illingworth},
  {Labb{\'e}}, {Leja}, {Lundgren}, {Magee}, {Marchesini}, {McCarthy},
  {Momcheva}, {Muzzin}, {Quadri}, {Steidel}, {Tal}, {Wake}, {Whitaker}, \&
  {Williams}}]{Brammer12}
{Brammer}, G.~B., {van Dokkum}, P.~G., {Franx}, M., {et~al.} 2012, \apjs, 200,
  13

\bibitem[{{Brunt} {et~al.}(2009){Brunt}, {Heyer}, \& {Mac Low}}]{Brunt09}
{Brunt}, C.~M., {Heyer}, M.~H., \& {Mac Low}, M.~M. 2009, \aap, 504, 883

\bibitem[{{Bruzual} \& {Charlot}(2003)}]{Bruzual03}
{Bruzual}, G., \& {Charlot}, S. 2003, \mnras, 344, 1000

\bibitem[{{Burkert} {et~al.}(1992){Burkert}, {Truran}, \&
  {Hensler}}]{Burkert92}
{Burkert}, A., {Truran}, J.~W., \& {Hensler}, G. 1992, \apj, 391, 651

\bibitem[{{Burkert} {et~al.}(2010){Burkert}, {Genzel}, {Bouch{\'e}}, {Cresci},
  {Khochfar}, {Sommer-Larsen}, {Sternberg}, {Naab}, {F{\"o}rster Schreiber},
  {Tacconi}, {Shapiro}, {Hicks}, {Lutz}, {Davies}, {Buschkamp}, \&
  {Genel}}]{Burkert10}
{Burkert}, A., {Genzel}, R., {Bouch{\'e}}, N., {et~al.} 2010, \apj, 725, 2324

\bibitem[{{Burkert} {et~al.}(2016){Burkert}, {F{\"o}rster Schreiber}, {Genzel},
  {Lang}, {Tacconi}, {Wisnioski}, {Wuyts}, {Bandara}, {Beifiori}, {Bender},
  {Brammer}, {Chan}, {Davies}, {Dekel}, {Fabricius}, {Fossati}, {Kulkarni},
  {Lutz}, {Mendel}, {Momcheva}, {Nelson}, {Naab}, {Renzini}, {Saglia},
  {Sharples}, {Sternberg}, {Wilman}, \& {Wuyts}}]{Burkert16}
{Burkert}, A., {F{\"o}rster Schreiber}, N.~M., {Genzel}, R., {et~al.} 2016,
  \apj, 826, 214

\bibitem[{{Cacciato} {et~al.}(2012){Cacciato}, {Dekel}, \&
  {Genel}}]{Cacciato12}
{Cacciato}, M., {Dekel}, A., \& {Genel}, S. 2012, \mnras, 421, 818

\bibitem[{{Cald{\'u}-Primo} \& {Schruba}(2016)}]{CalduPrimo16}
{Cald{\'u}-Primo}, A., \& {Schruba}, A. 2016, \aj, 151, 34

\bibitem[{{Cald{\'u}-Primo} {et~al.}(2013){Cald{\'u}-Primo}, {Schruba},
  {Walter}, {Leroy}, {Sandstrom}, {de Blok}, {Ianjamasimanana}, \&
  {Mogotsi}}]{CalduPrimo13}
{Cald{\'u}-Primo}, A., {Schruba}, A., {Walter}, F., {et~al.} 2013, \aj, 146,
  150

\bibitem[{{Calzetti} {et~al.}(2000){Calzetti}, {Armus}, {Bohlin}, {Kinney},
  {Koornneef}, \& {Storchi-Bergmann}}]{Calzetti00}
{Calzetti}, D., {Armus}, L., {Bohlin}, R.~C., {et~al.} 2000, \apj, 533, 682

\bibitem[{{Cappellari} {et~al.}(2013){Cappellari}, {Scott}, {Alatalo}, {Blitz},
  {Bois}, {Bournaud}, {Bureau}, {Crocker}, {Davies}, {Davis}, {de Zeeuw},
  {Duc}, {Emsellem}, {Khochfar}, {Krajnovi{\'c}}, {Kuntschner}, {McDermid},
  {Morganti}, {Naab}, {Oosterloo}, {Sarzi}, {Serra}, {Weijmans}, \&
  {Young}}]{Cappellari13}
{Cappellari}, M., {Scott}, N., {Alatalo}, K., {et~al.} 2013, \mnras, 432, 1709

\bibitem[{{Ceverino} {et~al.}(2010){Ceverino}, {Dekel}, \&
  {Bournaud}}]{Ceverino10}
{Ceverino}, D., {Dekel}, A., \& {Bournaud}, F. 2010, \mnras, 404, 2151

\bibitem[{{Chabrier}(2003)}]{Chabrier03}
{Chabrier}, G. 2003, Publications of the Astronomical Society of the Pacific,
  115, 763

\bibitem[{{Contini} {et~al.}(2012){Contini}, {Garilli}, {Le F{\`e}vre},
  {Kissler-Patig}, {Amram}, {Epinat}, {Moultaka}, {Paioro}, {Queyrel}, {Tasca},
  {Tresse}, {Vergani}, {L{\'o}pez-Sanjuan}, \& {Perez-Montero}}]{Contini12}
{Contini}, T., {Garilli}, B., {Le F{\`e}vre}, O., {et~al.} 2012, \aap, 539, A91

\bibitem[{{Cresci} {et~al.}(2009){Cresci}, {Hicks}, {Genzel}, {Schreiber},
  {Davies}, {Bouch{\'e}}, {Buschkamp}, {Genel}, {Shapiro}, {Tacconi},
  {Sommer-Larsen}, {Burkert}, {Eisenhauer}, {Gerhard}, {Lutz}, {Naab},
  {Sternberg}, {Cimatti}, {Daddi}, {Erb}, {Kurk}, {Lilly}, {Renzini},
  {Shapley}, {Steidel}, \& {Caputi}}]{Cresci09}
{Cresci}, G., {Hicks}, E.~K.~S., {Genzel}, R., {et~al.} 2009, \apj, 697, 115

\bibitem[{{Dalcanton} \& {Bernstein}(2002)}]{Dalcanton02}
{Dalcanton}, J.~J., \& {Bernstein}, R.~A. 2002, \aj, 124, 1328

\bibitem[{{Danovich} {et~al.}(2015){Danovich}, {Dekel}, {Hahn}, {Ceverino}, \&
  {Primack}}]{Danovich15}
{Danovich}, M., {Dekel}, A., {Hahn}, O., {Ceverino}, D., \& {Primack}, J. 2015,
  \mnras, 449, 2087

\bibitem[{{Dav{\'e}} {et~al.}(2012){Dav{\'e}}, {Finlator}, \&
  {Oppenheimer}}]{Dave12}
{Dav{\'e}}, R., {Finlator}, K., \& {Oppenheimer}, B.~D. 2012, \mnras, 421, 98

\bibitem[{{Davies} {et~al.}(2011){Davies}, {F{\"o}rster Schreiber}, {Cresci},
  {Genzel}, {Bouch{\'e}}, {Burkert}, {Buschkamp}, {Genel}, {Hicks}, {Kurk},
  {Lutz}, {Newman}, {Shapiro}, {Sternberg}, {Tacconi}, \& {Wuyts}}]{Davies11}
{Davies}, R., {F{\"o}rster Schreiber}, N.~M., {Cresci}, G., {et~al.} 2011,
  \apj, 741, 69

\bibitem[{{Davies} {et~al.}(2009){Davies}, {Maciejewski}, {Hicks}, {Tacconi},
  {Genzel}, \& {Engel}}]{Davies09}
{Davies}, R.~I., {Maciejewski}, W., {Hicks}, E.~K.~S., {et~al.} 2009, \apj,
  702, 114

\bibitem[{{Davis} {et~al.}(2003){Davis}, {Faber}, {Newman}, {Phillips},
  {Ellis}, {Steidel}, {Conselice}, {Coil}, {Finkbeiner}, {Koo}, {Guhathakurta},
  {Weiner}, {Schiavon}, {Willmer}, {Kaiser}, {Luppino}, {Wirth}, {Connolly},
  {Eisenhardt}, {Cooper}, \& {Gerke}}]{Davis03}
{Davis}, M., {Faber}, S.~M., {Newman}, J., {et~al.} 2003, in Society of
  Photo-Optical Instrumentation Engineers (SPIE) Conference Series, Vol. 4834,
  Discoveries and Research Prospects from 6- to 10-Meter-Class Telescopes II,
  ed. P.~{Guhathakurta}, 161--172

\bibitem[{{Dekel} \& {Burkert}(2014)}]{Dekel14}
{Dekel}, A., \& {Burkert}, A. 2014, \mnras, 438, 1870

\bibitem[{{Dekel} {et~al.}(2009{\natexlab{a}}){Dekel}, {Sari}, \&
  {Ceverino}}]{Dekel09b}
{Dekel}, A., {Sari}, R., \& {Ceverino}, D. 2009{\natexlab{a}}, \apj, 703, 785

\bibitem[{{Dekel} {et~al.}(2009{\natexlab{b}}){Dekel}, {Birnboim}, {Engel},
  {Freundlich}, {Goerdt}, {Mumcuoglu}, {Neistein}, {Pichon}, {Teyssier}, \&
  {Zinger}}]{Dekel09}
{Dekel}, A., {Birnboim}, Y., {Engel}, G., {et~al.} 2009{\natexlab{b}}, \nat,
  457, 451

\bibitem[{{Di Teodoro} {et~al.}(2016){Di Teodoro}, {Fraternali}, \&
  {Miller}}]{DiTeodoro16}
{Di Teodoro}, E.~M., {Fraternali}, F., \& {Miller}, S.~H. 2016, \aap, 594, A77

\bibitem[{{Dib} {et~al.}(2006){Dib}, {Bell}, \& {Burkert}}]{Dib06}
{Dib}, S., {Bell}, E., \& {Burkert}, A. 2006, \apj, 638, 797

\bibitem[{{Dickey} {et~al.}(1990){Dickey}, {Hanson}, \& {Helou}}]{Dickey90}
{Dickey}, J.~M., {Hanson}, M.~M., \& {Helou}, G. 1990, \apj, 352, 522

\bibitem[{{Druard} {et~al.}(2014){Druard}, {Braine}, {Schuster}, {Schneider},
  {Gratier}, {Bontemps}, {Boquien}, {Combes}, {Corbelli}, {Henkel}, {Herpin},
  {Kramer}, {van der Tak}, \& {van der Werf}}]{Druard14}
{Druard}, C., {Braine}, J., {Schuster}, K.~F., {et~al.} 2014, \aap, 567, A118

\bibitem[{{Dutton} \& {Macci{\`o}}(2014)}]{Dutton14}
{Dutton}, A.~A., \& {Macci{\`o}}, A.~V. 2014, \mnras, 441, 3359

\bibitem[{{Elmegreen} \& {Burkert}(2010)}]{ElmegreenB10}
{Elmegreen}, B.~G., \& {Burkert}, A. 2010, \apj, 712, 294

\bibitem[{{Elmegreen} \& {Elmegreen}(2006)}]{ElmegreenB06}
{Elmegreen}, B.~G., \& {Elmegreen}, D.~M. 2006, \apj, 650, 644

\bibitem[{{Elmegreen} \& {Scalo}(2004)}]{ElmegreenB04}
{Elmegreen}, B.~G., \& {Scalo}, J. 2004, \araa, 42, 211

\bibitem[{{Elmegreen} {et~al.}(2007){Elmegreen}, {Elmegreen}, {Ravindranath},
  \& {Coe}}]{ElmegreenD07}
{Elmegreen}, D.~M., {Elmegreen}, B.~G., {Ravindranath}, S., \& {Coe}, D.~A.
  2007, \apj, 658, 763

\bibitem[{{Epinat} {et~al.}(2010){Epinat}, {Amram}, {Balkowski}, \&
  {Marcelin}}]{Epinat10}
{Epinat}, B., {Amram}, P., {Balkowski}, C., \& {Marcelin}, M. 2010, \mnras,
  401, 2113

\bibitem[{{Epinat} {et~al.}(2008){Epinat}, {Amram}, \& {Marcelin}}]{Epinat08}
{Epinat}, B., {Amram}, P., \& {Marcelin}, M. 2008, \mnras, 390, 466

\bibitem[{{Epinat} {et~al.}(2009){Epinat}, {Contini}, {Le F{\`e}vre},
  {Vergani}, {Garilli}, {Amram}, {Queyrel}, {Tasca}, \& {Tresse}}]{Epinat09}
{Epinat}, B., {Contini}, T., {Le F{\`e}vre}, O., {et~al.} 2009, \aap, 504, 789

\bibitem[{{Epinat} {et~al.}(2012){Epinat}, {Tasca}, {Amram}, {Contini}, {Le
  F{\`e}vre}, {Queyrel}, {Vergani}, {Garilli}, {Kissler-Patig}, {Moultaka},
  {Paioro}, {Tresse}, {Bournaud}, {L{\'o}pez-Sanjuan}, \& {Perret}}]{Epinat12}
{Epinat}, B., {Tasca}, L., {Amram}, P., {et~al.} 2012, \aap, 539, A92

\bibitem[{{Escala} \& {Larson}(2008)}]{Escala08}
{Escala}, A., \& {Larson}, R.~B. 2008, \apj, 685, L31

\bibitem[{{Fisher} {et~al.}(2019){Fisher}, {Bolatto}, {White}, {Glazebrook},
  {Abraham}, \& {Obreschkow}}]{Fisher19}
{Fisher}, D.~B., {Bolatto}, A.~D., {White}, H., {et~al.} 2019, \apj, 870, 46

\bibitem[{{Fisher} {et~al.}(2017){Fisher}, {Glazebrook}, {Abraham}, {Damjanov},
  {White}, {Obreschkow}, {Basset}, {Bekiaris}, {Wisnioski}, {Green}, \&
  {Bolatto}}]{Fisher17}
{Fisher}, D.~B., {Glazebrook}, K., {Abraham}, R.~G., {et~al.} 2017, \apj, 839,
  L5

\bibitem[{{F{\"o}rster Schreiber} {et~al.}(2006){F{\"o}rster Schreiber},
  {Genzel}, {Lehnert}, {Bouch{\'e}}, {Verma}, {Erb}, {Shapley}, {Steidel},
  {Davies}, {Lutz}, {Nesvadba}, {Tacconi}, {Eisenhauer}, {Abuter}, {Gilbert},
  {Gillessen}, \& {Sternberg}}]{FS06b}
{F{\"o}rster Schreiber}, N.~M., {Genzel}, R., {Lehnert}, M.~D., {et~al.} 2006,
  \apj, 645, 1062

\bibitem[{{F{\"o}rster Schreiber} {et~al.}(2009){F{\"o}rster Schreiber},
  {Genzel}, {Bouch{\'e}}, {Cresci}, {Davies}, {Buschkamp}, {Shapiro},
  {Tacconi}, {Hicks}, {Genel}, {Shapley}, {Erb}, {Steidel}, {Lutz},
  {Eisenhauer}, {Gillessen}, {Sternberg}, {Renzini}, {Cimatti}, {Daddi},
  {Kurk}, {Lilly}, {Kong}, {Lehnert}, {Nesvadba}, {Verma}, {McCracken},
  {Arimoto}, {Mignoli}, \& {Onodera}}]{FS09}
{F{\"o}rster Schreiber}, N.~M., {Genzel}, R., {Bouch{\'e}}, N., {et~al.} 2009,
  \apj, 706, 1364

\bibitem[{{F{\"o}rster Schreiber} {et~al.}(2014){F{\"o}rster Schreiber},
  {Genzel}, {Newman}, {Kurk}, {Lutz}, {Tacconi}, {Wuyts}, {Bandara}, {Burkert},
  {Buschkamp}, {Carollo}, {Cresci}, {Daddi}, {Davies}, {Eisenhauer}, {Hicks},
  {Lang}, {Lilly}, {Mainieri}, {Mancini}, {Naab}, {Peng}, {Renzini}, {Rosario},
  {Shapiro Griffin}, {Shapley}, {Sternberg}, {Tacchella}, {Vergani},
  {Wisnioski}, {Wuyts}, \& {Zamorani}}]{FS14}
{F{\"o}rster Schreiber}, N.~M., {Genzel}, R., {Newman}, S.~F., {et~al.} 2014,
  \apj, 787, 38

\bibitem[{{F{\"o}rster Schreiber} {et~al.}(2018){F{\"o}rster Schreiber},
  {Renzini}, {Mancini}, {Genzel}, {Bouch{\'e}}, {Cresci}, {Hicks}, {Lilly},
  {Peng}, {Burkert}, {Carollo}, {Cimatti}, {Daddi}, {Davies}, {Genel}, {Kurk},
  {Lang}, {Lutz}, {Mainieri}, {McCracken}, {Mignoli}, {Naab}, {Oesch},
  {Pozzetti}, {Scodeggio}, {Shapiro Griffin}, {Shapley}, {Sternberg},
  {Tacchella}, {Tacconi}, {Wuyts}, \& {Zamorani}}]{FS18}
{F{\"o}rster Schreiber}, N.~M., {Renzini}, A., {Mancini}, C., {et~al.} 2018,
  The Astrophysical Journal Supplement Series, 238, 21

\bibitem[{{Freundlich} {et~al.}(2018){Freundlich}, {Combes}, {Tacconi},
  {Genzel}, {Garcia-Burillo}, {Neri}, {Contini}, {Bolatto}, {Lilly},
  {Salom{\'e}}, {Bicalho}, {Boissier}, {Boone}, {Bouch{\'e}}, {Bournaud},
  {Burkert}, {Carollo}, {Cooper}, {Cox}, {Feruglio}, {Schreiber}, {Juneau},
  {Lippa}, {Lutz}, {Naab}, {Renzini}, {Saintonge}, {Sternberg}, {Walter},
  {Weiner}, {Wei{\ss}}, \& {Wuyts}}]{Freundlich19}
{Freundlich}, J., {Combes}, F., {Tacconi}, L.~J., {et~al.} 2018, arXiv
  e-prints, arXiv:1812.08180

\bibitem[{{Fukui} {et~al.}(2009){Fukui}, {Kawamura}, {Wong}, {Murai},
  {Iritani}, {Mizuno}, {Mizuno}, {Onishi}, {Hughes}, {Ott}, {Muller},
  {Staveley- Smith}, \& {Kim}}]{Fukui09}
{Fukui}, Y., {Kawamura}, A., {Wong}, T., {et~al.} 2009, \apj, 705, 144

\bibitem[{{Gatto} {et~al.}(2015){Gatto}, {Walch}, {Low}, {Naab}, {Girichidis},
  {Glover}, {W{\"u}nsch}, {Klessen}, {Clark}, {Baczynski}, {Peters},
  {Ostriker}, {Ib{\'a}{\~n}ez- Mej{\'\i}a}, \& {Haid}}]{Gatto15}
{Gatto}, A., {Walch}, S., {Low}, M. M.~M., {et~al.} 2015, \mnras, 449, 1057

\bibitem[{{Genel} {et~al.}(2012{\natexlab{a}}){Genel}, {Dekel}, \&
  {Cacciato}}]{Genel12b}
{Genel}, S., {Dekel}, A., \& {Cacciato}, M. 2012{\natexlab{a}}, \mnras, 425,
  788

\bibitem[{{Genel} {et~al.}(2012{\natexlab{b}}){Genel}, {Naab}, {Genzel},
  {F{\"o}rster Schreiber}, {Sternberg}, {Oser}, {Johansson}, {Dav{\'e}},
  {Oppenheimer}, \& {Burkert}}]{Genel12a}
{Genel}, S., {Naab}, T., {Genzel}, R., {et~al.} 2012{\natexlab{b}}, \apj, 745,
  11

\bibitem[{{Genzel} {et~al.}(2006){Genzel}, {Tacconi}, {Eisenhauer},
  {F{\"o}rster Schreiber}, {Cimatti}, {Daddi}, {Bouch{\'e}}, {Davies},
  {Lehnert}, {Lutz}, {Nesvadba}, {Verma}, {Abuter}, {Shapiro}, {Sternberg},
  {Renzini}, {Kong}, {Arimoto}, \& {Mignoli}}]{Genzel06}
{Genzel}, R., {Tacconi}, L.~J., {Eisenhauer}, F., {et~al.} 2006, \nat, 442, 786

\bibitem[{{Genzel} {et~al.}(2008){Genzel}, {Burkert}, {Bouch{\'e}}, {Cresci},
  {F{\"o}rster Schreiber}, {Shapley}, {Shapiro}, {Tacconi}, {Buschkamp},
  {Cimatti}, {Daddi}, {Davies}, {Eisenhauer}, {Erb}, {Genel}, {Gerhard},
  {Hicks}, {Lutz}, {Naab}, {Ott}, {Rabien}, {Renzini}, {Steidel}, {Sternberg},
  \& {Lilly}}]{Genzel08}
{Genzel}, R., {Burkert}, A., {Bouch{\'e}}, N., {et~al.} 2008, \apj, 687, 59

\bibitem[{{Genzel} {et~al.}(2011){Genzel}, {Newman}, {Jones}, {F{\"o}rster
  Schreiber}, {Shapiro}, {Genel}, {Lilly}, {Renzini}, {Tacconi}, {Bouch{\'e}},
  {Burkert}, {Cresci}, {Buschkamp}, {Carollo}, {Ceverino}, {Davies}, {Dekel},
  {Eisenhauer}, {Hicks}, {Kurk}, {Lutz}, {Mancini}, {Naab}, {Peng},
  {Sternberg}, {Vergani}, \& {Zamorani}}]{Genzel11}
{Genzel}, R., {Newman}, S., {Jones}, T., {et~al.} 2011, \apj, 733, 101

\bibitem[{{Genzel} {et~al.}(2013){Genzel}, {Tacconi}, {Kurk}, {Wuyts},
  {Combes}, {Freundlich}, {Bolatto}, {Cooper}, {Neri}, {Nordon}, {Bournaud},
  {Burkert}, {Comerford}, {Cox}, {Davis}, {F{\"o}rster Schreiber},
  {Garc{\'{\i}}a-Burillo}, {Gracia-Carpio}, {Lutz}, {Naab}, {Newman},
  {Saintonge}, {Shapiro Griffin}, {Shapley}, {Sternberg}, \&
  {Weiner}}]{Genzel13}
{Genzel}, R., {Tacconi}, L.~J., {Kurk}, J., {et~al.} 2013, \apj, 773, 68

\bibitem[{{Genzel} {et~al.}(2014){Genzel}, {F{\"o}rster Schreiber}, {Lang},
  {Tacchella}, {Tacconi}, {Wuyts}, {Bandara}, {Burkert}, {Buschkamp},
  {Carollo}, {Cresci}, {Davies}, {Eisenhauer}, {Hicks}, {Kurk}, {Lilly},
  {Lutz}, {Mancini}, {Naab}, {Newman}, {Peng}, {Renzini}, {Shapiro Griffin},
  {Sternberg}, {Vergani}, {Wisnioski}, {Wuyts}, \& {Zamorani}}]{Genzel14b}
{Genzel}, R., {F{\"o}rster Schreiber}, N.~M., {Lang}, P., {et~al.} 2014, \apj,
  785, 75

\bibitem[{{Genzel} {et~al.}(2017){Genzel}, {F{\"o}rster Schreiber},
  {{\"U}bler}, {Lang}, {Naab}, {Bender}, {Tacconi}, {Wisnioski}, {Wuyts},
  {Alexander}, {Beifiori}, {Belli}, {Brammer}, {Burkert}, {Carollo}, {Chan},
  {Davies}, {Fossati}, {Galametz}, {Genel}, {Gerhard}, {Lutz}, {Mendel},
  {Momcheva}, {Nelson}, {Renzini}, {Saglia}, {Sternberg}, {Tacchella},
  {Tadaki}, \& {Wilman}}]{Genzel17}
{Genzel}, R., {F{\"o}rster Schreiber}, N.~M., {{\"U}bler}, H., {et~al.} 2017,
  \nat, 543, 397

\bibitem[{{Gilmore} \& {Reid}(1983)}]{Gilmore83}
{Gilmore}, G., \& {Reid}, N. 1983, \mnras, 202, 1025

\bibitem[{{Girard} {et~al.}(2018){Girard}, {Dessauges-Zavadsky}, {Schaerer},
  {Cirasuolo}, {Turner}, {Cava}, {Rodr{\'\i}guez-Mu{\~n}oz}, {Richard}, \&
  {P{\'e}rez-Gonz{\'a}lez}}]{Girard18}
{Girard}, M., {Dessauges-Zavadsky}, M., {Schaerer}, D., {et~al.} 2018, \aap,
  613, A72

\bibitem[{{Glazebrook}(2013)}]{Glazebrook13}
{Glazebrook}, K. 2013, Publications of the Astronomical Society of Australia,
  30, e056

\bibitem[{{Gnerucci} {et~al.}(2011){Gnerucci}, {Marconi}, {Cresci}, {Maiolino},
  {Mannucci}, {Calura}, {Cimatti}, {Cocchia}, {Grazian}, {Matteucci}, {Nagao},
  {Pozzetti}, \& {Troncoso}}]{Gnerucci11}
{Gnerucci}, A., {Marconi}, A., {Cresci}, G., {et~al.} 2011, \aap, 528, A88

\bibitem[{{Goldreich} \& {Lynden-Bell}(1965)}]{Goldreich65}
{Goldreich}, P., \& {Lynden-Bell}, D. 1965, \mnras, 130, 97

\bibitem[{{Grand} {et~al.}(2016){Grand}, {Springel}, {G{\'o}mez}, {Marinacci},
  {Pakmor}, {Campbell}, \& {Jenkins}}]{Grand16}
{Grand}, R. J.~J., {Springel}, V., {G{\'o}mez}, F.~A., {et~al.} 2016, \mnras,
  459, 199

\bibitem[{{Green} {et~al.}(2010){Green}, {Glazebrook}, {McGregor}, {Abraham},
  {Poole}, {Damjanov}, {McCarthy}, {Colless}, \& {Sharp}}]{Green10}
{Green}, A.~W., {Glazebrook}, K., {McGregor}, P.~J., {et~al.} 2010, \nat, 467,
  684

\bibitem[{{Green} {et~al.}(2014){Green}, {Glazebrook}, {McGregor}, {Damjanov},
  {Wisnioski}, {Abraham}, {Colless}, {Sharp}, {Crain}, {Poole}, \&
  {McCarthy}}]{Green14}
---. 2014, \mnras, 437, 1070

\bibitem[{{Grogin} {et~al.}(2011){Grogin}, {Kocevski}, {Faber}, {Ferguson},
  {Koekemoer}, {Riess}, {Acquaviva}, {Alexander}, {Almaini}, {Ashby}, {Barden},
  {Bell}, {Bournaud}, {Brown}, {Caputi}, {Casertano}, {Cassata}, {Castellano},
  {Challis}, {Chary}, {Cheung}, {Cirasuolo}, {Conselice}, {Roshan Cooray},
  {Croton}, {Daddi}, {Dahlen}, {Dav{\'e}}, {de Mello}, {Dekel}, {Dickinson},
  {Dolch}, {Donley}, {Dunlop}, {Dutton}, {Elbaz}, {Fazio}, {Filippenko},
  {Finkelstein}, {Fontana}, {Gardner}, {Garnavich}, {Gawiser}, {Giavalisco},
  {Grazian}, {Guo}, {Hathi}, {H{\"a}ussler}, {Hopkins}, {Huang}, {Huang},
  {Jha}, {Kartaltepe}, {Kirshner}, {Koo}, {Lai}, {Lee}, {Li}, {Lotz}, {Lucas},
  {Madau}, {McCarthy}, {McGrath}, {McIntosh}, {McLure}, {Mobasher},
  {Moustakas}, {Mozena}, {Nandra}, {Newman}, {Niemi}, {Noeske}, {Papovich},
  {Pentericci}, {Pope}, {Primack}, {Rajan}, {Ravindranath}, {Reddy}, {Renzini},
  {Rix}, {Robaina}, {Rodney}, {Rosario}, {Rosati}, {Salimbeni}, {Scarlata},
  {Siana}, {Simard}, {Smidt}, {Somerville}, {Spinrad}, {Straughn}, {Strolger},
  {Telford}, {Teplitz}, {Trump}, {van der Wel}, {Villforth}, {Wechsler},
  {Weiner}, {Wiklind}, {Wild}, {Wilson}, {Wuyts}, {Yan}, \& {Yun}}]{Grogin11}
{Grogin}, N.~A., {Kocevski}, D.~D., {Faber}, S.~M., {et~al.} 2011, \apjs, 197,
  35

\bibitem[{{Heyer} \& {Dame}(2015)}]{Heyer15}
{Heyer}, M., \& {Dame}, T.~M. 2015, Annual Review of Astronomy and
  Astrophysics, 53, 583

\bibitem[{{Heyer} \& {Brunt}(2004)}]{Heyer04}
{Heyer}, M.~H., \& {Brunt}, C.~M. 2004, \apj, 615, L45

\bibitem[{{Hohl}(1971)}]{Hohl71}
{Hohl}, F. 1971, \apj, 168, 343

\bibitem[{{Hopkins} {et~al.}(2011){Hopkins}, {Quataert}, \&
  {Murray}}]{Hopkins11}
{Hopkins}, P.~F., {Quataert}, E., \& {Murray}, N. 2011, \mnras, 417, 950

\bibitem[{{Hung} {et~al.}(2019){Hung}, {Hayward}, {Yuan}, {Boylan-Kolchin},
  {Faucher-Gigu{\`e}re}, {Hopkins}, {Kere{\v{s}}}, {Murray}, \&
  {Wetzel}}]{Hung19}
{Hung}, C.-L., {Hayward}, C.~C., {Yuan}, T., {et~al.} 2019, \mnras, 482, 5125

\bibitem[{{Ianjamasimanana} {et~al.}(2012){Ianjamasimanana}, {de Blok},
  {Walter}, \& {Heald}}]{Ianjamasimanana12}
{Ianjamasimanana}, R., {de Blok}, W.~J.~G., {Walter}, F., \& {Heald}, G.~H.
  2012, \aj, 144, 96

\bibitem[{{Ianjamasimanana} {et~al.}(2015){Ianjamasimanana}, {de Blok},
  {Walter}, {Heald}, {Cald{\'u}-Primo}, \& {Jarrett}}]{Ianjamasimanana15}
{Ianjamasimanana}, R., {de Blok}, W.~J.~G., {Walter}, F., {et~al.} 2015, \aj,
  150, 47

\bibitem[{{Immeli} {et~al.}(2004{\natexlab{a}}){Immeli}, {Samland}, {Gerhard},
  \& {Westera}}]{Immeli04a}
{Immeli}, A., {Samland}, M., {Gerhard}, O., \& {Westera}, P.
  2004{\natexlab{a}}, \aap, 413, 547

\bibitem[{{Immeli} {et~al.}(2004{\natexlab{b}}){Immeli}, {Samland}, {Westera},
  \& {Gerhard}}]{Immeli04b}
{Immeli}, A., {Samland}, M., {Westera}, P., \& {Gerhard}, O.
  2004{\natexlab{b}}, \apj, 611, 20

\bibitem[{{Inoue} {et~al.}(2016){Inoue}, {Dekel}, {Mandelker}, {Ceverino},
  {Bournaud}, \& {Primack}}]{Inoue16}
{Inoue}, S., {Dekel}, A., {Mandelker}, N., {et~al.} 2016, \mnras, 456, 2052

\bibitem[{{Johnson} {et~al.}(2018){Johnson}, {Harrison}, {Swinbank}, {Tiley},
  {Stott}, {Bower}, {Smail}, {Bunker}, {Sobral}, {Turner}, {Best}, {Bureau},
  {Cirasuolo}, {Jarvis}, {Magdis}, {Sharples}, {Bland-Hawthorn}, {Catinella},
  {Cortese}, {Croom}, {Federrath}, {Glazebrook}, {Sweet}, {Bryant}, {Goodwin},
  {Konstantopoulos}, {Lawrence}, {Medling}, {Owers}, \& {Richards}}]{Johnson18}
{Johnson}, H.~L., {Harrison}, C.~M., {Swinbank}, A.~M., {et~al.} 2018, \mnras,
  474, 5076

\bibitem[{{Jones} {et~al.}(2010){Jones}, {Swinbank}, {Ellis}, {Richard}, \&
  {Stark}}]{Jones10}
{Jones}, T.~A., {Swinbank}, A.~M., {Ellis}, R.~S., {Richard}, J., \& {Stark},
  D.~P. 2010, \mnras, 404, 1247

\bibitem[{{Joung} {et~al.}(2009){Joung}, {Mac Low}, \& {Bryan}}]{Joung09}
{Joung}, M.~R., {Mac Low}, M.-M., \& {Bryan}, G.~L. 2009, \apj, 704, 137

\bibitem[{{Juri{\'c}} {et~al.}(2008){Juri{\'c}}, {Ivezi{\'c}}, {Brooks},
  {Lupton}, {Schlegel}, {Finkbeiner}, {Padmanabhan}, {Bond}, {Sesar},
  {Rockosi}, {Knapp}, {Gunn}, {Sumi}, {Schneider}, {Barentine}, {Brewington},
  {Brinkmann}, {Fukugita}, {Harvanek}, {Kleinman}, {Krzesinski}, {Long},
  {Neilsen}, {Nitta}, {Snedden}, \& {York}}]{Juric08}
{Juri{\'c}}, M., {Ivezi{\'c}}, {\v{Z}}., {Brooks}, A., {et~al.} 2008, \apj,
  673, 864

\bibitem[{{Kamphuis} \& {Sancisi}(1993)}]{Kamphuis93}
{Kamphuis}, J., \& {Sancisi}, R. 1993, \aap, 273, L31

\bibitem[{{Kassin} {et~al.}(2007){Kassin}, {Weiner}, {Faber}, {Koo}, {Lotz},
  {Diemand}, {Harker}, {Bundy}, {Metevier}, {Phillips}, {Cooper}, {Croton},
  {Konidaris}, {Noeske}, \& {Willmer}}]{Kassin07}
{Kassin}, S.~A., {Weiner}, B.~J., {Faber}, S.~M., {et~al.} 2007, \apjl, 660,
  L35

\bibitem[{{Kassin} {et~al.}(2012){Kassin}, {Weiner}, {Faber}, {Gardner},
  {Willmer}, {Coil}, {Cooper}, {Devriendt}, {Dutton}, {Guhathakurta}, {Koo},
  {Metevier}, {Noeske}, \& {Primack}}]{Kassin12}
---. 2012, \apj, 758, 106

\bibitem[{{Kelly}(2007)}]{Kelly07}
{Kelly}, B.~C. 2007, \apj, 665, 1489

\bibitem[{{Kim} \& {Ostriker}(2018)}]{Kim18}
{Kim}, C.-G., \& {Ostriker}, E.~C. 2018, \apj, 853, 173

\bibitem[{{Kim} {et~al.}(2013){Kim}, {Ostriker}, \& {Kim}}]{Kim13}
{Kim}, C.-G., {Ostriker}, E.~C., \& {Kim}, W.-T. 2013, \apj, 776, 1

\bibitem[{{Kim} \& {Ostriker}(2007)}]{Kim07}
{Kim}, W.-T., \& {Ostriker}, E.~C. 2007, \apj, 660, 1232

\bibitem[{{Koch} {et~al.}(2019){Koch}, {Rosolowsky}, {Schruba}, {Leroy},
  {Kepley}, {Braine}, {Dalcanton}, \& {Johnson}}]{Koch19}
{Koch}, E.~W., {Rosolowsky}, E.~W., {Schruba}, A., {et~al.} 2019, arXiv
  e-prints, arXiv:1902.05007

\bibitem[{{Koekemoer} {et~al.}(2011){Koekemoer}, {Faber}, {Ferguson}, {Grogin},
  {Kocevski}, {Koo}, {Lai}, {Lotz}, {Lucas}, {McGrath}, {Ogaz}, {Rajan},
  {Riess}, {Rodney}, {Strolger}, {Casertano}, {Castellano}, {Dahlen},
  {Dickinson}, {Dolch}, {Fontana}, {Giavalisco}, {Grazian}, {Guo}, {Hathi},
  {Huang}, {van der Wel}, {Yan}, {Acquaviva}, {Alexander}, {Almaini}, {Ashby},
  {Barden}, {Bell}, {Bournaud}, {Brown}, {Caputi}, {Cassata}, {Challis},
  {Chary}, {Cheung}, {Cirasuolo}, {Conselice}, {Roshan Cooray}, {Croton},
  {Daddi}, {Dav{\'e}}, {de Mello}, {de Ravel}, {Dekel}, {Donley}, {Dunlop},
  {Dutton}, {Elbaz}, {Fazio}, {Filippenko}, {Finkelstein}, {Frazer}, {Gardner},
  {Garnavich}, {Gawiser}, {Gruetzbauch}, {Hartley}, {H{\"a}ussler},
  {Herrington}, {Hopkins}, {Huang}, {Jha}, {Johnson}, {Kartaltepe},
  {Khostovan}, {Kirshner}, {Lani}, {Lee}, {Li}, {Madau}, {McCarthy},
  {McIntosh}, {McLure}, {McPartland}, {Mobasher}, {Moreira}, {Mortlock},
  {Moustakas}, {Mozena}, {Nandra}, {Newman}, {Nielsen}, {Niemi}, {Noeske},
  {Papovich}, {Pentericci}, {Pope}, {Primack}, {Ravindranath}, {Reddy},
  {Renzini}, {Rix}, {Robaina}, {Rosario}, {Rosati}, {Salimbeni}, {Scarlata},
  {Siana}, {Simard}, {Smidt}, {Snyder}, {Somerville}, {Spinrad}, {Straughn},
  {Telford}, {Teplitz}, {Trump}, {Vargas}, {Villforth}, {Wagner}, {Wandro},
  {Wechsler}, {Weiner}, {Wiklind}, {Wild}, {Wilson}, {Wuyts}, \&
  {Yun}}]{Koekemoer11}
{Koekemoer}, A.~M., {Faber}, S.~M., {Ferguson}, H.~C., {et~al.} 2011, \apjs,
  197, 36

\bibitem[{{Kriek} {et~al.}(2015){Kriek}, {Shapley}, {Reddy}, {Siana}, {Coil},
  {Mobasher}, {Freeman}, {de Groot}, {Price}, {Sanders}, {Shivaei}, {Brammer},
  {Momcheva}, {Skelton}, {van Dokkum}, {Whitaker}, {Aird}, {Azadi}, {Kassis},
  {Bullock}, {Conroy}, {Dav{\'e}}, {Kere{\v s}}, \& {Krumholz}}]{Kriek15}
{Kriek}, M., {Shapley}, A.~E., {Reddy}, N.~A., {et~al.} 2015, \apjs, 218, 15

\bibitem[{{Krumholz} \& {Burkert}(2010)}]{Krumholz10}
{Krumholz}, M., \& {Burkert}, A. 2010, \apj, 724, 895

\bibitem[{{Krumholz} \& {Burkhart}(2016)}]{Krumholz16}
{Krumholz}, M.~R., \& {Burkhart}, B. 2016, \mnras, 458, 1671

\bibitem[{{Krumholz} {et~al.}(2018){Krumholz}, {Burkhart}, {Forbes}, \&
  {Crocker}}]{Krumholz18}
{Krumholz}, M.~R., {Burkhart}, B., {Forbes}, J.~C., \& {Crocker}, R.~M. 2018,
  \mnras, 477, 2716

\bibitem[{{Labb{\'e}} {et~al.}(2003){Labb{\'e}}, {Rudnick}, {Franx}, {Daddi},
  {van Dokkum}, {F{\"o}rster Schreiber}, {Kuijken}, {Moorwood}, {Rix},
  {R{\"o}ttgering}, {Trujillo}, {van der Wel}, {van der Werf}, \& {van
  Starkenburg}}]{Labbe03}
{Labb{\'e}}, I., {Rudnick}, G., {Franx}, M., {et~al.} 2003, \apjl, 591, L95

\bibitem[{{Lang} {et~al.}(2014){Lang}, {Wuyts}, {Somerville}, {F{\"o}rster
  Schreiber}, {Genzel}, {Bell}, {Brammer}, {Dekel}, {Faber}, {Ferguson},
  {Grogin}, {Kocevski}, {Koekemoer}, {Lutz}, {McGrath}, {Momcheva}, {Nelson},
  {Primack}, {Rosario}, {Skelton}, {Tacconi}, {van Dokkum}, \&
  {Whitaker}}]{Lang14}
{Lang}, P., {Wuyts}, S., {Somerville}, R.~S., {et~al.} 2014, \apj, 788, 11

\bibitem[{{Lang} {et~al.}(2017){Lang}, {F{\"o}rster Schreiber}, {Genzel},
  {Wuyts}, {Wisnioski}, {Beifiori}, {Belli}, {Bender}, {Brammer}, {Burkert},
  {Chan}, {Davies}, {Fossati}, {Galametz}, {Kulkarni}, {Lutz}, {Mendel},
  {Momcheva}, {Naab}, {Nelson}, {Saglia}, {Seitz}, {Tacchella}, {Tacconi},
  {Tadaki}, {{\"U}bler}, {van Dokkum}, \& {Wilman}}]{Lang17}
{Lang}, P., {F{\"o}rster Schreiber}, N.~M., {Genzel}, R., {et~al.} 2017, \apj,
  840, 92

\bibitem[{{Larson}(1981)}]{Larson81}
{Larson}, R.~B. 1981, \mnras, 194, 809

\bibitem[{{Law} {et~al.}(2009){Law}, {Steidel}, {Erb}, {Larkin}, {Pettini},
  {Shapley}, \& {Wright}}]{Law09}
{Law}, D.~R., {Steidel}, C.~C., {Erb}, D.~K., {et~al.} 2009, \apj, 697, 2057

\bibitem[{{Leaman} {et~al.}(2017){Leaman}, {Mendel}, {Wisnioski}, {Brooks},
  {Beasley}, {Starkenburg}, {Martig}, {Battaglia}, {Christensen}, {Cole}, {de
  Boer}, \& {Wills}}]{Leaman17}
{Leaman}, R., {Mendel}, J.~T., {Wisnioski}, E., {et~al.} 2017, \mnras, 472,
  1879

\bibitem[{{Lehnert} {et~al.}(2013){Lehnert}, {Le Tiran}, {Nesvadba}, {van
  Driel}, {Boulanger}, \& {Di Matteo}}]{Lehnert13}
{Lehnert}, M.~D., {Le Tiran}, L., {Nesvadba}, N.~P.~H., {et~al.} 2013, \aap,
  555, A72

\bibitem[{{Lehnert} {et~al.}(2009){Lehnert}, {Nesvadba}, {Le Tiran}, {Di
  Matteo}, {van Driel}, {Douglas}, {Chemin}, \& {Bournaud}}]{Lehnert09}
{Lehnert}, M.~D., {Nesvadba}, N.~P.~H., {Le Tiran}, L., {et~al.} 2009, \apj,
  699, 1660

\bibitem[{{Leroy} {et~al.}(2008){Leroy}, {Walter}, {Brinks}, {Bigiel}, {de
  Blok}, {Madore}, \& {Thornley}}]{Leroy08}
{Leroy}, A.~K., {Walter}, F., {Brinks}, E., {et~al.} 2008, \aj, 136, 2782

\bibitem[{{Leroy} {et~al.}(2009){Leroy}, {Walter}, {Bigiel}, {Usero}, {Weiss},
  {Brinks}, {de Blok}, {Kennicutt}, {Schuster}, {Kramer}, {Wiesemeyer}, \&
  {Roussel}}]{Leroy09}
{Leroy}, A.~K., {Walter}, F., {Bigiel}, F., {et~al.} 2009, \aj, 137, 4670

\bibitem[{{Levy} {et~al.}(2018){Levy}, {Bolatto}, {Teuben}, {S{\'a}nchez},
  {Barrera-Ballesteros}, {Blitz}, {Colombo}, {Garc{\'\i}a-Benito},
  {Herrera-Camus}, {Husemann}, {Kalinova}, {Lan}, {Leung}, {Mast}, {Utomo},
  {van de Ven}, {Vogel}, \& {Wong}}]{Levy18}
{Levy}, R.~C., {Bolatto}, A.~D., {Teuben}, P., {et~al.} 2018, \apj, 860, 92

\bibitem[{{Lilly} {et~al.}(2013){Lilly}, {Carollo}, {Pipino}, {Renzini}, \&
  {Peng}}]{Lilly13}
{Lilly}, S.~J., {Carollo}, C.~M., {Pipino}, A., {Renzini}, A., \& {Peng}, Y.
  2013, \apj, 772, 119

\bibitem[{{Livermore} {et~al.}(2015){Livermore}, {Jones}, {Richard}, {Bower},
  {Swinbank}, {Yuan}, {Edge}, {Ellis}, {Kewley}, {Smail}, {Coppin}, \&
  {Ebeling}}]{Livermore15}
{Livermore}, R.~C., {Jones}, T.~A., {Richard}, J., {et~al.} 2015, \mnras, 450,
  1812

\bibitem[{{Lutz} {et~al.}(2011){Lutz}, {Poglitsch}, {Altieri}, {Andreani},
  {Aussel}, {Berta}, {Bongiovanni}, {Brisbin}, {Cava}, {Cepa}, {Cimatti},
  {Daddi}, {Dominguez-Sanchez}, {Elbaz}, {F{\"o}rster Schreiber}, {Genzel},
  {Grazian}, {Gruppioni}, {Harwit}, {Le Floc'h}, {Magdis}, {Magnelli},
  {Maiolino}, {Nordon}, {P{\'e}rez Garc{\'{\i}}a}, {Popesso}, {Pozzi},
  {Riguccini}, {Rodighiero}, {Saintonge}, {Sanchez Portal}, {Santini}, {Shao},
  {Sturm}, {Tacconi}, {Valtchanov}, {Wetzstein}, \& {Wieprecht}}]{Lutz11}
{Lutz}, D., {Poglitsch}, A., {Altieri}, B., {et~al.} 2011, \aap, 532, A90

\bibitem[{{Mac Low} \& {Klessen}(2004)}]{MacLow04}
{Mac Low}, M.-M., \& {Klessen}, R.~S. 2004, Reviews of Modern Physics, 76, 125

\bibitem[{{Mac Low} {et~al.}(1998){Mac Low}, {Klessen}, {Burkert}, \&
  {Smith}}]{MacLow98}
{Mac Low}, M.-M., {Klessen}, R.~S., {Burkert}, A., \& {Smith}, M.~D. 1998,
  \prl, 80, 2754

\bibitem[{{Mac Low} {et~al.}(1989){Mac Low}, {McCray}, \& {Norman}}]{MacLow89}
{Mac Low}, M.-M., {McCray}, R., \& {Norman}, M.~L. 1989, \apj, 337, 141

\bibitem[{{Magnelli} {et~al.}(2013){Magnelli}, {Popesso}, {Berta}, {Pozzi},
  {Elbaz}, {Lutz}, {Dickinson}, {Altieri}, {Andreani}, {Aussel},
  {B{\'e}thermin}, {Bongiovanni}, {Cepa}, {Charmandaris}, {Chary}, {Cimatti},
  {Daddi}, {F{\"o}rster Schreiber}, {Genzel}, {Gruppioni}, {Harwit}, {Hwang},
  {Ivison}, {Magdis}, {Maiolino}, {Murphy}, {Nordon}, {Pannella}, {P{\'e}rez
  Garc{\'{\i}}a}, {Poglitsch}, {Rosario}, {Sanchez-Portal}, {Santini}, {Scott},
  {Sturm}, {Tacconi}, \& {Valtchanov}}]{Magnelli13}
{Magnelli}, B., {Popesso}, P., {Berta}, S., {et~al.} 2013, \aap, 553, A132

\bibitem[{{Mandelker} {et~al.}(2014){Mandelker}, {Dekel}, {Ceverino}, {Tweed},
  {Moody}, \& {Primack}}]{Mandelker14}
{Mandelker}, N., {Dekel}, A., {Ceverino}, D., {et~al.} 2014, \mnras, 443, 3675

\bibitem[{{Mason} {et~al.}(2017){Mason}, {Treu}, {Fontana}, {Jones},
  {Morishita}, {Amorin}, {Brada{\v{c}}}, {Quinn Finney}, {Grillo}, {Henry},
  {Hoag}, {Huang}, {Schmidt}, {Trenti}, \& {Vulcani}}]{Mason17}
{Mason}, C.~A., {Treu}, T., {Fontana}, A., {et~al.} 2017, \apj, 838, 14

\bibitem[{{McKee} \& {Ostriker}(2007)}]{McKee07}
{McKee}, C.~F., \& {Ostriker}, E.~C. 2007, Annual Review of Astronomy and
  Astrophysics, 45, 565

\bibitem[{{McKee} {et~al.}(2015){McKee}, {Parravano}, \&
  {Hollenbach}}]{McKee15}
{McKee}, C.~F., {Parravano}, A., \& {Hollenbach}, D.~J. 2015, \apj, 814, 13

\bibitem[{{Meng} {et~al.}(2018){Meng}, {Gnedin}, \& {Li}}]{Meng18}
{Meng}, X., {Gnedin}, O., \& {Li}, H. 2018, arXiv e-prints, arXiv:1810.06647

\bibitem[{{Meurer} {et~al.}(1996){Meurer}, {Carignan}, {Beaulieu}, \&
  {Freeman}}]{Meurer96}
{Meurer}, G.~R., {Carignan}, C., {Beaulieu}, S.~F., \& {Freeman}, K.~C. 1996,
  \aj, 111, 1551

\bibitem[{{Miller} {et~al.}(2012){Miller}, {Ellis}, {Sullivan}, {Bundy},
  {Newman}, \& {Treu}}]{Miller12}
{Miller}, S.~H., {Ellis}, R.~S., {Sullivan}, M., {et~al.} 2012, \apj, 753, 74

\bibitem[{{Mogotsi} {et~al.}(2016){Mogotsi}, {de Blok}, {Cald{\'u}-Primo},
  {Walter}, {Ianjamasimanana}, \& {Leroy}}]{Mogotsi16}
{Mogotsi}, K.~M., {de Blok}, W.~J.~G., {Cald{\'u}-Primo}, A., {et~al.} 2016,
  \aj, 151, 15

\bibitem[{{Moiseev} {et~al.}(2015){Moiseev}, {Tikhonov}, \&
  {Klypin}}]{Moiseev15}
{Moiseev}, A.~V., {Tikhonov}, A.~V., \& {Klypin}, A. 2015, \mnras, 449, 3568

\bibitem[{{Momcheva} {et~al.}(2016){Momcheva}, {Brammer}, {van Dokkum},
  {Skelton}, {Whitaker}, {Nelson}, {Fumagalli}, {Maseda}, {Leja}, {Franx},
  {Rix}, {Bezanson}, {Da Cunha}, {Dickey}, {F{\"o}rster Schreiber},
  {Illingworth}, {Kriek}, {Labb{\'e}}, {Ulf Lange}, {Lundgren}, {Magee},
  {Marchesini}, {Oesch}, {Pacifici}, {Patel}, {Price}, {Tal}, {Wake}, {van der
  Wel}, \& {Wuyts}}]{Momcheva16}
{Momcheva}, I.~G., {Brammer}, G.~B., {van Dokkum}, P.~G., {et~al.} 2016, \apjs,
  225, 27

\bibitem[{{Moster} {et~al.}(2018){Moster}, {Naab}, \& {White}}]{Moster18}
{Moster}, B.~P., {Naab}, T., \& {White}, S. D.~M. 2018, \mnras, 477, 1822

\bibitem[{{Naab} \& {Ostriker}(2017)}]{Naab17}
{Naab}, T., \& {Ostriker}, J.~P. 2017, Annual Review of Astronomy and
  Astrophysics, 55, 59

\bibitem[{{Navarro} {et~al.}(1996){Navarro}, {Frenk}, \& {White}}]{NFW96}
{Navarro}, J.~F., {Frenk}, C.~S., \& {White}, S.~D.~M. 1996, \apj, 462, 563

\bibitem[{{Nelson} {et~al.}(2016){Nelson}, {van Dokkum}, {F{\"o}rster
  Schreiber}, {Franx}, {Brammer}, {Momcheva}, {Wuyts}, {Whitaker}, {Skelton},
  {Fumagalli}, {Hayward}, {Kriek}, {Labb{\'e}}, {Leja}, {Rix}, {Tacconi}, {van
  der Wel}, {van den Bosch}, {Oesch}, {Dickey}, \& {Ulf Lange}}]{NelsonE16b}
{Nelson}, E.~J., {van Dokkum}, P.~G., {F{\"o}rster Schreiber}, N.~M., {et~al.}
  2016, \apj, 828, 27

\bibitem[{{Newman} {et~al.}(2013){Newman}, {Genzel}, {F{\"o}rster Schreiber},
  {Shapiro Griffin}, {Mancini}, {Lilly}, {Renzini}, {Bouch{\'e}}, {Burkert},
  {Buschkamp}, {Carollo}, {Cresci}, {Davies}, {Eisenhauer}, {Genel}, {Hicks},
  {Kurk}, {Lutz}, {Naab}, {Peng}, {Sternberg}, {Tacconi}, {Wuyts}, {Zamorani},
  \& {Vergani}}]{Newman13}
{Newman}, S.~F., {Genzel}, R., {F{\"o}rster Schreiber}, N.~M., {et~al.} 2013,
  \apj, 767, 104

\bibitem[{{Noguchi}(1999)}]{Noguchi99}
{Noguchi}, M. 1999, \apj, 514, 77

\bibitem[{{Noordermeer}(2008)}]{Noordermeer08}
{Noordermeer}, E. 2008, \mnras, 385, 1359

\bibitem[{{Obreschkow} {et~al.}(2015){Obreschkow}, {Glazebrook}, {Bassett},
  {Fisher}, {Abraham}, {Wisnioski}, {Green}, {McGregor}, {Damjanov}, {Popping},
  \& {J{\o}rgensen}}]{Obreschkow15}
{Obreschkow}, D., {Glazebrook}, K., {Bassett}, R., {et~al.} 2015, \apj, 815, 97

\bibitem[{{Ostriker} \& {Shetty}(2011)}]{OstrikerE11}
{Ostriker}, E.~C., \& {Shetty}, R. 2011, \apj, 731, 41

\bibitem[{{Patr{\'\i}cio} {et~al.}(2018){Patr{\'\i}cio}, {Richard}, {Carton},
  {Contini}, {Epinat}, {Brinchmann}, {Schmidt}, {Krajnovi{\'c}}, {Bouch{\'e}},
  {Weilbacher}, {Pell{\'o}}, {Caruana}, {Maseda}, {Finley}, {Bauer},
  {Martinez}, {Mahler}, {Lagattuta}, {Cl{\'e}ment}, {Soucail}, \&
  {Wisotzki}}]{Patricio18}
{Patr{\'\i}cio}, V., {Richard}, J., {Carton}, D., {et~al.} 2018, \mnras, 477,
  18

\bibitem[{{Petric} \& {Rupen}(2007)}]{Petric07}
{Petric}, A.~O., \& {Rupen}, M.~P. 2007, \aj, 134, 1952

\bibitem[{{Pillepich} {et~al.}(2019){Pillepich}, {Nelson}, {Springel},
  {Pakmor}, {Torrey}, {Weinberger}, {Vogelsberger}, {Marinacci}, {Genel}, {van
  der Wel}, \& {Hernquist}}]{Pillepich19}
{Pillepich}, A., {Nelson}, D., {Springel}, V., {et~al.} 2019, arXiv e-prints,
  arXiv:1902.05553

\bibitem[{{Price} {et~al.}(2016){Price}, {Kriek}, {Shapley}, {Reddy},
  {Freeman}, {Coil}, {de Groot}, {Shivaei}, {Siana}, {Azadi}, {Barro},
  {Mobasher}, {Sanders}, \& {Zick}}]{Price16}
{Price}, S.~H., {Kriek}, M., {Shapley}, A.~E., {et~al.} 2016, \apj, 819, 80

\bibitem[{{Price} {et~al.}(2019){Price}, {Kriek}, {Barro}, {Shapley}, {Reddy},
  {Freeman}, {Coil}, {Shivaei}, {Azadi}, {de Groot}, {Siana}, {Mobasher}, {Sand
  ers}, {Leung}, {Fetherolf}, {Zick}, {{\"U}bler}, \& {Schreiber}}]{Price19}
{Price}, S.~H., {Kriek}, M., {Barro}, G., {et~al.} 2019, arXiv e-prints,
  arXiv:1902.09554

\bibitem[{{Rathaus} \& {Sternberg}(2016)}]{Rathaus16}
{Rathaus}, B., \& {Sternberg}, A. 2016, \mnras, 458, 3168

\bibitem[{{Rix} \& {Bovy}(2013)}]{Rix13}
{Rix}, H.-W., \& {Bovy}, J. 2013, Astronomy and Astrophysics Review, 21, 61

\bibitem[{{Saintonge} {et~al.}(2013){Saintonge}, {Lutz}, {Genzel}, {Magnelli},
  {Nordon}, {Tacconi}, {Baker}, {Bandara}, {Berta}, {F{\"o}rster Schreiber},
  {Poglitsch}, {Sturm}, {Wuyts}, \& {Wuyts}}]{Saintonge13}
{Saintonge}, A., {Lutz}, D., {Genzel}, R., {et~al.} 2013, \apj, 778, 2

\bibitem[{{Sharples} {et~al.}(2013){Sharples}, {Bender}, {Agudo Berbel},
  {Bezawada}, {Castillo}, {Cirasuolo}, {Davidson}, {Davies}, {Dubbeldam},
  {Fairley}, {Finger}, {F{\"o}rster Schreiber}, {Gonte}, {Hess}, {Jung},
  {Lewis}, {Lizon}, {Muschielok}, {Pasquini}, {Pirard}, {Popovic}, {Ramsay},
  {Rees}, {Richter}, {Riquelme}, {Rodrigues}, {Saviane}, {Schlichter},
  {Schmidtobreick}, {Segovia}, {Smette}, {Szeifert}, {van Kesteren}, {Wegner},
  \& {Wiezorrek}}]{Sharples13}
{Sharples}, R., {Bender}, R., {Agudo Berbel}, A., {et~al.} 2013, The Messenger,
  151, 21

\bibitem[{{Sharples} {et~al.}(2004){Sharples}, {Bender}, {Lehnert}, {Ramsay
  Howat}, {Bremer}, {Davies}, {Genzel}, {Hofmann}, {Ivison}, {Saglia}, \&
  {Thatte}}]{Sharples04}
{Sharples}, R.~M., {Bender}, R., {Lehnert}, M.~D., {et~al.} 2004, in Society of
  Photo-Optical Instrumentation Engineers (SPIE) Conference Series, Vol. 5492,
  Ground-based Instrumentation for Astronomy, ed. A.~F.~M. {Moorwood} \&
  M.~{Iye}, 1179--1186

\bibitem[{{Shetty} \& {Ostriker}(2012)}]{Shetty12}
{Shetty}, R., \& {Ostriker}, E.~C. 2012, \apj, 754, 2

\bibitem[{{Shields}(1990)}]{Shields90}
{Shields}, G.~A. 1990, Annual Review of Astronomy and Astrophysics, 28, 525

\bibitem[{{Silk}(2001)}]{Silk01}
{Silk}, J. 2001, \mnras, 324, 313

\bibitem[{{Simons} {et~al.}(2016){Simons}, {Kassin}, {Trump}, {Weiner},
  {Heckman}, {Barro}, {Koo}, {Guo}, {Pacifici}, {Koekemoer}, \&
  {Stephens}}]{Simons16}
{Simons}, R.~C., {Kassin}, S.~A., {Trump}, J.~R., {et~al.} 2016, \apj, 830, 14

\bibitem[{{Simons} {et~al.}(2017){Simons}, {Kassin}, {Weiner}, {Faber},
  {Trump}, {Heckman}, {Koo}, {Pacifici}, {Primack}, {Snyder}, \& {de la
  Vega}}]{Simons17}
{Simons}, R.~C., {Kassin}, S.~A., {Weiner}, B.~J., {et~al.} 2017, \apj, 843, 46

\bibitem[{{Skelton} {et~al.}(2014){Skelton}, {Whitaker}, {Momcheva}, {Brammer},
  {van Dokkum}, {Labb{\'e}}, {Franx}, {van der Wel}, {Bezanson}, {Da Cunha},
  {Fumagalli}, {F{\"o}rster Schreiber}, {Kriek}, {Leja}, {Lundgren}, {Magee},
  {Marchesini}, {Maseda}, {Nelson}, {Oesch}, {Pacifici}, {Patel}, {Price},
  {Rix}, {Tal}, {Wake}, \& {Wuyts}}]{Skelton14}
{Skelton}, R.~E., {Whitaker}, K.~E., {Momcheva}, I.~G., {et~al.} 2014, \apjs,
  214, 24

\bibitem[{{Stone} {et~al.}(1998){Stone}, {Ostriker}, \& {Gammie}}]{Stone98}
{Stone}, J.~M., {Ostriker}, E.~C., \& {Gammie}, C.~F. 1998, \apj, 508, L99

\bibitem[{{Stott} {et~al.}(2016){Stott}, {Swinbank}, {Johnson}, {Tiley},
  {Magdis}, {Bower}, {Bunker}, {Bureau}, {Harrison}, {Jarvis}, {Sharples},
  {Smail}, {Sobral}, {Best}, \& {Cirasuolo}}]{Stott16}
{Stott}, J.~P., {Swinbank}, A.~M., {Johnson}, H.~L., {et~al.} 2016, \mnras,
  457, 1888

\bibitem[{{Sun} {et~al.}(2018){Sun}, {Leroy}, {Schruba}, {Rosolowsky},
  {Hughes}, {Kruijssen}, {Meidt}, {Schinnerer}, {Blanc}, {Bigiel}, {Bolatto},
  {Chevance}, {Groves}, {Herrera}, {Hygate}, {Pety}, {Querejeta}, {Usero}, \&
  {Utomo}}]{Sun18}
{Sun}, J., {Leroy}, A.~K., {Schruba}, A., {et~al.} 2018, \apj, 860, 172

\bibitem[{{Swinbank} {et~al.}(2012{\natexlab{a}}){Swinbank}, {Smail}, {Sobral},
  {Theuns}, {Best}, \& {Geach}}]{Swinbank12b}
{Swinbank}, A.~M., {Smail}, I., {Sobral}, D., {et~al.} 2012{\natexlab{a}},
  \apj, 760, 130

\bibitem[{{Swinbank} {et~al.}(2012{\natexlab{b}}){Swinbank}, {Sobral}, {Smail},
  {Geach}, {Best}, {McCarthy}, {Crain}, \& {Theuns}}]{Swinbank12}
{Swinbank}, A.~M., {Sobral}, D., {Smail}, I., {et~al.} 2012{\natexlab{b}},
  \mnras, 426, 935

\bibitem[{{Swinbank} {et~al.}(2011){Swinbank}, {Papadopoulos}, {Cox}, {Krips},
  {Ivison}, {Smail}, {Thomson}, {Neri}, {Richard}, \& {Ebeling}}]{Swinbank11}
{Swinbank}, A.~M., {Papadopoulos}, P.~P., {Cox}, P., {et~al.} 2011, \apj, 742,
  11

\bibitem[{{Swinbank} {et~al.}(2017){Swinbank}, {Harrison}, {Trayford},
  {Schaller}, {Smail}, {Schaye}, {Theuns}, {Smit}, {Alexander}, {Bacon},
  {Bower}, {Contini}, {Crain}, {de Breuck}, {Decarli}, {Epinat}, {Fumagalli},
  {Furlong}, {Galametz}, {Johnson}, {Lagos}, {Richard}, {Vernet}, {Sharples},
  {Sobral}, \& {Stott}}]{Swinbank17}
{Swinbank}, A.~M., {Harrison}, C.~M., {Trayford}, J., {et~al.} 2017, \mnras,
  467, 3140

\bibitem[{{Tacchella} {et~al.}(2015{\natexlab{a}}){Tacchella}, {Carollo},
  {Renzini}, {Schreiber}, {Lang}, {Wuyts}, {Cresci}, {Dekel}, {Genzel},
  {Lilly}, {Mancini}, {Newman}, {Onodera}, {Shapley}, {Tacconi}, {Woo}, \&
  {Zamorani}}]{Tacchella15}
{Tacchella}, S., {Carollo}, C.~M., {Renzini}, A., {et~al.} 2015{\natexlab{a}},
  Science, 348, 314

\bibitem[{{Tacchella} {et~al.}(2015{\natexlab{b}}){Tacchella}, {Lang},
  {Carollo}, {F{\"o}rster Schreiber}, {Renzini}, {Shapley}, {Wuyts}, {Cresci},
  {Genzel}, {Lilly}, {Mancini}, {Newman}, {Tacconi}, {Zamorani}, {Davies},
  {Kurk}, \& {Pozzetti}}]{Tacchella15a}
{Tacchella}, S., {Lang}, P., {Carollo}, C.~M., {et~al.} 2015{\natexlab{b}},
  \apj, 802, 101

\bibitem[{{Tacconi} {et~al.}(2013){Tacconi}, {Neri}, {Genzel}, {Combes},
  {Bolatto}, {Cooper}, {Wuyts}, {Bournaud}, {Burkert}, {Comerford}, {Cox},
  {Davis}, {F{\"o}rster Schreiber}, {Garc{\'\i}a-Burillo}, {Gracia-Carpio},
  {Lutz}, {Naab}, {Newman}, {Omont}, {Saintonge}, {Shapiro Griffin}, {Shapley},
  {Sternberg}, \& {Weiner}}]{Tacconi13}
{Tacconi}, L.~J., {Neri}, R., {Genzel}, R., {et~al.} 2013, \apj, 768, 74

\bibitem[{{Tacconi} {et~al.}(2018){Tacconi}, {Genzel}, {Saintonge}, {Combes},
  {Garc{\'{\i}}a-Burillo}, {Neri}, {Bolatto}, {Contini}, {F{\"o}rster
  Schreiber}, {Lilly}, {Lutz}, {Wuyts}, {Accurso}, {Boissier}, {Boone},
  {Bouch{\'e}}, {Bournaud}, {Burkert}, {Carollo}, {Cooper}, {Cox}, {Feruglio},
  {Freundlich}, {Herrera-Camus}, {Juneau}, {Lippa}, {Naab}, {Renzini},
  {Salome}, {Sternberg}, {Tadaki}, {{\"U}bler}, {Walter}, {Weiner}, \&
  {Weiss}}]{Tacconi18}
{Tacconi}, L.~J., {Genzel}, R., {Saintonge}, A., {et~al.} 2018, \apj, 853, 179

\bibitem[{{Tadaki} {et~al.}(2018){Tadaki}, {Iono}, {Yun}, {Aretxaga},
  {Hatsukade}, {Hughes}, {Ikarashi}, {Izumi}, {Kawabe}, {Kohno}, {Lee},
  {Matsuda}, {Nakanishi}, {Saito}, {Tamura}, {Ueda}, {Umehata}, {Wilson},
  {Michiyama}, {Ando}, \& {Kamieneski}}]{Tadaki18}
{Tadaki}, K., {Iono}, D., {Yun}, M.~S., {et~al.} 2018, \nat, 560, 613

\bibitem[{{Tamburro} {et~al.}(2009){Tamburro}, {Rix}, {Leroy}, {Mac Low},
  {Walter}, {Kennicutt}, {Brinks}, \& {de Blok}}]{Tamburro09}
{Tamburro}, D., {Rix}, H.~W., {Leroy}, A.~K., {et~al.} 2009, \aj, 137, 4424

\bibitem[{{Toomre}(1964)}]{Toomre64}
{Toomre}, A. 1964, \apj, 139, 1217

\bibitem[{{Tully} \& {Fisher}(1977)}]{Tully77}
{Tully}, R.~B., \& {Fisher}, J.~R. 1977, \aap, 54, 661

\bibitem[{{Turner} {et~al.}(2017){Turner}, {Cirasuolo}, {Harrison}, {McLure},
  {Dunlop}, {Swinbank}, {Johnson}, {Sobral}, {Matthee}, \&
  {Sharples}}]{Turner17}
{Turner}, O.~J., {Cirasuolo}, M., {Harrison}, C.~M., {et~al.} 2017, \mnras,
  471, 1280

\bibitem[{{{\"U}bler} {et~al.}(2017){{\"U}bler}, {F{\"o}rster Schreiber},
  {Genzel}, {Wisnioski}, {Wuyts}, {Lang}, {Naab}, {Burkert}, {van Dokkum},
  {Tacconi}, {Wilman}, {Fossati}, {Mendel}, {Beifiori}, {Belli}, {Bender},
  {Brammer}, {Chan}, {Davies}, {Fabricius}, {Galametz}, {Lutz}, {Momcheva},
  {Nelson}, {Saglia}, {Seitz}, \& {Tadaki}}]{Uebler17}
{{\"U}bler}, H., {F{\"o}rster Schreiber}, N.~M., {Genzel}, R., {et~al.} 2017,
  \apj, 842, 121

\bibitem[{{{\"U}bler} {et~al.}(2018){{\"U}bler}, {Genzel}, {Tacconi},
  {F{\"o}rster Schreiber}, {Neri}, {Contursi}, {Belli}, {Nelson}, {Lang},
  {Shimizu}, {Davies}, {Herrera-Camus}, {Lutz}, {Plewa}, {Price}, {Schuster},
  {Sternberg}, {Tadaki}, {Wisnioski}, \& {Wuyts}}]{Uebler18}
{{\"U}bler}, H., {Genzel}, R., {Tacconi}, L.~J., {et~al.} 2018, \apjl, 854, L24

\bibitem[{{Utomo} {et~al.}(2019){Utomo}, {Blitz}, \& {Falgarone}}]{Utomo19}
{Utomo}, D., {Blitz}, L., \& {Falgarone}, E. 2019, \apj, 871, 17

\bibitem[{{van der Kruit} \& {Freeman}(1984)}]{vdKruit84}
{van der Kruit}, P.~C., \& {Freeman}, K.~C. 1984, \apj, 278, 81

\bibitem[{{van der Kruit} \& {Freeman}(2011)}]{vdKruit11}
---. 2011, \araa, 49, 301

\bibitem[{{van der Wel} {et~al.}(2012){van der Wel}, {Bell}, {H{\"a}ussler},
  {McGrath}, {Chang}, {Guo}, {McIntosh}, {Rix}, {Barden}, {Cheung}, {Faber},
  {Ferguson}, {Galametz}, {Grogin}, {Hartley}, {Kartaltepe}, {Kocevski},
  {Koekemoer}, {Lotz}, {Mozena}, {Peth}, \& {Peng}}]{vdWel12}
{van der Wel}, A., {Bell}, E.~F., {H{\"a}ussler}, B., {et~al.} 2012, \apjs,
  203, 24

\bibitem[{{van der Wel} {et~al.}(2014{\natexlab{a}}){van der Wel}, {Franx},
  {van Dokkum}, {Skelton}, {Momcheva}, {Whitaker}, {Brammer}, {Bell}, {Rix},
  {Wuyts}, {Ferguson}, {Holden}, {Barro}, {Koekemoer}, {Chang}, {McGrath},
  {H{\"a}ussler}, {Dekel}, {Behroozi}, {Fumagalli}, {Leja}, {Lundgren},
  {Maseda}, {Nelson}, {Wake}, {Patel}, {Labb{\'e}}, {Faber}, {Grogin}, \&
  {Kocevski}}]{vdWel14a}
{van der Wel}, A., {Franx}, M., {van Dokkum}, P.~G., {et~al.}
  2014{\natexlab{a}}, \apj, 788, 28

\bibitem[{{van der Wel} {et~al.}(2014{\natexlab{b}}){van der Wel}, {Chang},
  {Bell}, {Holden}, {Ferguson}, {Giavalisco}, {Rix}, {Skelton}, {Whitaker},
  {Momcheva}, {Brammer}, {Kassin}, {Martig}, {Dekel}, {Ceverino}, {Koo},
  {Mozena}, {van Dokkum}, {Franx}, {Faber}, \& {Primack}}]{vdWel14b}
{van der Wel}, A., {Chang}, Y.-Y., {Bell}, E.~F., {et~al.} 2014{\natexlab{b}},
  \apj, 792, L6

\bibitem[{{Varidel} {et~al.}(2016){Varidel}, {Pracy}, {Croom}, {Owers}, \&
  {Sadler}}]{Varidel16}
{Varidel}, M., {Pracy}, M., {Croom}, S., {Owers}, M.~S., \& {Sadler}, E. 2016,
  Publications of the Astronomical Society of Australia, 33, e006

\bibitem[{{Varidel} {et~al.}(2019){Varidel}, {Croom}, {Lewis}, {Brewer}, {Di
  Teodoro}, {Bland -Hawthorn}, {Bryant}, {Federrath}, {Foster}, {Glazebrook},
  {Goodwin}, {Groves}, {Hopkins}, {Lawrence}, {L{\'o}pez-S{\'a}nchez},
  {Medling}, {Owers}, {Richards}, {Scalzo}, {Scott}, {Sweet}, {Taranu}, \& {van
  de Sande}}]{Varidel19}
{Varidel}, M.~R., {Croom}, S.~M., {Lewis}, G.~F., {et~al.} 2019, arXiv
  e-prints, arXiv:1903.03121

\bibitem[{{Walter} {et~al.}(2008){Walter}, {Brinks}, {de Blok}, {Bigiel},
  {Kennicutt}, {Thornley}, \& {Leroy}}]{Walter08}
{Walter}, F., {Brinks}, E., {de Blok}, W.~J.~G., {et~al.} 2008, \aj, 136, 2563

\bibitem[{{Wang} \& {Silk}(1994)}]{Wang94}
{Wang}, B., \& {Silk}, J. 1994, \apj, 427, 759

\bibitem[{{Wang} {et~al.}(2010){Wang}, {Klessen}, {Dullemond}, {van den Bosch},
  \& {Fuchs}}]{Wang10}
{Wang}, H.-H., {Klessen}, R.~S., {Dullemond}, C.~P., {van den Bosch}, F.~C., \&
  {Fuchs}, B. 2010, \mnras, 407, 705

\bibitem[{{Weiner} {et~al.}(2006){Weiner}, {Willmer}, {Faber}, {Melbourne},
  {Kassin}, {Phillips}, {Harker}, {Metevier}, {Vogt}, \& {Koo}}]{Weiner06a}
{Weiner}, B.~J., {Willmer}, C.~N.~A., {Faber}, S.~M., {et~al.} 2006, \apj, 653,
  1027

\bibitem[{{Whitaker} {et~al.}(2014){Whitaker}, {Franx}, {Leja}, {van Dokkum},
  {Henry}, {Skelton}, {Fumagalli}, {Momcheva}, {Brammer}, {Labb{\'e}},
  {Nelson}, \& {Rigby}}]{Whitaker14}
{Whitaker}, K.~E., {Franx}, M., {Leja}, J., {et~al.} 2014, \apj, 795, 104

\bibitem[{{White} {et~al.}(2017){White}, {Fisher}, {Murray}, {Glazebrook},
  {Abraham}, {Bolatto}, {Green}, {Mentuch Cooper}, \& {Obreschkow}}]{White17}
{White}, H.~A., {Fisher}, D.~B., {Murray}, N., {et~al.} 2017, \apj, 846, 35

\bibitem[{{Wilson} {et~al.}(2011){Wilson}, {Warren}, {Irwin}, {Knapen},
  {Israel}, {Serjeant}, {Attewell}, {Bendo}, {Brinks}, {Butner}, {Clements},
  {Leech}, {Matthews}, {M{\"u}hle}, {Mortier}, {Parkin}, {Petitpas}, {Tan},
  {Tilanus}, {Usero}, {Vaccari}, {van der Werf}, {Wiegert}, \&
  {Zhu}}]{Wilson11}
{Wilson}, C.~D., {Warren}, B.~E., {Irwin}, J., {et~al.} 2011, \mnras, 410, 1409

\bibitem[{{Wisnioski} {et~al.}(2012){Wisnioski}, {Glazebrook}, {Blake},
  {Poole}, {Green}, {Wyder}, \& {Martin}}]{Wisnioski12}
{Wisnioski}, E., {Glazebrook}, K., {Blake}, C., {et~al.} 2012, \mnras, 422,
  3339

\bibitem[{{Wisnioski} {et~al.}(2011){Wisnioski}, {Glazebrook}, {Blake},
  {Wyder}, {Martin}, {Poole}, {Sharp}, {Couch}, {Kacprzak}, {Brough},
  {Colless}, {Contreras}, {Croom}, {Croton}, {Davis}, {Drinkwater}, {Forster},
  {Gilbank}, {Gladders}, {Jelliffe}, {Jurek}, {Li}, {Madore}, {Pimbblet},
  {Pracy}, {Woods}, \& {Yee}}]{Wisnioski11}
---. 2011, \mnras, 417, 2601

\bibitem[{{Wisnioski} {et~al.}(2015){Wisnioski}, {F{\"o}rster Schreiber},
  {Wuyts}, {Wuyts}, {Bandara}, {Wilman}, {Genzel}, {Bender}, {Davies},
  {Fossati}, {Lang}, {Mendel}, {Beifiori}, {Brammer}, {Chan}, {Fabricius},
  {Fudamoto}, {Kulkarni}, {Kurk}, {Lutz}, {Nelson}, {Momcheva}, {Rosario},
  {Saglia}, {Seitz}, {Tacconi}, \& {van Dokkum}}]{Wisnioski15}
{Wisnioski}, E., {F{\"o}rster Schreiber}, N.~M., {Wuyts}, S., {et~al.} 2015,
  \apj, 799, 209

\bibitem[{{Wisnioski} {et~al.}(2018){Wisnioski}, {Mendel}, {F{\"o}rster
  Schreiber}, {Genzel}, {Wilman}, {Wuyts}, {Belli}, {Beifiori}, {Bender},
  {Brammer}, {Chan}, {Davies}, {Davies}, {Fabricius}, {Fossati}, {Galametz},
  {Lang}, {Lutz}, {Nelson}, {Momcheva}, {Rosario}, {Saglia}, {Tacconi},
  {Tadaki}, {{\"U}bler}, \& {van Dokkum}}]{Wisnioski18}
{Wisnioski}, E., {Mendel}, J.~T., {F{\"o}rster Schreiber}, N.~M., {et~al.}
  2018, \apj, 855, 97

\bibitem[{{Wong} {et~al.}(2009){Wong}, {Hughes}, {Fukui}, {Kawamura}, {Mizuno},
  {Ott}, {Muller}, {Pineda}, {Welty}, {Kim}, {Mizuno}, {Murai}, \&
  {Onishi}}]{Wong09}
{Wong}, T., {Hughes}, A., {Fukui}, Y., {et~al.} 2009, \apj, 696, 370

\bibitem[{{Wuyts} {et~al.}(2011){Wuyts}, {F{\"o}rster Schreiber}, {Lutz},
  {Nordon}, {Berta}, {Altieri}, {Andreani}, {Aussel}, {Bongiovanni}, {Cepa},
  {Cimatti}, {Daddi}, {Elbaz}, {Genzel}, {Koekemoer}, {Magnelli}, {Maiolino},
  {McGrath}, {P{\'e}rez Garc{\'{\i}}a}, {Poglitsch}, {Popesso}, {Pozzi},
  {Sanchez-Portal}, {Sturm}, {Tacconi}, \& {Valtchanov}}]{WuytsS11a}
{Wuyts}, S., {F{\"o}rster Schreiber}, N.~M., {Lutz}, D., {et~al.} 2011, \apj,
  738, 106

\bibitem[{{Wuyts} {et~al.}(2012){Wuyts}, {F{\"o}rster Schreiber}, {Genzel},
  {Guo}, {Barro}, {Bell}, {Dekel}, {Faber}, {Ferguson}, {Giavalisco}, {Grogin},
  {Hathi}, {Huang}, {Kocevski}, {Koekemoer}, {Koo}, {Lotz}, {Lutz}, {McGrath},
  {Newman}, {Rosario}, {Saintonge}, {Tacconi}, {Weiner}, \& {van der
  Wel}}]{WuytsS12}
{Wuyts}, S., {F{\"o}rster Schreiber}, N.~M., {Genzel}, R., {et~al.} 2012, \apj,
  753, 114

\bibitem[{{Wuyts} {et~al.}(2016){Wuyts}, {F{\"o}rster Schreiber}, {Wisnioski},
  {Genzel}, {Burkert}, {Bandara}, {Beifiori}, {Belli}, {Bender}, {Brammer},
  {Chan}, {Davies}, {Fossati}, {Galametz}, {Kulkarni}, {Lang}, {Lutz},
  {Mendel}, {Momcheva}, {Naab}, {Nelson}, {Saglia}, {Seitz}, {Tacconi},
  {Tadaki}, {{\"U}bler}, {van Dokkum}, {Wilman}, \& {Wuyts}}]{WuytsS16}
{Wuyts}, S., {F{\"o}rster Schreiber}, N.~M., {Wisnioski}, E., {et~al.} 2016,
  \apj, 831, 149

\bibitem[{{Yoachim} \& {Dalcanton}(2006)}]{Yoachim06}
{Yoachim}, P., \& {Dalcanton}, J.~J. 2006, \aj, 131, 226

\bibitem[{{Zhou} {et~al.}(2017){Zhou}, {Federrath}, {Yuan}, {Bian}, {Medling},
  {Shi}, {Bland- Hawthorn}, {Bryant}, {Brough}, {Catinella}, {Croom},
  {Goodwin}, {Goldstein}, {Green}, {Konstantopoulos}, {Lawrence}, {Owers},
  {Richards}, \& {Sanchez}}]{Zhou17}
{Zhou}, L., {Federrath}, C., {Yuan}, T., {et~al.} 2017, \mnras, 470, 4573

\end{thebibliography}



\end{document}